\documentclass {aastex63}

\shorttitle{HIGH-RESOLUTION X-RAY SPECTROSCOPY OF SNR 1987A}
\shortauthors{Ravi et al.}

\graphicspath{{./}{figures/}}

\usepackage{tablefootnote}
\usepackage{longtable}
\usepackage{xcolor}
\usepackage{graphics}
\usepackage{epsf}
\usepackage{ragged2e}

\begin{document}

\title{Spectral Evolution of the X-Ray Remnant of SN 1987A: A High-Resolution \textit{Chandra} HETG Study}

\correspondingauthor{Aravind Pazhayath Ravi}
\email{aravind.pazhayathravi@mavs.uta.edu}

\author{Aravind P. Ravi}
\affiliation{Department of Physics, University of Texas at Arlington, Box 19059, 
Arlington, TX 76019, USA} 

\author{Sangwook Park}
\affiliation{Department of Physics, University of Texas at Arlington,
 Box 19059, 
Arlington, TX 76019, USA} 

\author{Svetozar A. Zhekov}
\affiliation{Institute of Astronomy and National Astronomical Observatory (Bulgarian Academy of Sciences), \\
 72 Tsarigradsko Chaussee Blvd.,
Sofia 1784, Bulgaria}

\author{Marco Miceli}
\affiliation{Dipartimento di Fisica e Chimica, Universita degli Studi di Palermo,
 Piazza del Parlamento 1,
90134 Palermo, Italy}
\affiliation{INAF-Osservatorio Astronomico di Palermo,
 Piazza del Parlamento 1,
90134 Palermo, Italy}

\author{Salvatore Orlando}
\affiliation{INAF-Osservatorio Astronomico di Palermo,
 Piazza del Parlamento 1,
90134 Palermo, Italy}

\author{Kari A. Frank}
\affiliation{CIERA, Northwestern University,
 1800 Sherman Ave., Evanston, IL 60201, USA}

\author{David N. Burrows}
\affiliation{Department of Astronomy and Astrophysics, The Pennsylvannia State University,
University Park, PA 16802, USA}

\begin{abstract}

Based on observations with the
\textit{Chandra} X-ray Observatory, we present the latest spectral evolution of the X-ray
remnant of SN 1987A (SNR 1987A). We present a high-resolution
spectroscopic analysis using our new deep ($\sim$312 ks) \textit{Chandra} HETG
observation taken in March 2018, as well as archival \textit{Chandra} gratings spectroscopic data taken in 2004, 2007, and 2011 with similarly deep exposures ($\sim$170 - 350 ks). We perform detailed spectral model fits to quantify changing plasma conditions over the last 14 years. Recent changes in electron temperatures and volume emission measures suggest that the shocks moving through the inner ring have started interacting with less dense circumstellar material, probably beyond the inner ring. We find significant changes in the X-ray line flux ratios (among H- and He-like Si and Mg ions) in 2018, consistent with changes in the thermal conditions of the X-ray emitting plasma that we infer based on the broadband spectral analysis. Post-shock electron temperatures suggested by line flux ratios are in the range $\sim$0.8 - 2.5 keV as of 2018. We do not yet observe any evidence of substantial abundance enhancement, suggesting that the X-ray emission component from the reverse-shocked metal-rich ejecta is not yet significant in the observed X-ray spectrum.
\end{abstract}


\section{Introduction} 

Supernova (SN) 1987A is a core collapse SN discovered on 24 February 1987 in the Large Magellanic Cloud (LMC). It is the nearest (distance $\sim$51 kpc) and hence the brightest observed SN since Kepler’s SN in 1604 AD (\citealp{doi:10.1146/annurev.aa.27.090189.003213, doi:10.1146/annurev.aa.31.090193.001135, doi:10.1146/annurev-astro-082615-105405} for detailed reviews). Owing to its proximity, SN 1987A can be detected and resolved even \small$>$30 years after the explosion. It is a unique astrophysical laboratory for studying the birth of a supernova remnant (SNR) and a neutron star. The excellent spatial and spectral resolution of \textit{Chandra} has been used to study the photometric, morphological, and spectroscopic evolution of SNR 1987A over the last two decades. As part of our on-going \textit{Chandra} monitoring program, we have observed SNR 1987A roughly every 6 months for the past 21 years (total 48 observations as of March 2021). We have presented previous results from our \textit{Chandra} monitoring campaign of SNR 1987A in the literature \citep[e.g.,][]{2000ApJ...543L.149B, 2002ApJ...567..314P, 2004ApJ...610..275P, 2005ApJ...634L..73P, 2006ApJ...646.1001P, 2011ApJ...733L..35P, 2009ApJ...703.1752R, 2010MNRAS.407.1157Z, 2013ApJ...764...11H, 2016ApJ...829...40F, 2020ApJ...899...21B}.

X-ray emission from SNR 1987A has been dominated by the shock interaction with the dense ``inner ring", also called the equatorial ring (ER), which is the dense circumstellar material (CSM) produced by the stellar winds from the massive progenitor star during its late stages of stellar evolution before the SN explosion \citep{1991ApJ...379..659L,1991ApJ...380..575L,1995ApJ...452..680B}. As the shock entered the main body of the ER around 2004 ($\sim$6200 days after the SN), the observed soft X-ray flux (in the 0.5 - 2.0 keV energy band) showed a dramatic increase \citep{2005ApJ...634L..73P}. There was a corresponding deceleration in the expansion rate of the X-ray remnant around the same time \citep{2009ApJ...703.1752R}. Until $\sim$2015 ($\sim$10500 days after the SN), the X-ray expansion rate of the ER has stayed constant at $\sim$1600 km s$^{-1}$ \citep{2016ApJ...829...40F}. Though this regular monitoring program with \textit{Chandra} has provided us with some insights into the spatial and thermal distribution of the shocked gas, the moderate spectral resolution of the Advanced CCD Imaging Spectrometer \citep[ACIS;][]{2003SPIE.4851...28G} does not allow us to study the detailed spectral evolution of SNR 1987A.

There are two diffraction grating spectrometers aboard \textit{Chandra}, i.e., High Energy Transmission Grating \citep[HETG;][]{2000ApJ...539L..41C} and Low Energy Transmission Grating \citep[LETG;][]{2000ApJ...530L.111B}. While they are of similar designs, the HETG and LETG have higher responses at relatively shorter ($\lambda <$ 12 \AA{}) and longer ($\lambda >$ 12 \AA{}) wavelengths, respectively. Spectral resolutions of these grating spectrometers are an order of magnitude higher than that of the ACIS. Thus, the \textit{Chandra} HETG/LETG provide an excellent opportunity for a detailed spectroscopic study of SNR 1987A, complementing our high-resolution imaging observations with the ACIS. The first such observation was performed in October 1999 using HETG. X-ray emission was found to be dominated by shock-heated gas having an electron temperature \textit{kT}  $\sim$ 2.6 keV \citep{2002ApJ...574..166M}. Due to poor photon statistics ($\sim$850 total counts in the first-order dispersed spectrum), only a composite line profile was analyzed with those data, for which a radial velocity of  $v_{r}$ $\sim$3500 km s$^{-1}$ along the line of sight was estimated for the X-ray emitting hot gas \citep{2002ApJ...574..166M}. Between 2004 and 2011, there have been four deep high-resolution \textit{Chandra} gratings spectroscopic observations of SNR 1987A using HETG and LETG \citep{2005ApJ...628L.127Z, 2006ApJ...645..293Z, 2009ApJ...692.1190Z, 2008ApJ...676L.131D,  2012ApJ...752..103D}. We refer to them hereafter as LETG 2004, HETG 2007, LETG 2007, and HETG 2011.

Based on the LETG 2004 data, strong emission lines from H- and He-like ions were identified \citep{2005ApJ...628L.127Z}. For each He-like ion, the resonance (\textit{r}), intercombination (\textit{i}), and forbidden (\textit{f}) lines correspond to the electron transitions, 1\textit{s}2\textit{p} $^{1}P_{1}$ $\rightarrow$ 1\textit{s}$^{2}$ $^{1}S_{0}$, 1\textit{s}2\textit{p} $^{3}P_{1}$ and 1\textit{s}2\textit{p} $^{3}P_{2}$ $\rightarrow$ 1\textit{s}$^{2}$ $^{1}S_{0}$, and 1\textit{s}2\textit{s} $^{3}S_{1}$ $\rightarrow$ 1\textit{s}$^{2}$ $^{1}S_{0}$, respectively \citep{1969MNRAS.145..241G}. Collectively we refer to these triplet lines as He$\alpha$, hereafter. Similarly, we refer to the doublet lines from H-like ions as Ly$\alpha$, hereafter. \cite{2005ApJ...628L.127Z} measured line widths, shifts and fluxes of strong emission lines from electron transitions corresponding to Ly$\alpha$ and He$\alpha$ in abundant chemical elements (e.g., Ne, Mg, Si). They inferred that the observed X-ray emission originates from the hot gas interacting with shocks with velocities in the range of 300 - 1700 km s$^{-1}$. Such an ensemble of shocks was the result of the blast wave interacting with the complex density distribution through the ER. The broadband X-ray spectrum was fitted with a two-component shock model to derive thermal conditions of the X-ray emitting plasma \citep{2006ApJ...645..293Z}. The estimated electron temperatures based on the spectral model fits to the broadband spectrum of the LETG 2007 data was higher than the ion temperatures inferred from the individual emission line widths \citep{2009ApJ...692.1190Z}. These high electron temperatures were attributed to the X-ray emission coming from gas that has been shocked multiple times, both by the blast wave and the shocks reflected off the clumpy ER of SNR 1987A \citep{2009ApJ...692.1190Z}. Based on the HETG 2011 dataset, the observed X-ray spectra were modeled as the weighted sum of the nonequilibrium ionization (NEI) emission from two simple 1-D hydrodynamic simulations to reproduce all the observed radii and light curves \citep{2012ApJ...752..103D}.

SNR 1987A has also been observed several times with \textit{XMM-Newton} \citep{2006A&A...460..811H, 2008ApJ...676..361H, 2010A&A...515A...5S, 2012A&A...548L...3M}. The \textit{XMM-Newton} data showed evidence for Fe K lines ($E \sim$ 6.6 keV), whose candidate origins might include the reverse-shocked Fe-rich ejecta \citep[e.g.,][]{Sun_2021}. Recent 3-D hydrodynamic (HD) and magnetohydrodynamic (MHD) models of SN 1987A also suggested that the reverse shock would soon ($\sim$35 years after the SN) begin heating the heavy elements contained in the stellar ejecta \citep{2015ApJ...810..168O, 2019A&A...622A..73O, 2020A&A...636A..22O}. Our \textit{Chandra} ACIS monitoring observations have suggested that the blast wave has now started leaving the dense ER \citep{2016ApJ...829...40F}. Meanwhile, recent optical observations with the Hubble Space Telescope (\textit{HST}) have observed diffuse spot-like emission features since $\sim$2013 (day $\sim$9500), outside the ER \citep{2015ApJ...806L..19F,2019ApJ...886..147L}. Infrared observations with \textit{Spitzer} (at 3.6 and 4.5 $\mu m$) have also noted that the emission from ER started declining since $\sim$2010 (day $\sim$8500) \citep{2016AJ....151...62A, 2020ApJ...890....2A}. At radio wavelengths, a re-acceleration of the blast wave since $\sim$2012 (day $\sim$9300) was reported \citep{2018ApJ...867...65C}. These multi-wavelength observations show that the remnant of SN 1987A is now entering a new phase, where the blast wave has started probing outside the ER, while the reverse shock may be closing in on the metal-rich ejecta towards the center. Recently, \textit{NuSTAR} observations of SNR 1987A have showed evidence for hard X-ray emission up to \textit{E} $\sim$ 20 keV \citep{Greco_2021, 2021ApJ...916...76A}. While its origins are still unclear, either nonthermal \citep{Greco_2021} or thermal \citep{2021ApJ...916...76A} origins have been suggested.

In this work, we report on the results from the high-resolution \textit{Chandra} HETG spectroscopy of SNR 1987A based on new observations taken in March 2018 with a deep $\sim$312 ks exposure. In Section \ref{2}, we describe our observations and methods employed for reducing the data. In Section \ref{3}, we describe our spectral models for the X-ray emitting gas, present analysis methods and our fit results for both the broadband spectra and the individual line profiles. In Section \ref{4}, we discuss a physical picture of the evolution, based on our results from Section \ref{3}. Finally in Section \ref{5}, we present our conclusions.\\

\section{Observations and Data Reduction \label{2}} 

In 2018 we observed SNR 1987A for $\sim$312 ks with the HETG spectrometer aboard \textit{Chandra} (HETG 2018, hereafter). Utilizing the ACIS-S + HETG configuration, our observation was split over a series of 13 sequences (\textit{Chandra} ObsIDs: 20322, 20323, 20927, 21037, 21038, 21042, 21043, 21044, 21049, 21050, 21051, 21052, 21053) between March 19 and April 2. The roll angles were roughly between 254$^\circ$ and  264$^\circ$ (which are similar to the previous deep HETG observations of SNR 1987A). Therefore, the dispersion axis was aligned approximately with the minor axis (north - south) of the (apparently) elliptical ER of SNR 1987A.

We generated the grating ancillary response functions (gARFs) and grating redistribution matrix files (gRMFs) for all spectra using the \verb|chandra_repro| script, part of the CIAO 4.13 data analysis software \citep{2006SPIE.6270E..1VF}, along with \textit{Chandra} calibration database CALDB v.4.9.4. After the standard HETG data reduction, the total effective exposure is $\sim$312 ks. Using the CIAO script \verb|tgextract|, we extracted the positive and negative first-order HETG spectra for each of the thirteen observations as PHA II files (level = 2). We then extracted the source spectra from a narrow rectangular region, with a cross-dispersion angular width of 4$^{\prime \prime}$.7, while the background spectra were extracted from much wider adjacent rectangular regions with corresponding angular widths of 19$^{\prime \prime}$. We combined the resultant source spectra from each individual ObsID into one spectrum, each for the positive and negative first-order HETG arms. We analyzed the positively- and negatively-dispersed spectra separately, since SNR 1987A is a resolved X-ray source and spatial-spectral effects along the dispersion direction are significant \citep{2005ApJ...628L.127Z}. The HETG contains two sets of gratings, namely the Medium Energy Grating (MEG) and the High Energy Grating (HEG). The MEG intercepts rays from outer X-ray mirror shells and is optimized roughly for the wavelength range 2.5 - 30 \AA{}. The HEG intercepts rays from the two inner shells of the mirror assembly and is optimized roughly for the wavelength range 1.2 - 15.0 \AA{}.

\begin{deluxetable*}{lccccccccc}
\tabletypesize{\footnotesize}
\tablecolumns{4}
\tablecaption{Observation Log}
\tablewidth{0pt}
\tablehead{\colhead{Epoch} & \colhead{Grating} & \colhead{Total Effective Exposure} & \colhead{Total Counts} & \colhead{Total Counts}& \colhead{Total Counts}& \colhead{Age}\\
                           &  & (ks) & (first-order MEG)& (first-order LEG)& (first-order HEG)& (days since SN)  }
 
\startdata
August-September 2004 & LETG & 287 & ... & 14447 & ... & $\sim$6399 \\
March 2007 & HETG & 354 & 29149 & ... & 9822 & $\sim$7339 \\
September 2007 & LETG & 284 & ... & 37484 & ... & $\sim$7504 \\
March 2011 & HETG & 177 & 28770 & ... & 10146 & $\sim$8796 \\
March 2018$^{\dagger}$ & HETG & 312 & 41606 & ... & 16917 & $\sim$11351 \\[3 pt]
\enddata
\tablenotetext{}{Note: Deep \textit{Chandra} observations of the X-ray remnant of SN 1987A used in this work.\\
$^{\dagger}$ Our new observation. \label{tab:1}}
\end{deluxetable*}

We detect a total of 58523 counts (background subtracted) in the first-order dispersed spectrum in the 2.5 - 15 \AA{} range: 22233 counts in MEG $+1$, 19373 counts in MEG $-1$, 7745 counts in HEG $+1$ and 9172 counts in HEG $-1$. Differences in photon statistics between individual gratings (HEG and MEG) are due to differential sensitivities of HEG and MEG in this wavelength range. While MEG has significantly higher responses at wavelengths for emission lines of our interest (e.g., $\lambda$ $\sim$ 6.0 - 9.5 \AA{}; He$\alpha$ and Ly$\alpha$ lines from Si and Mg ions) than HEG, we utilize both MEG and HEG spectra in our analysis to achieve the maximum photon count statistics. We rebinned all the individual broadband spectra to achieve a minimum of 30 counts per energy channel. Such a binning does not reduce energy resolution and allows for $\chi^{2}$ statistics. For our X-ray spectral model fits, we use version 12.10.1f of XSPEC software package \citep{1996ASPC..101...17A}. In Figure \ref{fig:1}, we show the broadband first-order HETG (MEG $\pm$1) spectrum of SNR 1987A as of March 2018. Differences in photon count statistics between the positive and negative first-order MEG grating spectra are due to differences in their effective areas at a given wavelength. We also reprocessed all the previous deep HETG and LETG observation data that were published in the literature \citep{2005ApJ...628L.127Z, 2006ApJ...645..293Z, 2009ApJ...692.1190Z, 2008ApJ...676L.131D,  2012ApJ...752..103D}, utilizing CALDB v.4.9.4 for a self-consistent study of the temporal changes in the X-ray emitting plasma characteristics. We summarize our \textit{Chandra} HETG and LETG data sets in Table \ref{tab:1}.\\

\section{Analysis and Results \label{3}} 

\subsection{Broadband Spectral Model Fits \label{3.1}}

Based on the previous \textit{Chandra} data it has been amply shown that spectral model fits with two characteristic components are required to adequately describe the observed \textit{Chandra} gratings spectrum of SNR 1987A \citep[e.g.,][]{2006ApJ...645..293Z, 2008ApJ...676L.131D, 2009ApJ...692.1190Z}.
\begin{figure*}[!ht]
\begin{center}
\begin{tabular}{cc}
\includegraphics[width=0.5\textwidth]{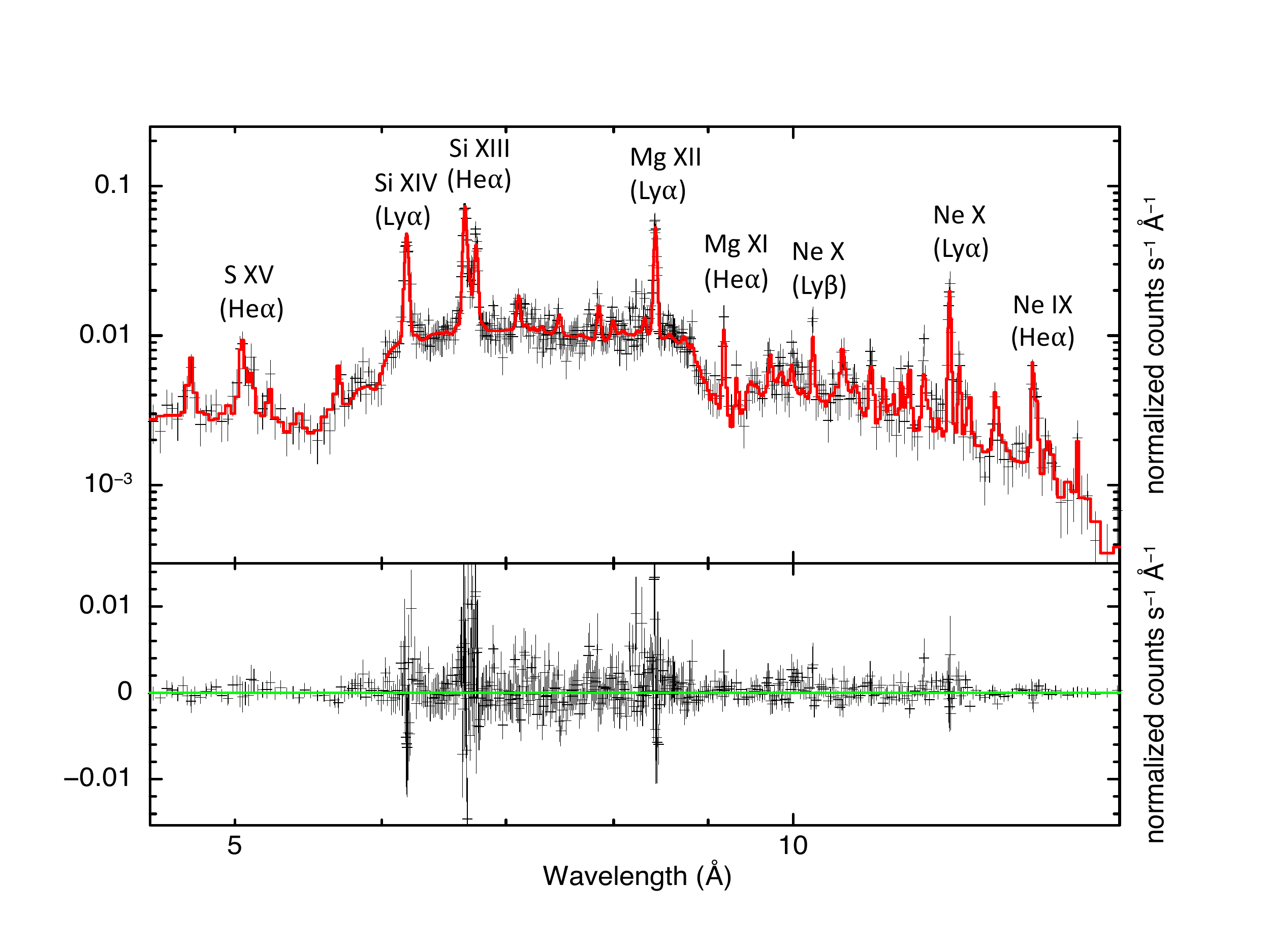} & \includegraphics[width=0.5\textwidth]{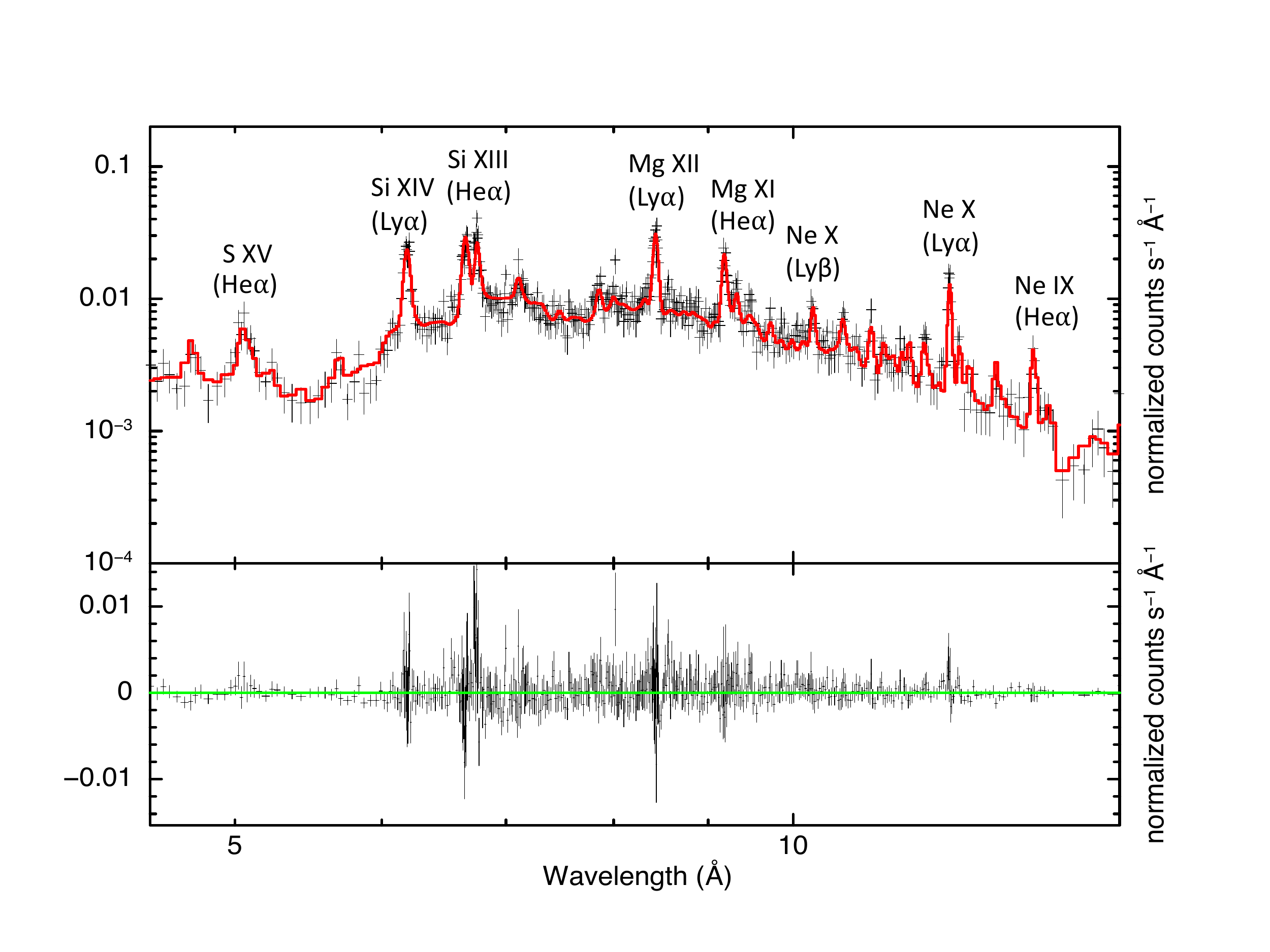} \\

(a) MEG +1: March 2018 & (b) MEG -1: March 2018 \\[3pt]
\end{tabular}
\caption{High-resolution first-order dispersed X-ray spectra of SNR 1987A (broadband: 4.5 - 15 \AA{}) extracted from our new deep \textit{Chandra} observation of March 2018 using HETG (Table \ref{tab:1}). MEG $\pm$1 spectra are shown in panels (a) and (b), respectively. In both panels, the spectral data are represented by black crosses. Our best-fit broadband spectral model (solid red line) is overlaid and strong emission lines are labelled in both panels. In (a) and (b), the bottom panel shows the residuals from the best-fit model. A stark dip in photon counts for the positive arm spectrum (a) around 9.2 \AA{} is caused by a chip gap between ACIS-S2 and ACIS-S3.} \label{fig:1}
\end{center}
\end{figure*}
Adopting such an approach, we performed simultaneous spectral fitting for all the deep HETG (MEG, HEG) and LETG (LEG) first-order dispersed spectra taken at five different epochs including our new data (Table \ref{tab:1}). 

Significant non-equilibrium effects have been recognized in the X-ray emitting plasma of SNR 1987A based on previous \textit{Chandra} grating spectral analyses \citep{2006ApJ...645..293Z, 2008ApJ...676L.131D, 2009ApJ...692.1190Z}. We model the X-ray emission spectrum with a two-component plane-parallel shock model (\verb|vpshock| in XSPEC) that takes into account the non-equilibrium ionization (NEI) effects in a hot optically-thin plasma \citep{2001ApJ...548..820B}. We adopt the NEI version 3.0 in XSPEC, based on ATOMDB v3.0.9  \footnote{ http://www.atomdb.org/}. In our latest HETG 2018 broadband spectra (Figure \ref{fig:1}), we do not find any significant emission features corresponding to trace elements. This is consistent with previous analyses of the X-ray emission from SNR 1987A between 2004 and 2007 \citep{2006ApJ...645..293Z, 2008ApJ...676L.131D, 2009ApJ...692.1190Z}, where X-ray emission from these trace elements were negligible. This is also noted with \textit{XMM-Newton} data from 2012 - 2017 \citep{2021ApJ...916...76A}. Thus, we set the abundances of all trace elements with \verb|vpshock| (version ATOMDB v3.0.9) to zero, using the XSPEC command \verb|NEI_TRACE_ABUND|. In our spectral model fits, we varied the electron temperature (\textit{kT}) and the ionization age ($\tau = n_{e}t$, where $n_{e}$ is the post-shock electron density and $t$ is the time since the gas first entered the shock). We also keep the normalization parameter free, which corresponds to the volume emission measures (\textit{EM}). We also convolved the physical shock model with a Gaussian profile (\verb|gsmooth|) to represent the broadening of the spectral lines due to the bulk and turbulent motion of the X-ray emitting plasma \citep[see][for a detailed discussion]{2012ApJ...752..103D}. Within \verb|gsmooth|, both the magnitude of the smoothing measured by the Gaussian kernel and the $\alpha$-index (exponent for \verb|gsmooth| energy dependence) are kept free.

We fit the total foreground absorption column by applying two-component multiplicative absorption models, \verb|tbabs| and \verb|tbnew| \footnote{https://pulsar.sternwarte.uni-erlangen.de/wilms/research/tbabs/}. The absorption component \verb|tbabs| accounts for the Galactic absorption column $N_\mathrm{H,Gal}$, and it is fixed at $N_\mathrm{H,Gal} =$ 6 $\times$ 10$^{20}$ cm$^{-2}$ \citep{doi:10.1146/annurev.aa.28.090190.001243}. The second absorption component, \verb|tbnew|, accounts for the LMC absorption column, $N_\mathrm{H,LMC}$ and is varied in the fits. We use \verb|tbnew| as the absorption column for LMC as it allows us to set individual elemental abundances associated with $N_\mathrm{H,LMC}$ at their respective LMC values. Recent measurements of the LMC abundances based on the X-ray spectral analysis of SNRs \citep{2016AJ....151..161S, 2016A&A...585A.162M} suggest $\sim$50$\%$ lower LMC abundance values for O, Ne, Mg, and Fe than the previously estimated values \citep{1998ApJ...505..732H}. Our derived shock parameters are consistent (within statistical uncertainties), assuming either set of these LMC abundances in \verb|tbnew|. We tie $N_\mathrm{H,LMC}$ among all the fitted spectra, assuming no temporal variation in the LMC column for SNR 1987A. We obtain a best-fit total absorption column density, $N_\mathrm{H} = N_\mathrm{H,Gal} + N_\mathrm{H,LMC}$ = 2.17$^{+0.22}_{-0.22}$ $\times$ 10$^{21}$ cm$^{-2}$. This value is comparable with estimates of $N_\mathrm{H}$ obtained by other X-ray analyses:  2.35$^{+0.09}_{-0.08}$ $\times$ 10$^{21}$ cm$^{-2}$ \citep{2006ApJ...646.1001P}, 1.44$^{+0.16}_{-0.12}$ $\times$ 10$^{21}$ cm$^{-2}$ \citep{2009ApJ...692.1190Z}, and 2.60$^{+0.05}_{-0.05}$ $\times$ 10$^{21}$ cm$^{-2}$ \citep{2021ApJ...916...76A}. We note that more recent HI surveys have suggested much higher $N_\mathrm{H,Gal}$ values towards LMC, i.e., $\sim$ (2.5 - 4) $\times$ 10$^{21}$ cm$^{-2}$ \citep {2005A&A...440..775K, 2013MNRAS.431..394W, 2016A&A...594A.116H}. Adopting these high Galactic column, the overall fits are equally good, but it results in a negligible LMC column. The implied negligible LMC column does not appear to be reasonable because optical extinction estimates show that the LMC contribution is greater than the Milky Way contribution \citep{1990AJ.....99.1483F, 2011ApJ...743..186F}. A detailed analysis of the contribution of the Galactic $N_\mathrm{H}$ towards total absorption column density is beyond the scope of our work. While this issue was similarly outlined by \cite{2021ApJ...916...76A} in relation to their analysis of the \textit{XMM-Newton} data of SNR 1987A, we note that the total columns (Galactic + LMC) are generally consistent between either values of the Galactic column, and thus that the best-fit parameters in our spectral model fits (i.e., electron temperatures, ionization age, abundances, and volume emission measures) are not significantly affected (within statistical uncertainties).

We note that the low energy emission (E \small$\leq$ 2 keV) is likely to be dominated by radiative shocks \citep{2002ApJ...572..906P, 2008A&A...492..481G}. For a given temperature, these radiative shocks will be softer than adiabatic shocks \citep{2006A&A...449..171N}. This, along with uncertainties in the measured absorption column and the decreasing low-energy sensitivity of the ACIS detectors, adds to the systematic uncertainties in our analysis. Additional systematic uncertainties arise from the two-temperature fit to our broadband spectra, which is a crude approximation of a complex multi-temperature shocked medium. In Section \ref{4.4}, we compare our inferred results from the broadband spectral fits with a continuous distribution of plasma characteristics from self-consistent MHD simulations. We note that our data are somewhat insensitive to high energy/temperature components (compared to \textit{XMM-Newton} or \textit{NuSTAR} observations) due to the relatively low sensitivity of \textit{Chandra} grating spectra above $\sim$6 keV, which also contributes to systematic uncertainties, particularly regarding any high-energy components.

\begin{deluxetable*}{lccccc}~
\tabletypesize{\footnotesize}
\tabletypesize{\scriptsize}

\tablecolumns{6}
\tablecaption{Simultaneous Two-Shock Fit Results}
\tablewidth{0pt}
\vspace{-0.1cm}
\tablehead{\colhead{Parameters} & \colhead{LETG 2004} & \colhead{HETG 2007} & \colhead{LETG 2007} & \colhead{HETG 2011} & \colhead{HETG 2018}}

\startdata
$\chi^{2}/$dof &  \ldots & \ldots &  \ldots & \ldots &  3454/5181  \\ 
$N_\mathrm{H}$($10^{21}$ cm$^{-2}$ )$^{a}$ & 2.17 & 2.17 & 2.17 & 2.17 & 2.17 $^{+0.22}_{-0.22}$  \\
$kT_{soft}$ (keV) &  0.54 $^{+0.04}_{-0.05}$   &  0.61 $^{+0.02}_{-0.03}$  & 0.57 $^{+0.02}_{-0.01}$  & 0.62 $^{+0.03}_{-0.03}$  &  0.86 $^{+0.07}_{-0.08}$ \\
$kT_{hard}$ (keV) &  2.49 $^{+0.46}_{-0.28}$   &  2.25 $^{+0.30}_{-0.21}$  & 1.88 $^{+0.14}_{-0.13}$  & 1.96 $^{+0.18}_{-0.15}$ &  2.41 $^{+0.20}_{-0.16}$ \\ 
$\tau_{soft}$ $^{b}$  &  3.29 $^{+1.25}_{-0.77}$  & 3.36 $^{+0.58}_{-0.41}$  & 3.80 $^{+0.59}_{-0.47}$   &  7.51  $^{+2.17}_{-1.53}$ & 6.20  $^{+1.84}_{-1.42}$ \\ 
$\tau_{hard}$ $^{b}$  &  1.10 $^{+0.50}_{-0.25}$  & 2.43 $^{+0.63}_{-0.51}$  & 2.27 $^{+0.46}_{-0.36}$   &  3.20  $^{+0.72}_{-0.34}$ & 5.99  $^{+1.15}_{-0.50}$ \\ 
$EM_{soft}$ $^{c}$  &  4.46 $^{+0.34}_{-0.46}$  & 12.58 $^{+0.49}_{-0.43}$ & 14.41 $^{+0.77}_{-0.93}$ & 23.59 $^{+0.96}_{-1.92}$ & 18.26 $^{+1.58}_{-1.61}$  \\ 
$EM_{hard}$ $^{c}$ &  2.26 $^{+0.12}_{-0.43}$  & 5.02 $^{+0.58}_{-0.40}$ & 7.13 $^{+0.58}_{-0.45}$ & 13.51 $^{+1.05}_{-1.20}$ & 17.85 $^{+0.96}_{-1.27}$  \\
He (fixed)  &  1.98 & 1.98 & 1.98 & 1.98 & 1.98 \\
C (fixed)  &   0.12 & 0.12 & 0.12 & 0.12 & 0.12 \\
N (fixed)   &   0.92 & 0.92 & 0.92 & 0.92 & 0.92 \\
O (fixed)   &   0.14 & 0.14 & 0.14 & 0.14 & 0.14  \\
Ne   &  0.34 & 0.34  & 0.34 & 0.34 & 0.34 $^{+0.01}_{-0.01}$   \\
Mg   &    0.25 & 0.25  & 0.25 & 0.25 &  0.25 $^{+0.01}_{-0.01}$ \\
Si   &    0.36 & 0.36  & 0.36 & 0.36 &  0.36 $^{+0.01}_{-0.01}$\\
S   &      0.40 & 0.40  & 0.40 & 0.40 & 0.40 $^{+0.04}_{-0.04}$ \\
Ar (fixed)  &    0.776  & 0.776  & 0.776  &  0.776 & 0.776\\
Ca (fixed) &   0.354 & 0.354  & 0.354  & 0.354 &  0.354\\
Fe   &    0.194 & 0.194  & 0.194 & 0.194 &  0.19 $^{+0.01}_{-0.01}$\\
Ni  (fixed) &    0.662 & 0.662  & 0.662 & 0.662  &  0.662\\ [3 pt]
\enddata
\tablenotetext{}{\justify{Note: 90$\%$ confidence intervals are provided as error bars and all abundances are expressed as ratios to the \textbf{aspl} solar abundance table.  The two shock components are represented by soft and hard labels. \label{tab:2}}\\
\\
\hspace{0.1in}$^{a}$ $N_\mathrm{H} = N_\mathrm{H,Gal} + N_\mathrm{H,LMC}$\\
\\
\hspace{0.1in}$^{b}$ Ionization age ($\tau = n_{e}t$) in units of 10$^{11}$ cm$^{-3}$ s.\\
\\
\hspace{0.1in}$^{c}$ $EM = \int n_{e} n_{H} dV$ in units of 10$^{58}$ cm$^{-3}$ for an adopted distance of 51 kpc.\\}
\end{deluxetable*}

\begin{figure*}[htbp]
\centering
\begin{longtable}{cc}
  \includegraphics[width=0.5\textwidth,height=7.0cm]{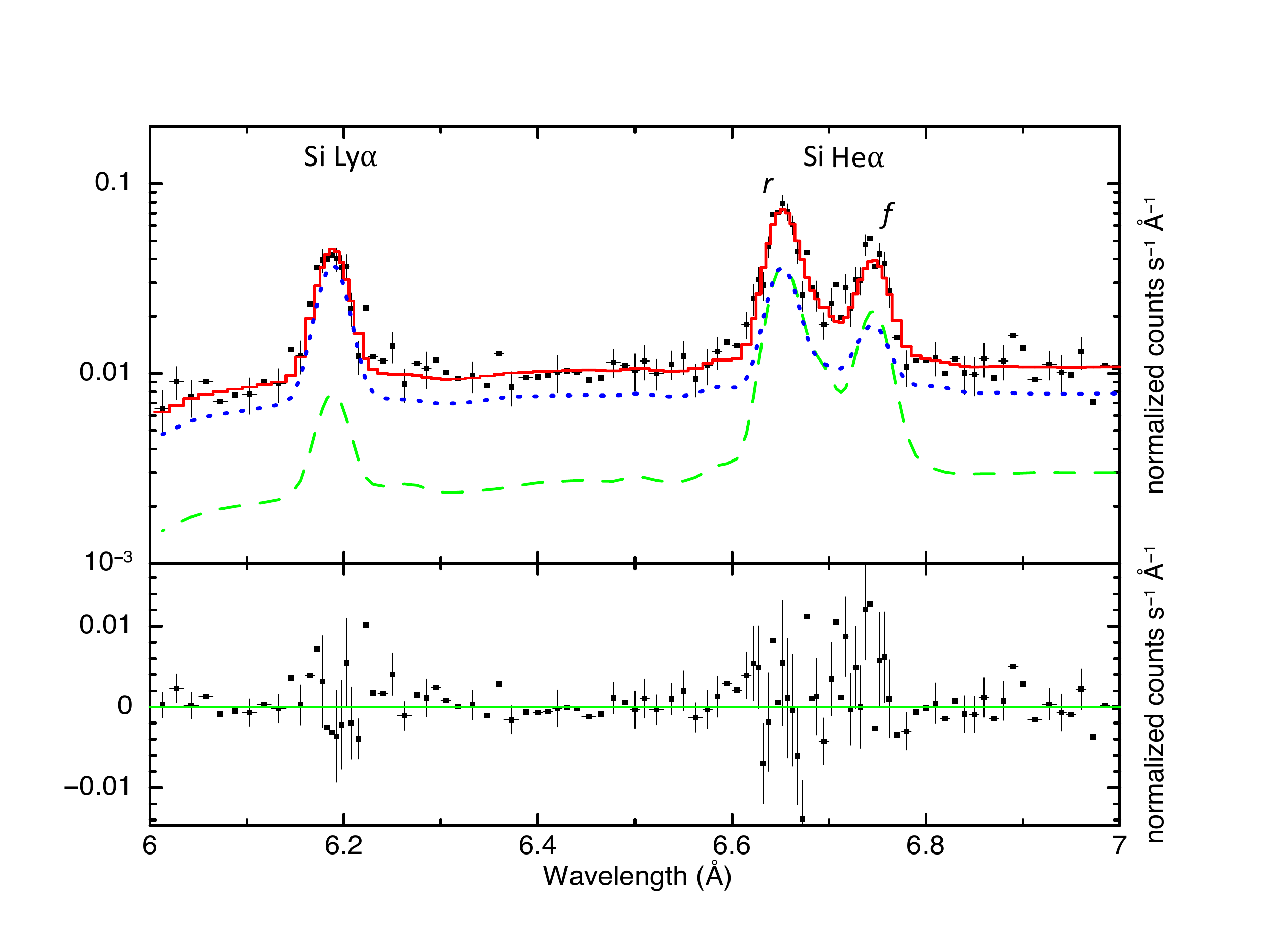} & \includegraphics[width=0.5\textwidth,height=7.0cm]{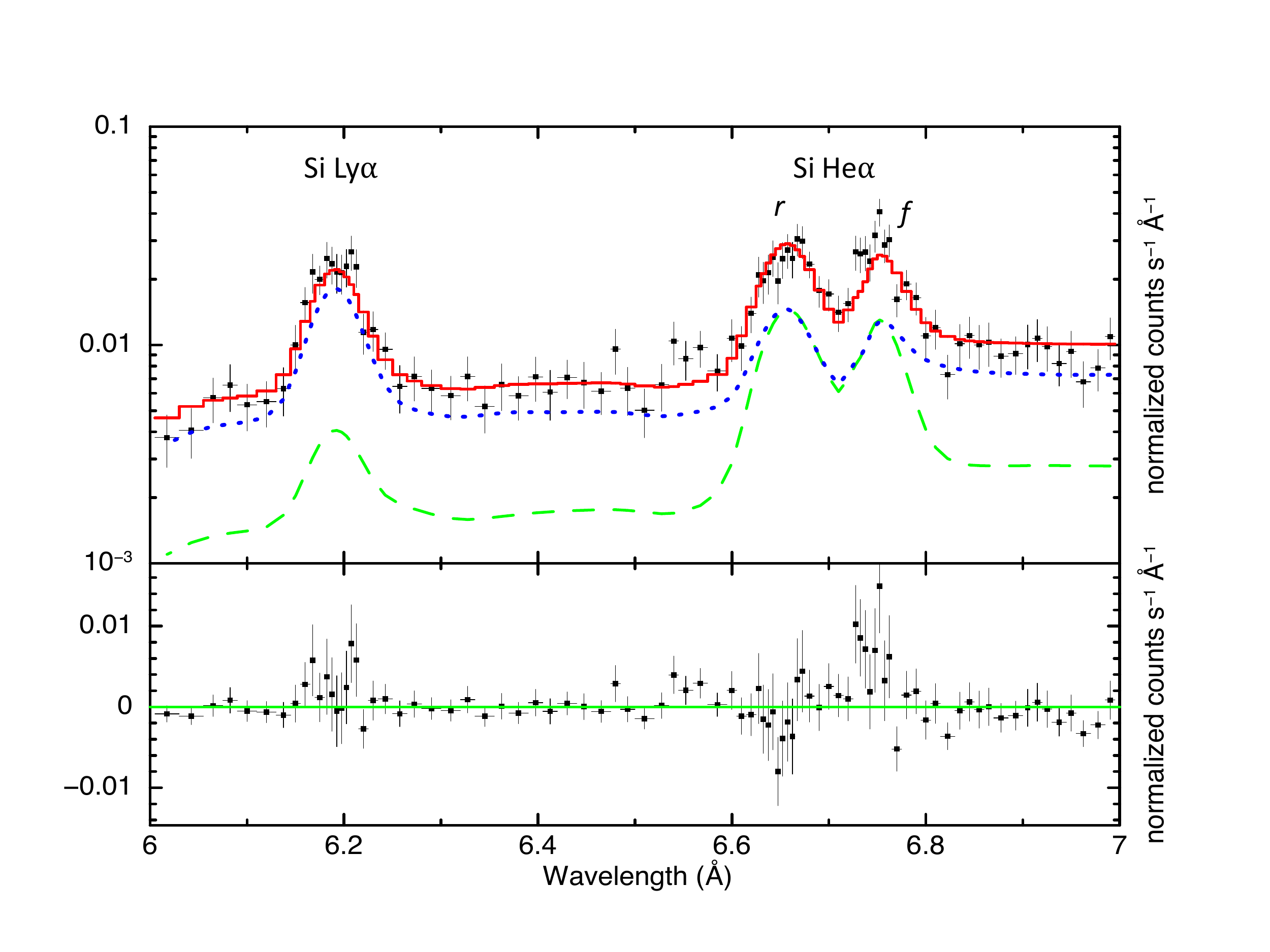}  \\
(a) Si: MEG +1 2018 & (b) Si: MEG -1 2018 \\
  \includegraphics[width=0.5\textwidth,height=7.0cm]{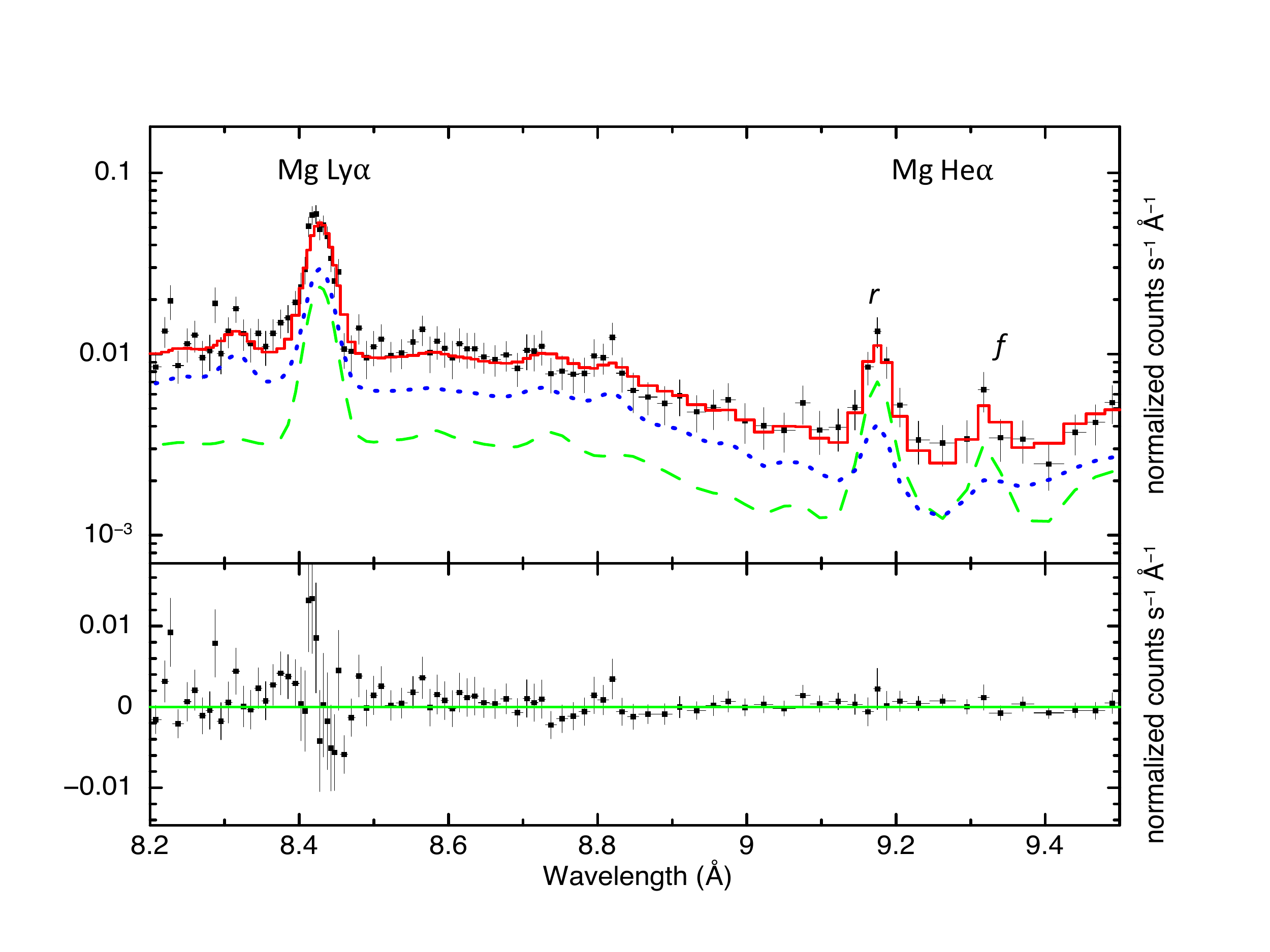} & \includegraphics[width=0.5\textwidth,height=7.0cm]{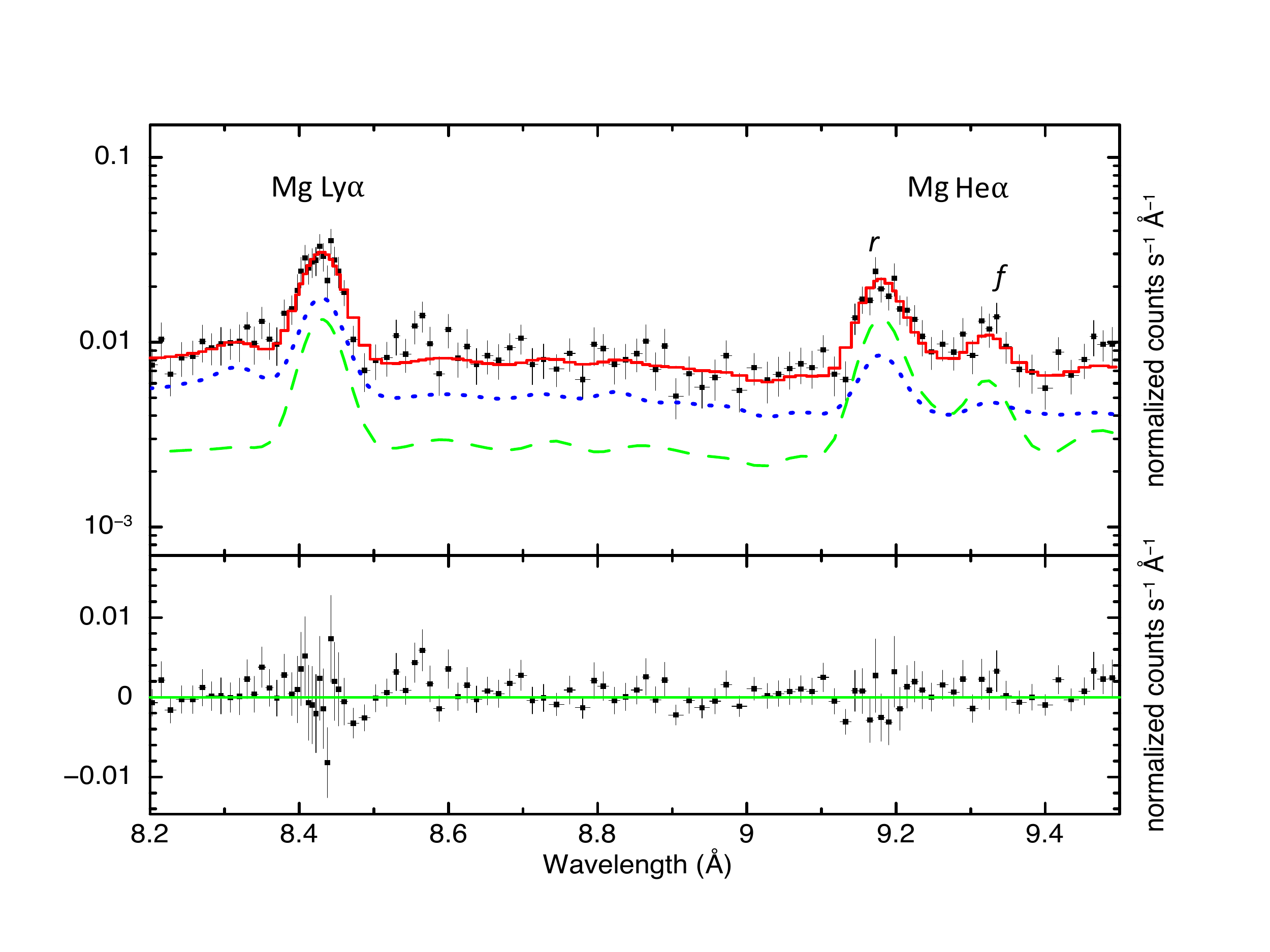}  \\
(c) Mg: MEG +1 2018 & (d) Mg: MEG -1 2018 \\
   \includegraphics[width=0.5\textwidth,height=7.0cm]{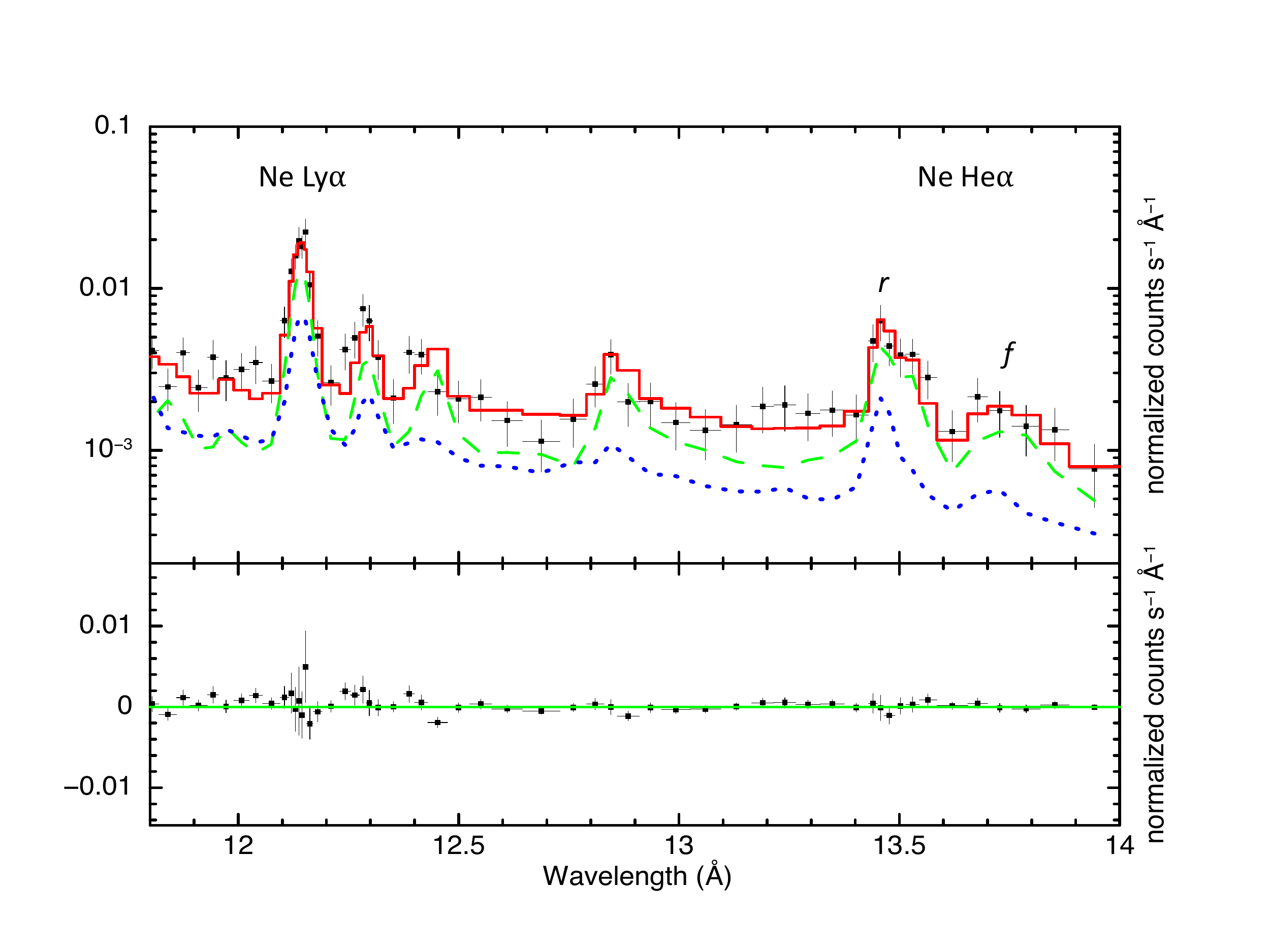} & \includegraphics[width=0.5\textwidth,height=7.0cm]{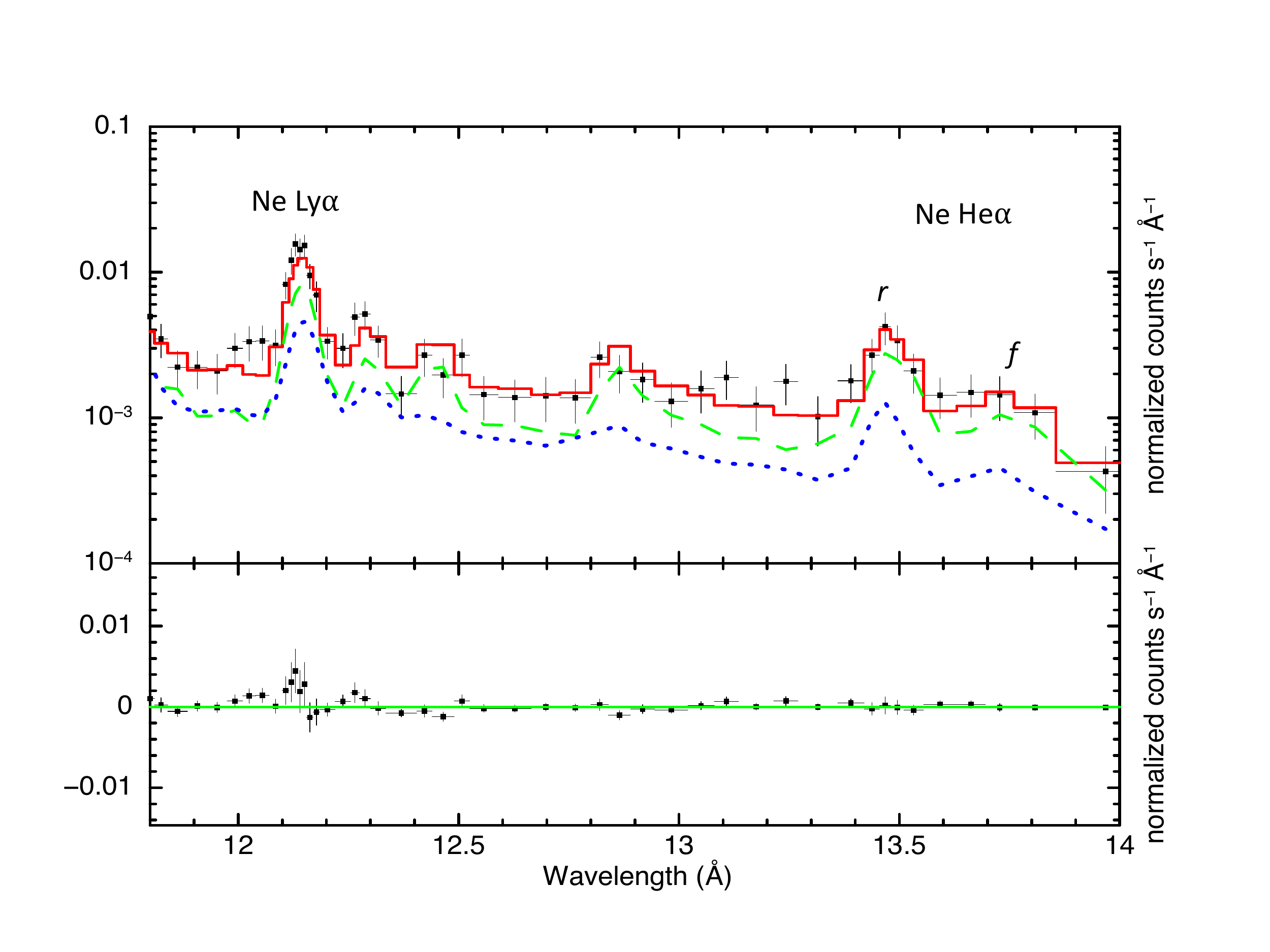}   \\
(e) Ne: MEG +1 2018 & (f) Ne: MEG -1 2018 \\[3pt]
\end{longtable}
\caption{Zoom-in views of the HETG 2018 spectrum of SNR 1987A for strong emission lines of Si, Mg, and Ne. For emission lines from He-like ions, the resolved resonance (\textit{r}) and forbidden (\textit{f}) lines are marked. In all panels, the best-fit two-component spectral model is overlaid (red curve). The best-fit soft (green dashed lines) and hard component (blue dotted lines) models are overlaid in each panel. The bottom plot in each panel shows residuals from the best-fit model.\label{fig:2}}

\end{figure*}

Previous works have shown no evidence for significant temporal changes of the elemental abundances in the observed X-ray spectrum of SNR 1987A \citep{2006ApJ...645..293Z, 2008ApJ...676L.131D, 2009ApJ...692.1190Z, 2020ApJ...899...21B}. Based on our spectral model fits to each individual spectrum taken at every epoch between 2004 and 2018, we confirm that the estimated abundances are consistent with each other within statistical uncertainties. Thus, we assumed constant abundances for Ne, Mg, Si, S, and Fe, for which various emission lines from the K- and L-shell electron transitions are evident in the observed spectra of HETG 2018 (Figure \ref{fig:1}), and tied the abundance of each element together for joint fits to all five data sets (LETG 2004, HETG 2007, LETG 2007, HETG 2011, and HETG 2018) in order to obtain the best-constrained abundance values for all epochs between 2004 and 2018. With \textit{XMM-Newton} data, \cite{Sun_2021} reported decreasing average abundances of N, O, Ne, and Mg over a similar period as our analysis. They have suggested that this may reflect the different chemical compositions between the two plasma components. However, based on our deep \textit{Chandra} HETG data, we find no evidence for a statistically significant evolution of elemental abundances in either of our soft or hard component spectral model fits.

The solar abundance table \verb|angr| \citep{1989GeCoA..53..197A} was adopted in all our previous \textit{Chandra} spectral analyses of SNR 1987A \citep{2002ApJ...574..166M, 2002ApJ...567..314P, 2004ApJ...610..275P, 2006ApJ...646.1001P, 2006ApJ...645..293Z, 2008ApJ...676L.131D, 2009ApJ...692.1190Z}. In this work we use an updated solar abundance table, \verb|aspl| \citep{2009ARA&A..47..481A}. We found that this change in the assumed solar abundances does not significantly affect our results. The best-fit values of the spectral model parameters are consistent within statistical uncertainties between model fits, adopting these two different solar abundance tables. Thus, hereafter, we quote all elemental abundance values with respect to the \verb|aspl| solar abundances. We fix other elemental abundances for SNR 1987A at: He = 1.98 \citep{2010ApJ...717.1140M}, C = 0.12 \citep{1996ApJ...461..993F}, N = 0.92, O = 0.14 \citep{2009ApJ...692.1190Z}. We also fix Ar = 0.776, Ca = 0.354, and Ni = 0.662 at their respective LMC values \citep{1992ApJ...384..508R}. We tie the elemental abundances between the soft and hard component shock models for all epochs.

In Table \ref{tab:2}, we show the results from our simultaneous two-component shock model fits to the MEG and HEG ($\pm1$) spectra from HETG 2007, HETG 2011, HETG 2018 and LEG ($\pm1$) spectra from LETG 2004, LETG 2007. In Figure \ref{fig:1}, we present our broadband best-fit model overlaid over HETG 2018 (MEG $\pm$1). We also present our best-fit spectral model in several sub-bands around strong emission lines from Si, Mg, and Ne for these data sets (Figure \ref{fig:2}). Between 2004 and 2011, the observed electron temperature (\textit{kT}) for the soft component showed no significant changes (within uncertainties), while for the hard component it gradually decreased from $\sim$2.5 keV to $\sim$1.8 keV. In contrast, between 2011 and 2018, both the soft and hard component \textit{kT} have increased from $\sim$0.62 keV to $\sim$0.86 keV ($\sim$38$\%$) and from $\sim$1.96 keV to $\sim$2.41 keV ($\sim$23$\%$), respectively. Electron temperatures and ionization ages measured in Table \ref{tab:2} for epochs of LETG 2004, HETG 2007, LETG 2007 are consistent (within statistical uncertainties) with previous measurements \citep{2006ApJ...645..293Z, 2008ApJ...676L.131D, 2009ApJ...692.1190Z}.  From such a broadband spectral model fit, our best-fit elemental abundances are: Ne = 0.34 $\pm$ 0.01, Mg = 0.25 $\pm$ 0.01, Si =  0.36 $\pm$ 0.01, S =  0.40 $\pm$ 0.04, and Fe =  0.19 $\pm$ 0.01. These best-fit abundances are consistent (within statistical uncertainties) with those for the ER as measured in the previous \textit{Chandra} gratings spectral analyses of SNR 1987A \citep{2006ApJ...645..293Z, 2008ApJ...676L.131D, 2012ApJ...752..103D, 2009ApJ...692.1190Z}.\\

\subsection{Individual Emission Line Fits \label{3.2}}
\begin{figure*}[htbp]
\centering
\begin{tabular}{ccc}
   \includegraphics[width=0.33\textwidth]{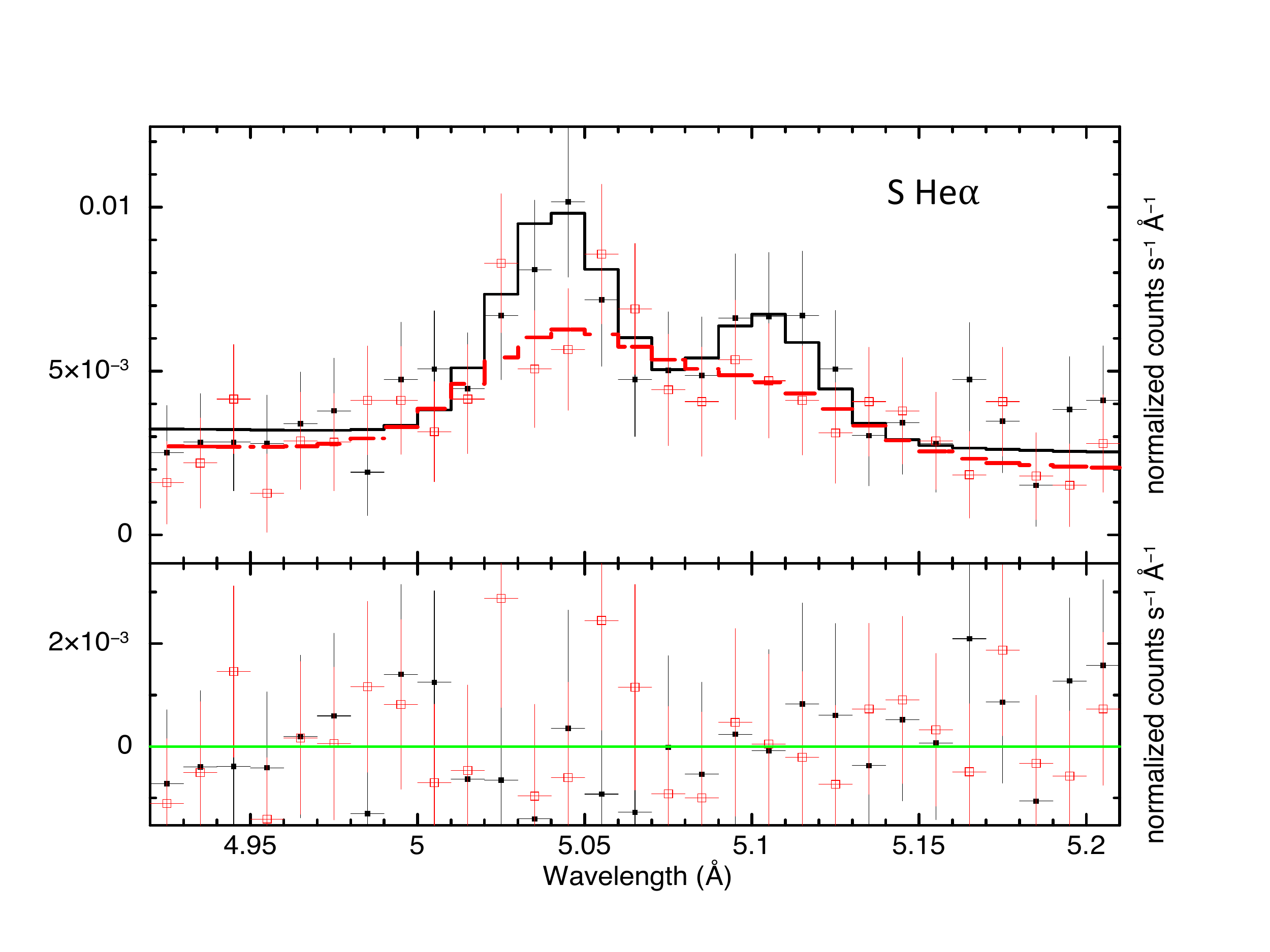} &
   \includegraphics[width=0.33\textwidth]{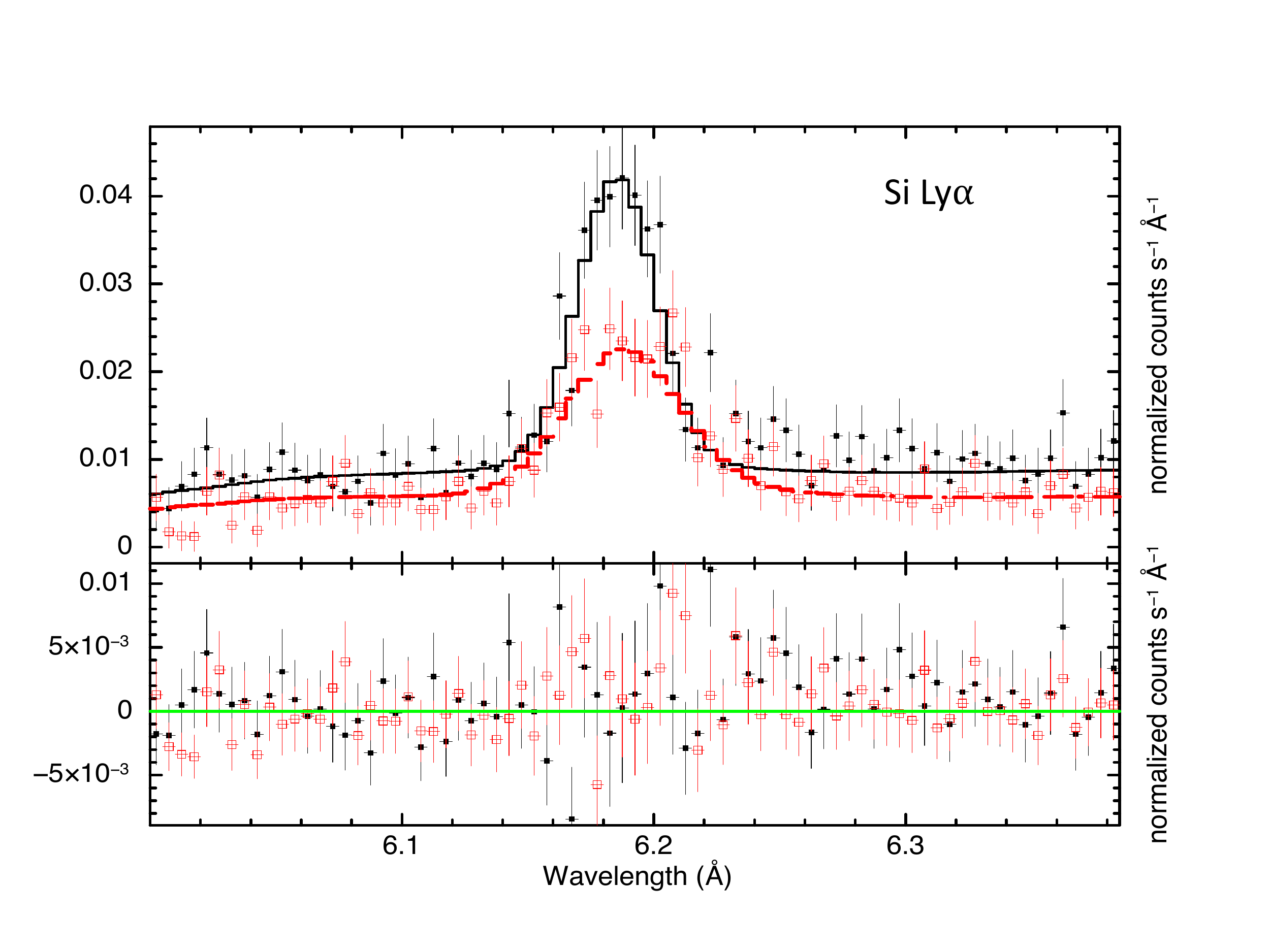} & 
   \includegraphics[width=0.33\textwidth]{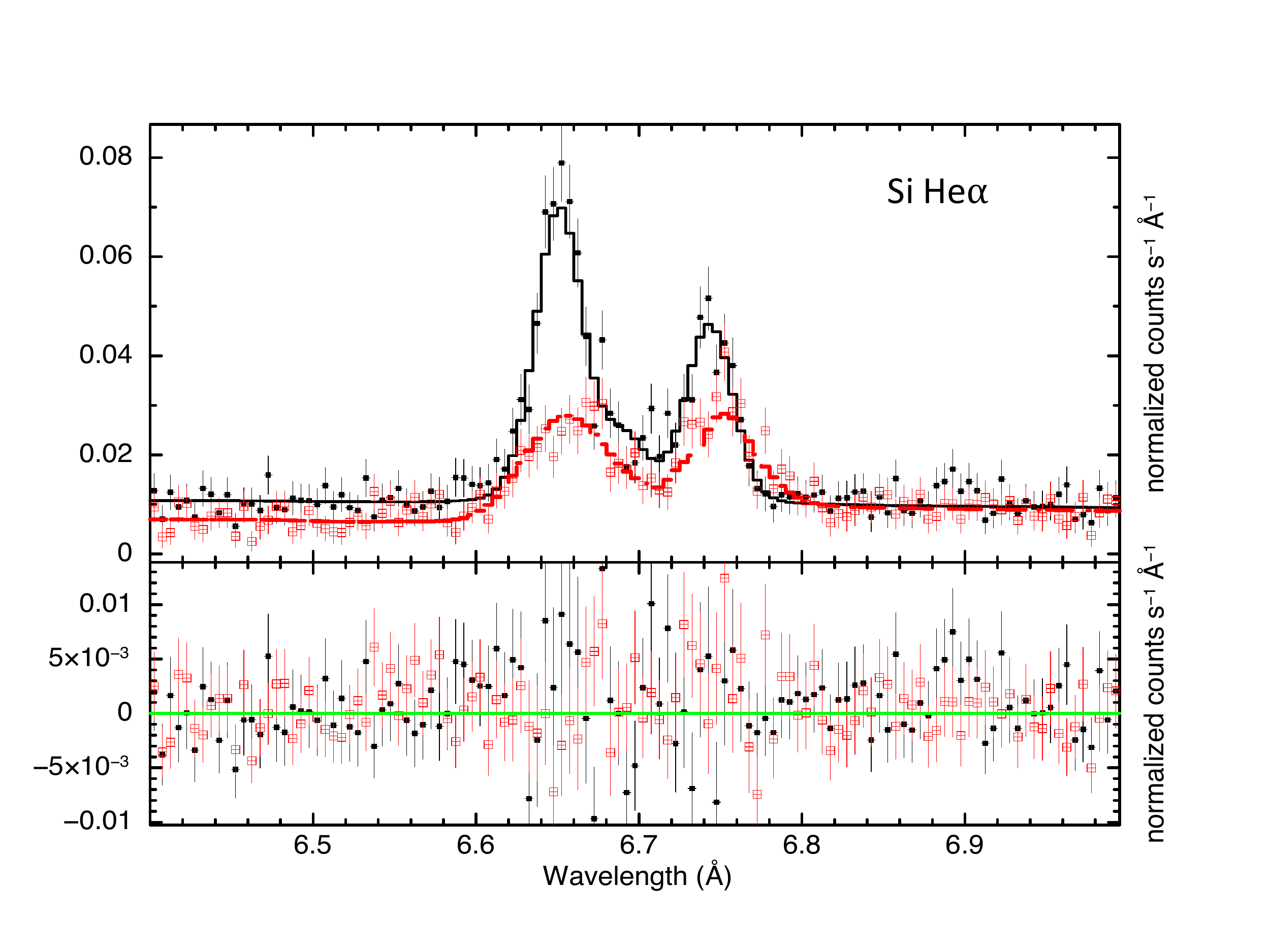} \\ 
  (a) MEG 2018: S He$\alpha$ & (b) MEG 2018: Si Ly$\alpha$ & (c) MEG 2018: Si He$\alpha$
\end{tabular}
\begin{tabular}{ccc}
   \includegraphics[width=0.33\textwidth]{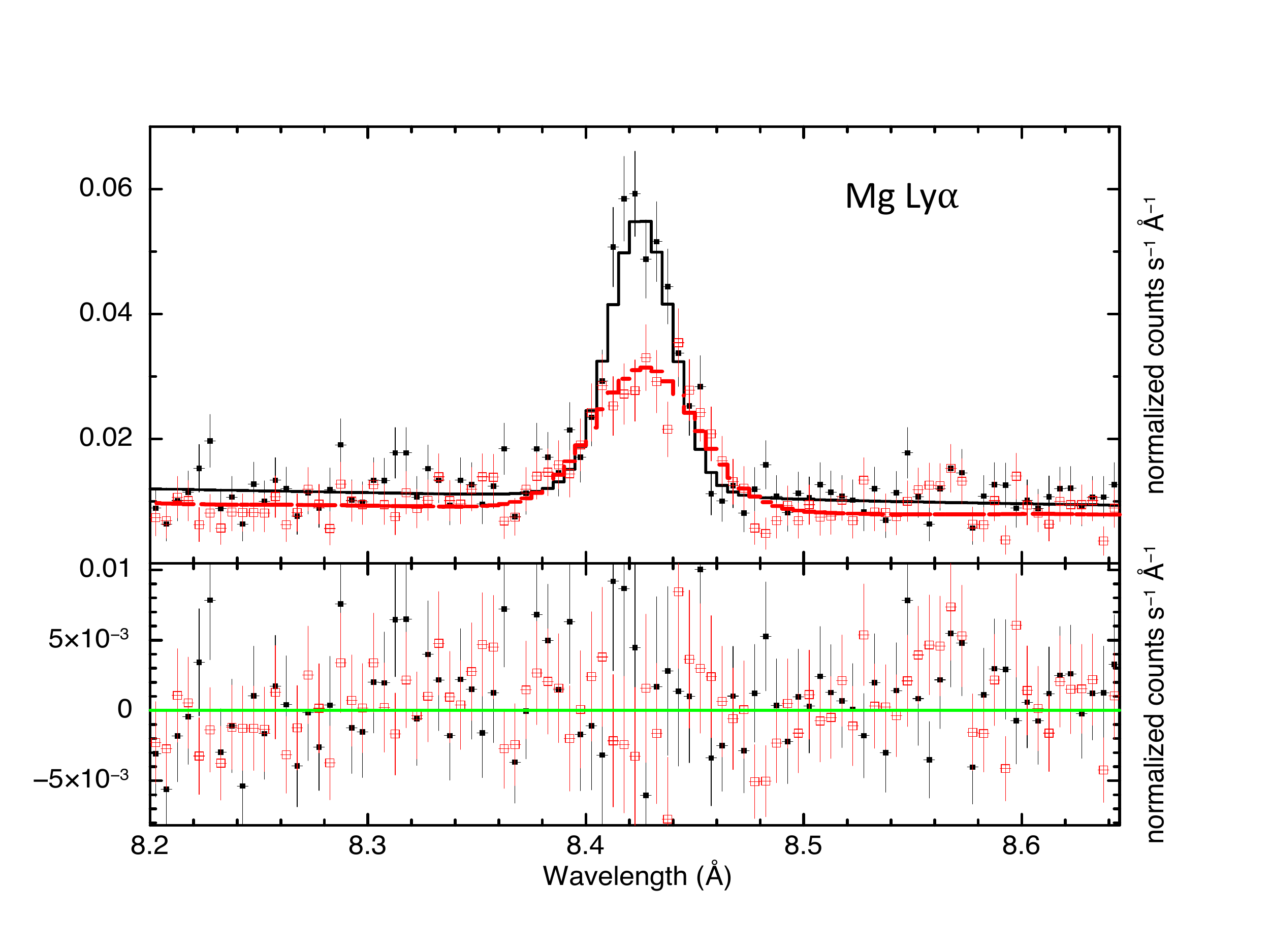} &
   \includegraphics[width=0.33\textwidth]{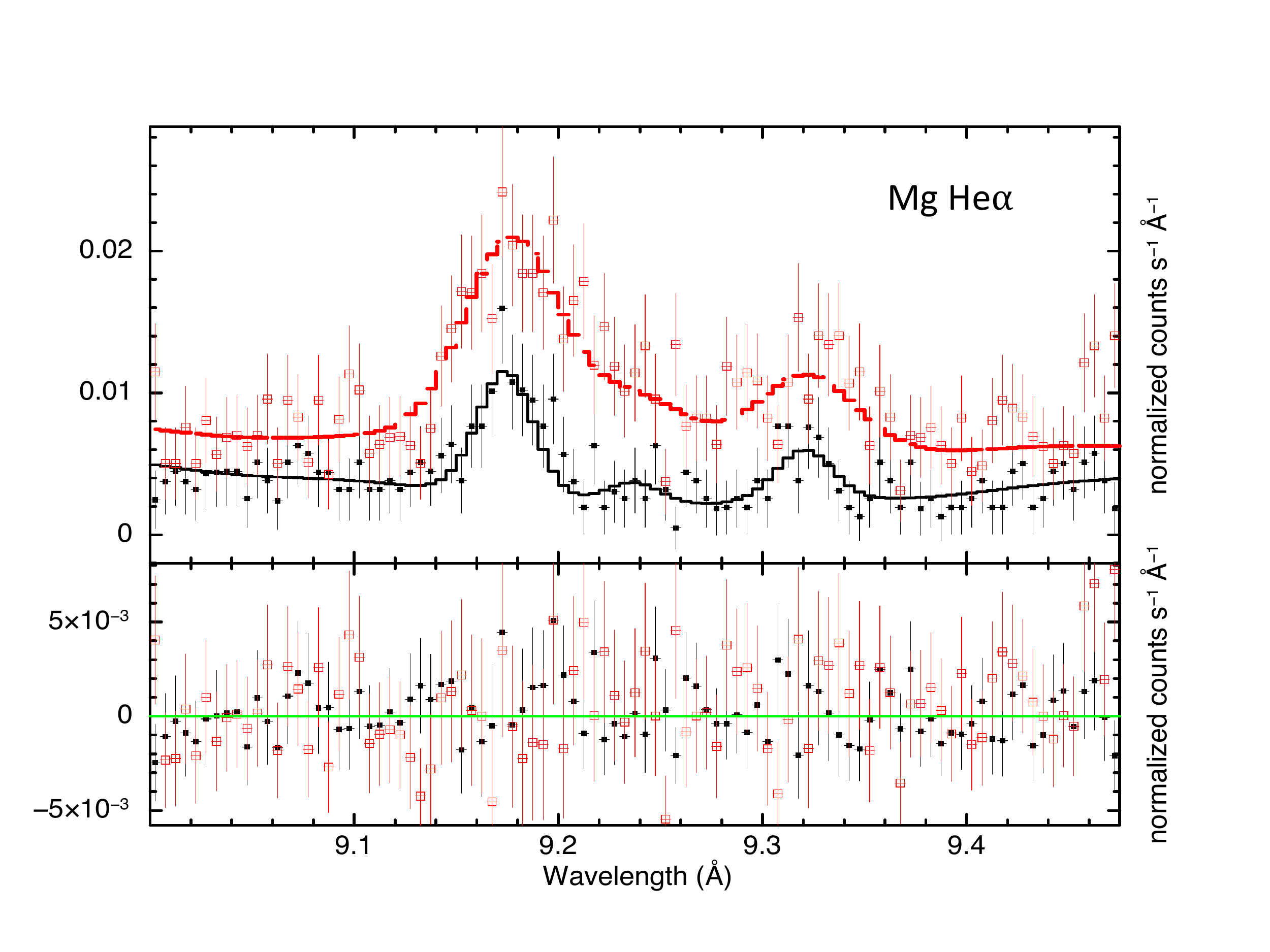} &  
   \includegraphics[width=0.33\textwidth]{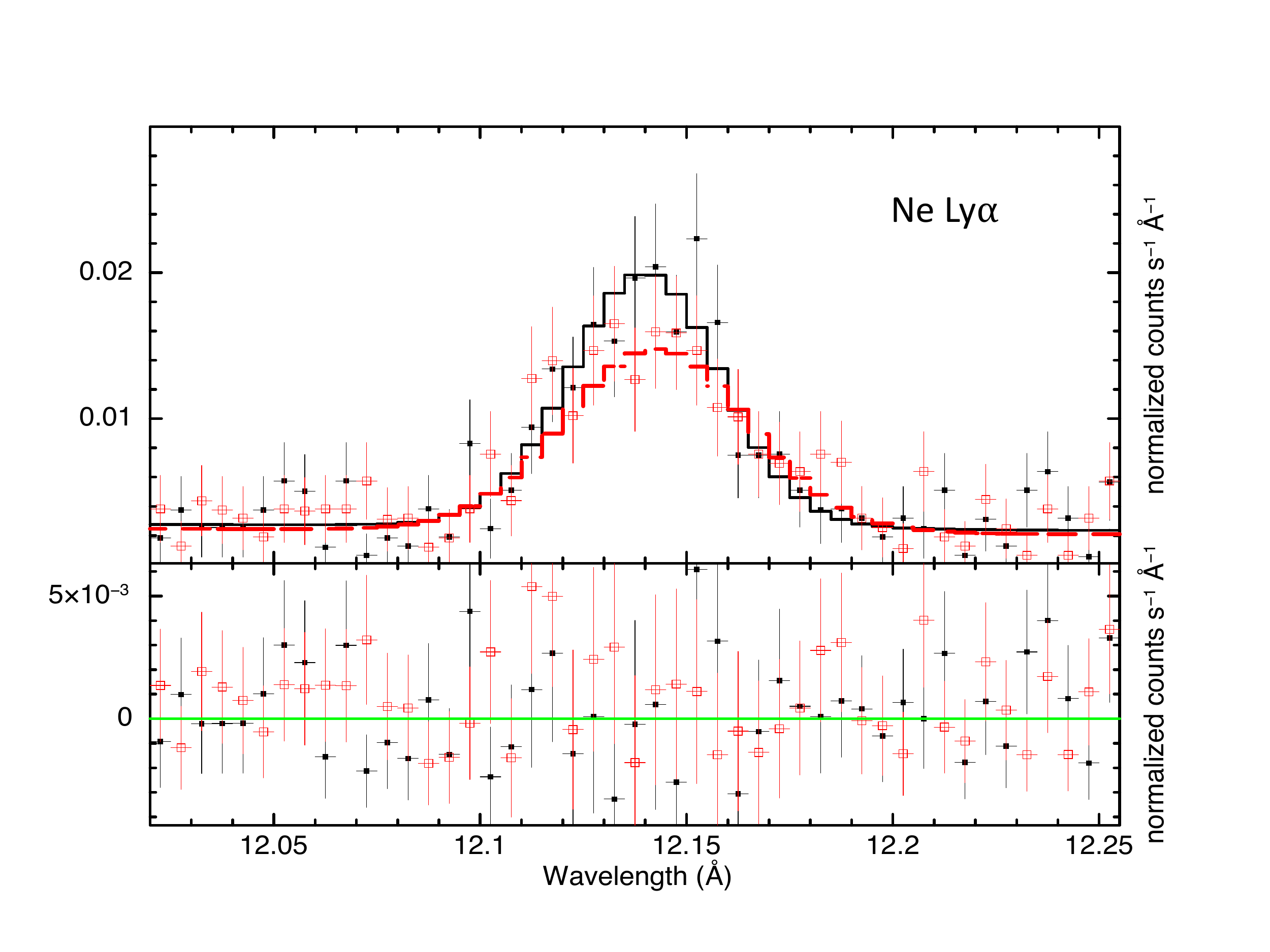} \\ 
   (d)MEG 2018: Mg Ly$\alpha$ & (e)  MEG 2018: Mg He$\alpha$ & (f) MEG 2018: Ne Ly$\alpha$
\end{tabular}
\begin{tabular}{ccc}
   \includegraphics[width=0.33\textwidth]{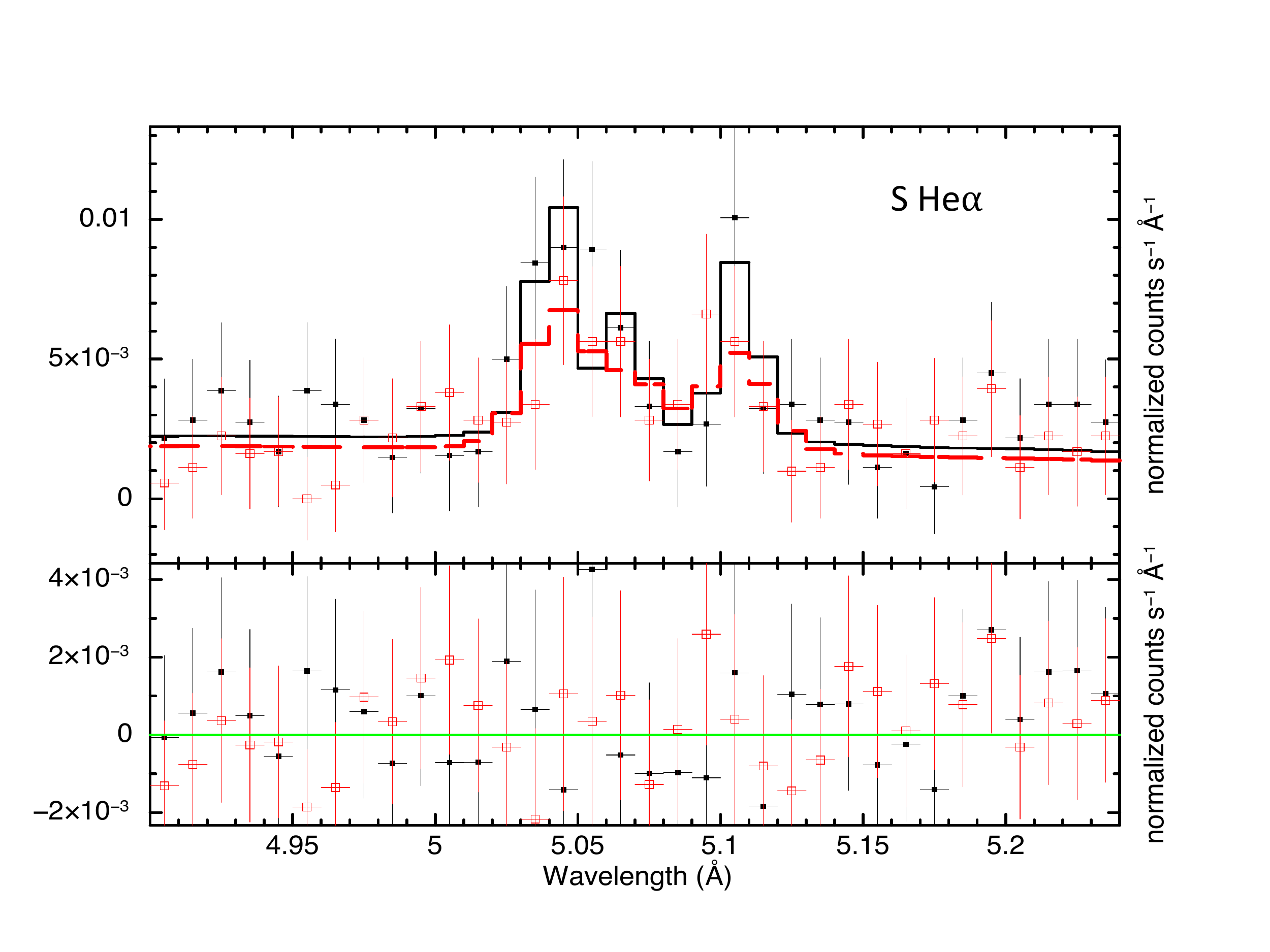} &
   \includegraphics[width=0.33\textwidth]{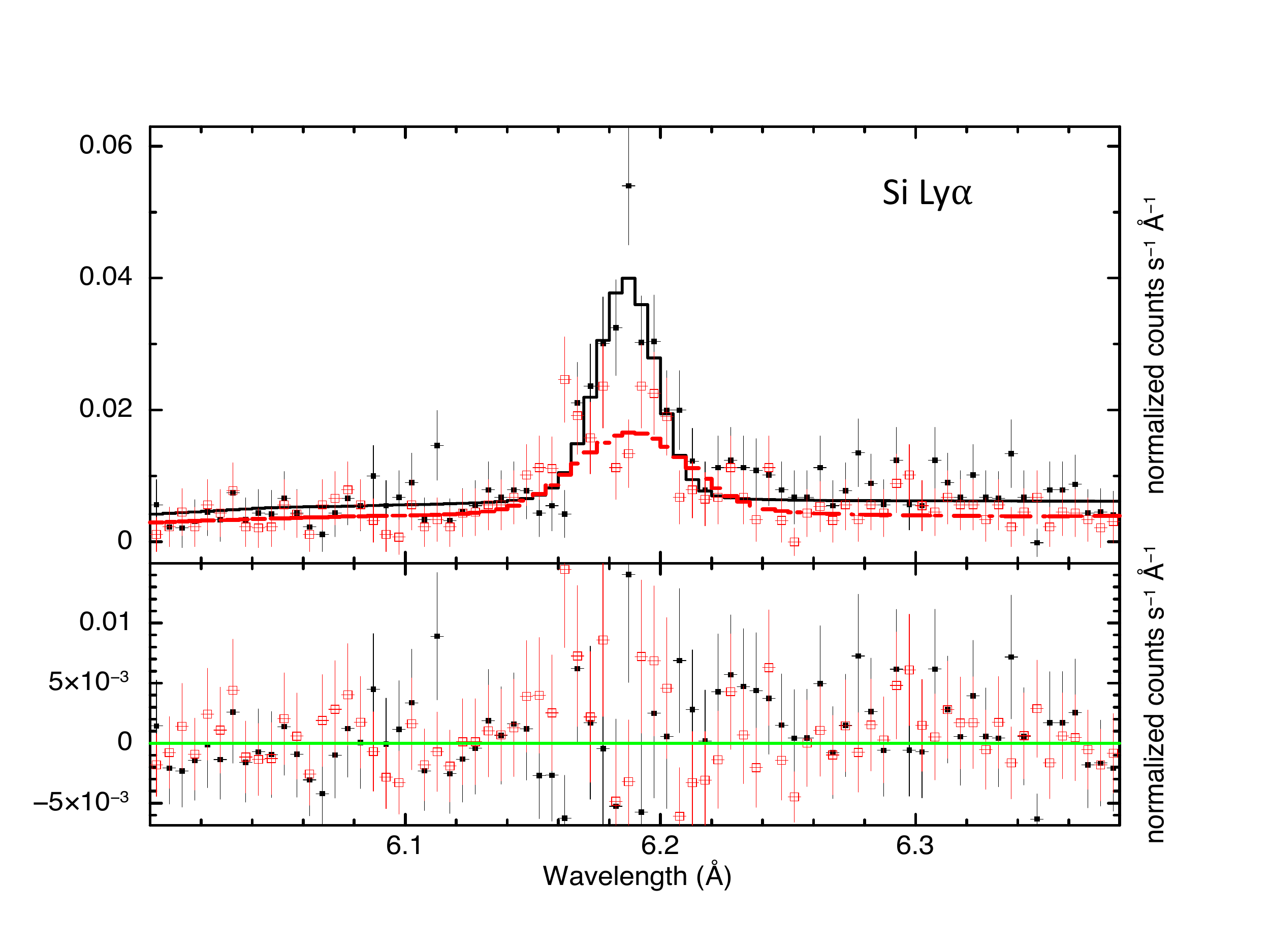} &  
   \includegraphics[width=0.33\textwidth]{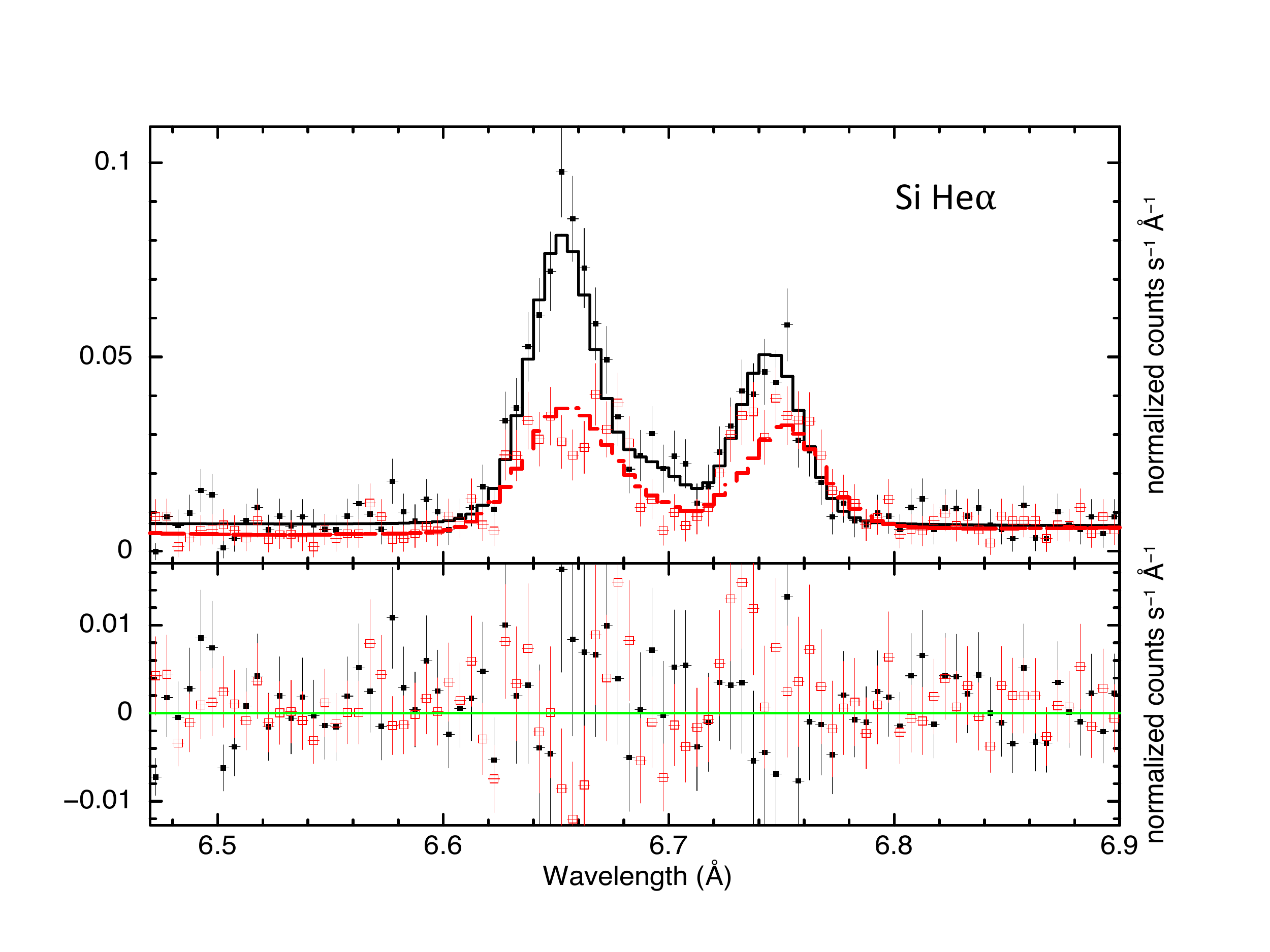} \\ 
  (g) MEG 2011: S  He$\alpha$& (h) MEG 2011: Si Ly$\alpha$  & (i) MEG 2011: Si He$\alpha$
\end{tabular}
\begin{tabular}{ccc}
   \includegraphics[width=0.33\textwidth]{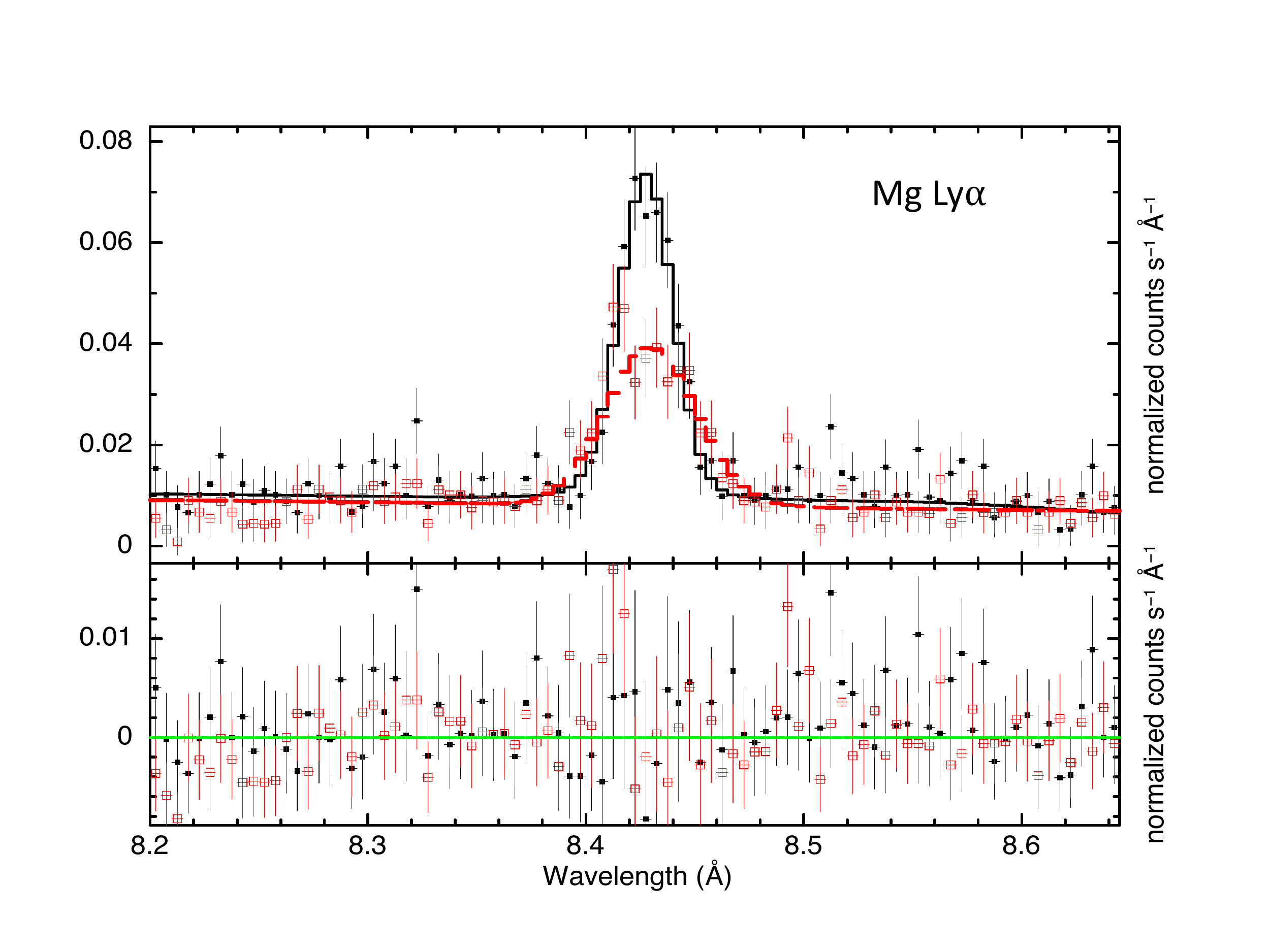} &
   \includegraphics[width=0.33\textwidth]{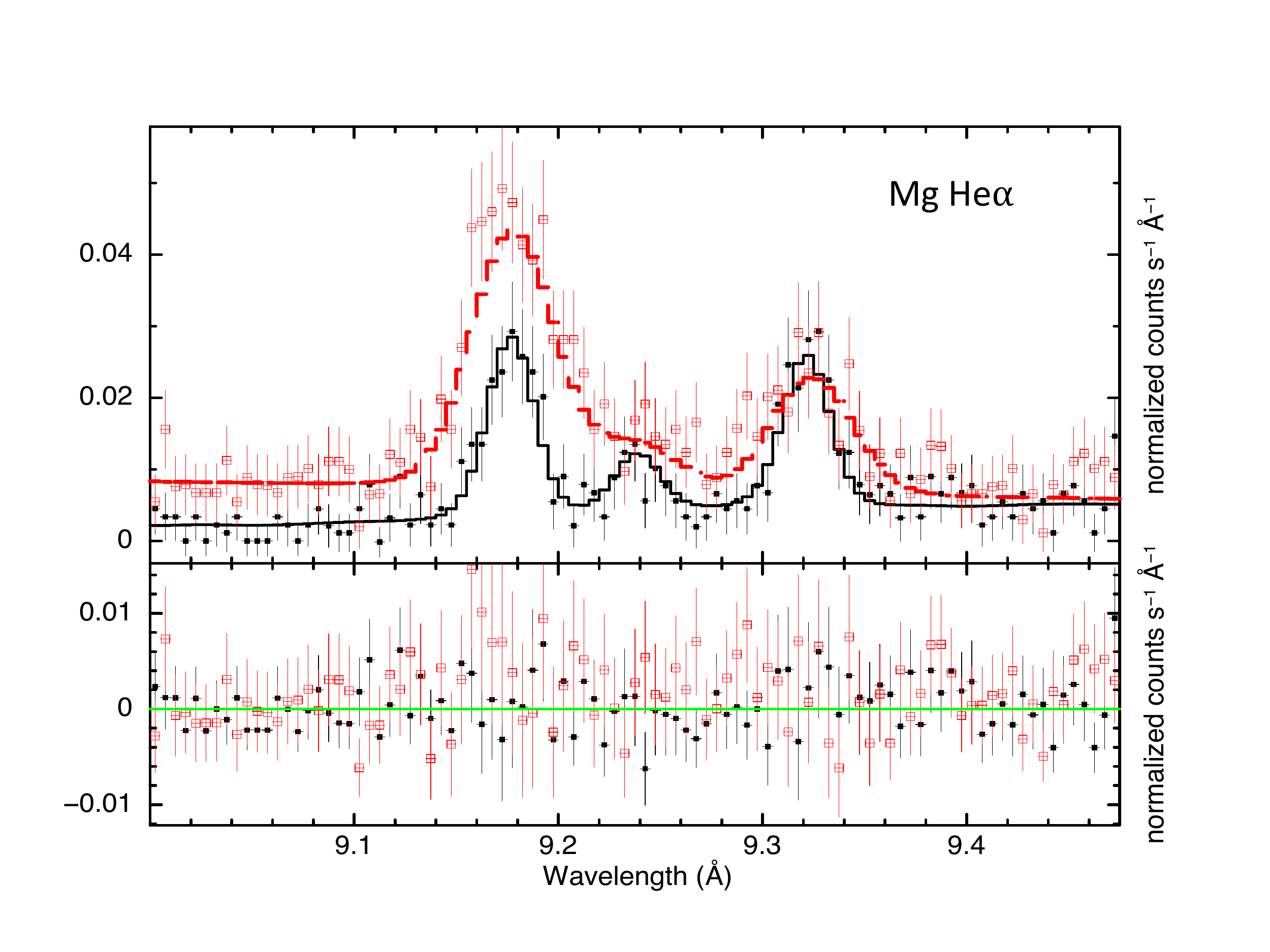} &  
   \includegraphics[width=0.33\textwidth]{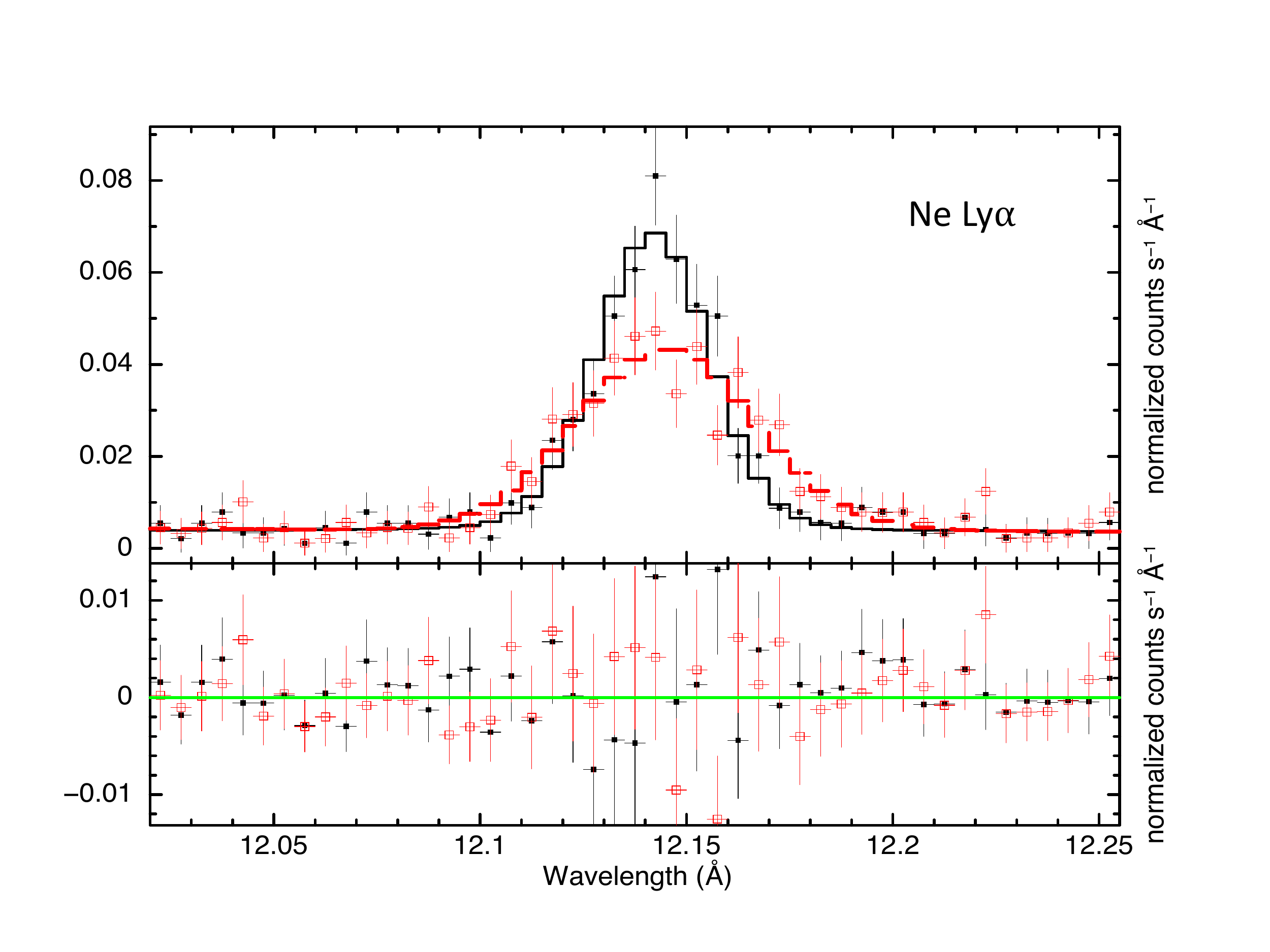} \\ 
   (j)MEG 2011: Mg Ly$\alpha$  & (k)  MEG 2011: Mg He$\alpha$ & (l) MEG 2011: Ne Ly$\alpha$
\end{tabular}

\caption{Examples of line profiles in the X-ray spectrum of SNR 1987A. Panels a - f and g - l show MEG 2018 and MEG 2011 data, respectively. In all panels, black is MEG +1 and red is MEG -1. The best-fit Gaussian models for $\pm$1 grating spectra are overlaid in all panels with black solid lines and red dashed lines, respectively. \label{fig:3}}

\end{figure*}

High-resolution HETG/LETG spectroscopy provides an excellent opportunity for line diagnostics to investigate the thermal condition of the X-ray emitting hot gas, complementary to the broadband spectral model fits. In the broadband spectrum (Figure \ref{fig:1}), we identify several emission lines from He-like and/or H-like ions of S, Si, Mg and Ne.  We fit the He$\alpha$ triplet lines (from He-like ions) with three Gaussian profiles and an underlying constant continuum, in the ``narrow" bands around the emission lines. Similarly, for the Ly$\alpha$ and Ly$\beta$ doublet lines (from H-like ions), we use two Gaussian profiles and an underlying constant continuum. For strong He$\alpha$ lines of Si and Mg we clearly resolve resonance (\textit{r}), and forbidden (\textit{f}) lines, while their intercombination (\textit{i}) lines are faint, and are not clearly distinguishable (Figure \ref{fig:3}). Thus, we allow the line ratios (\textit{f}/\textit{r} and \textit{i}/\textit{r}) to vary in the fits. For these emission lines from H- and He-like ions, we fixed the line centers at their lab measured values from ATOMDB v3.0.9. We include a redshift parameter for each emission line to account for the line-of-sight velocity effects on the line center wavelengths. This redshift parameter is free for each individual Ly$\alpha$, Ly$\beta$, and He$\alpha$ multiplet, to account for any possible differential line shifts between them. Thus, free parameters in our  He$\alpha$ line model are the FWHM, normalizations for Gaussians, \textit{f}/\textit{r} and \textit{i}/\textit{r} ratios, continuum normalizations, and redshifts. For our models of the Ly$\alpha$ and Ly$\beta$ lines, we assumed an equal flux from each of the sub-component lines based on ATOMDB v3.0.9, because our HETG data cannot resolve them individually. Free parameters are the FWHM, Gaussian and continuum normalizations, and redshifts.
\setcounter{table}{2}
\begin{deluxetable*}{lccccccc}~
\tabletypesize{\footnotesize}
\tabletypesize{\scriptsize}
\tablecolumns{4}
\tablecaption{SNR 1987A: Line Fluxes}
\tablewidth{0pt}
\tablehead{\colhead{Lines} & \colhead{Ionization State}&  \colhead{$\lambda ^{a}_{lab}$} & \colhead{Flux$^{b}$} & \colhead{Flux$^{b}$} &  \colhead{Flux$^{b}$} & \colhead{Flux$^{b}$} & \colhead{Flux$^{b}$} \\
                        &   & (\AA{}) & LETG 2004  & HETG 2007 & LETG 2007 & HETG 2011 & HETG 2018}

\startdata
S He$\alpha$ & XV & 5.039 &  3.5 $\pm$ 1.5   & 8.7 $\pm$ 2.6 & 6.0 $\pm$ 2.1 & 20.6 $\pm$ 3.5 & 27.3 $\pm$ 3.6 \\ 
Si Ly$\alpha$ & XIV & 6.180 & 2.5 $\pm$ 0.6    & 6.0 $\pm$ 0.7 & 11.4 $\pm$ 1.5 & 18.3 $\pm$ 1.5 & 28.1 $\pm$ 1.4 \\
Si He$\alpha$  & XIII & 6.648 & 13.8 $\pm$ 1.2  & 34.6 $\pm$ 1.3 & 46.0 $\pm$ 2.1  & 79.4 $\pm$ 2.7 & 76.5 $\pm$ 2.2 \\
Mg Ly$\alpha$ & XII& 8.419 &  4.9 $\pm$ 2.1  & 12.6 $\pm$ 0.8 & 17.8 $\pm$ 2.1 & 35.6 $\pm$ 1.8 & 37.4 $\pm$ 1.7 \\
Mg He$\alpha$ & XI& 9.169 & 19.5 $\pm$ 2.6  & 47.5 $\pm$ 2.4 & 59.4 $\pm$ 5.2 & 82.1 $\pm$ 4.6 & 53.6 $\pm$ 4.4\\
Ne Ly$\beta$ & X& 10.238 & ... & 12.4 $\pm$ 2.3 & 4.8 $\pm$ 3.2 & 21.2 $\pm$ 3.8 & 17.9 $\pm$ 4.4 \\
Ne Ly$\alpha$ & X& 12.138 & 42.0 $\pm$ 3.1  & 86.0 $\pm$ 3.6 & 107.4 $\pm$ 5.1 & 155.3 $\pm$ 7.3 & 114.0 $\pm$ 8.5 \\
Ne He$\alpha$ & IX& 13.447 & 69.5 $\pm$ 4.7  & 146.5 $\pm$ 8.1  & 188.5 $\pm$ 8.7 & 186.4 $\pm$ 15.5 & 128.6 $\pm$ 28.1 \\[3pt]
\enddata
\tablenotetext{}{$^{a}$ The laboratory wavelength of the strongest component within the multiplet.\\
$^{b}$ The observed total multiplet flux in units of $10^{-6}$ photons cm$^{-2}$ s$^{-1}$ and associated $1\sigma$ uncertainties. \label{tab:3}\\}
\end{deluxetable*}

Based on the best-fit redshift parameter, we find that the centroid shifts for all the strong emission lines are consistent (within statistical uncertainties) with the optical estimates of SNR 1987A redshift of $\sim$286 km s$^{-1}$ \citep{2008A&A...492..481G}. We also find that the fluxes between positive and negative arms are consistent within statistical uncertainties. Thus, we fit all four of the HETG $\pm$1 (MEG $\pm$1 and HEG $\pm$1) spectra simultaneously, sharing normalizations, line centers, line flux ratios and redshifts. For the LETG spectra, we adopt a similar method for the LEG $\pm$1 spectral fits. We untied their FWHMs to account for differential widths that are caused by the tilted orientation of the ER with respect to the plane of the sky \citep[inclination angle $i=$  44 - 45$^\circ$;][]{1995ApJ...439..730P} with the northern rim towards Earth. In Figure \ref{fig:3} we show example best-fit line profiles from our Gaussian model fitting for He$\alpha$ and Ly$\alpha$ lines from Si, Mg, S, and Ne ions.
\begin{figure*}[htbp]
\centering
\begin{tabular}{cc}
   \includegraphics[width=0.5\textwidth]{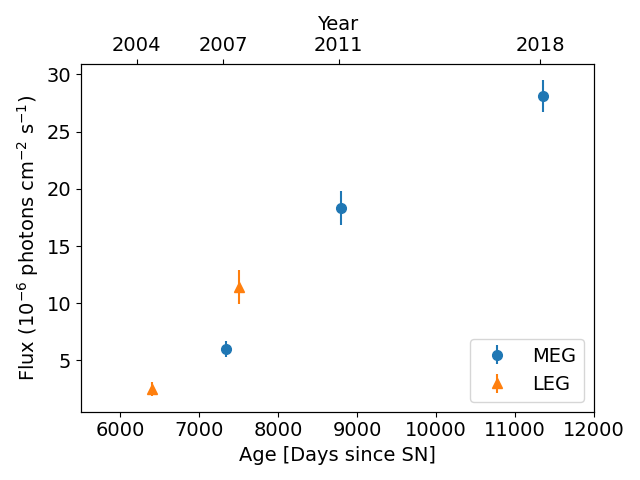} &
   \includegraphics[width=0.5\textwidth]{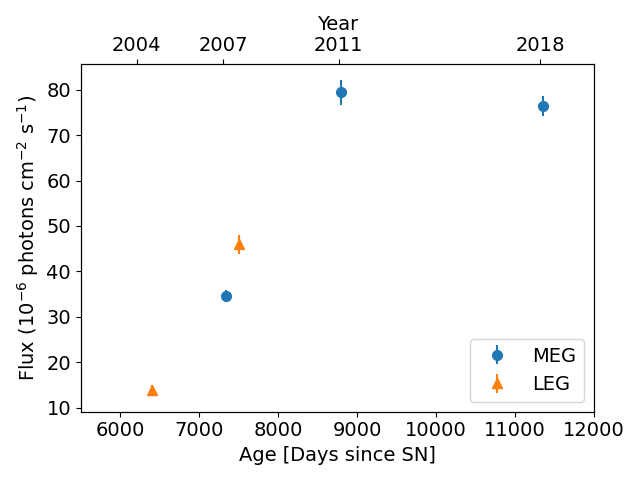} \\
   (a) Si (Ly$\alpha$) & (b) Si (He$\alpha$)\\ 
   \includegraphics[width=0.5\textwidth]{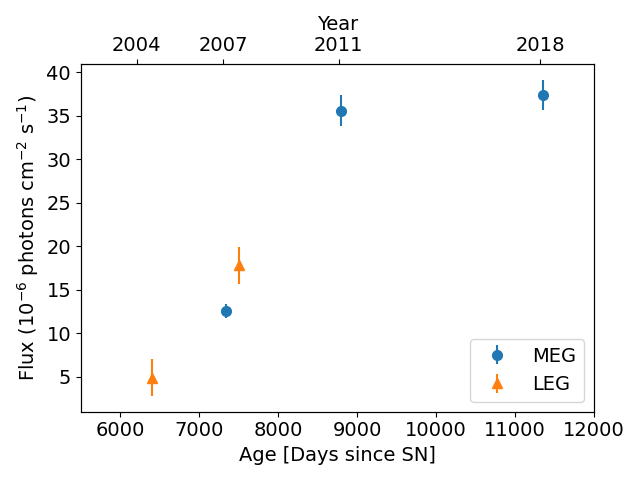} &
   \includegraphics[width=0.5\textwidth]{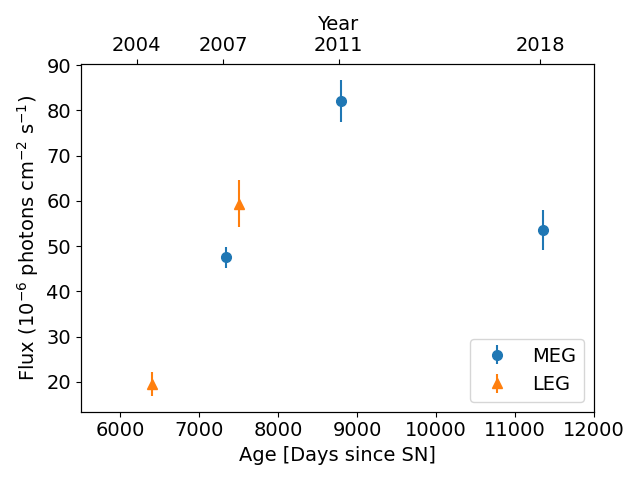} \\
   (c) Mg (Ly$\alpha$) & (d) Mg (He$\alpha$) \\ 
   \\
   \includegraphics[width=0.5\textwidth]{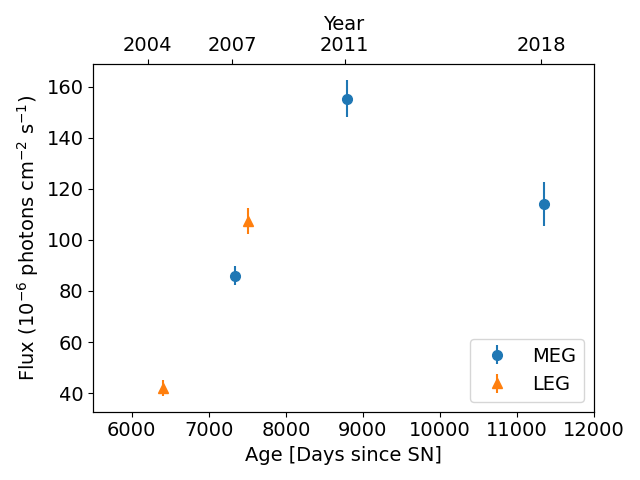} &
   \includegraphics[width=0.5\textwidth]{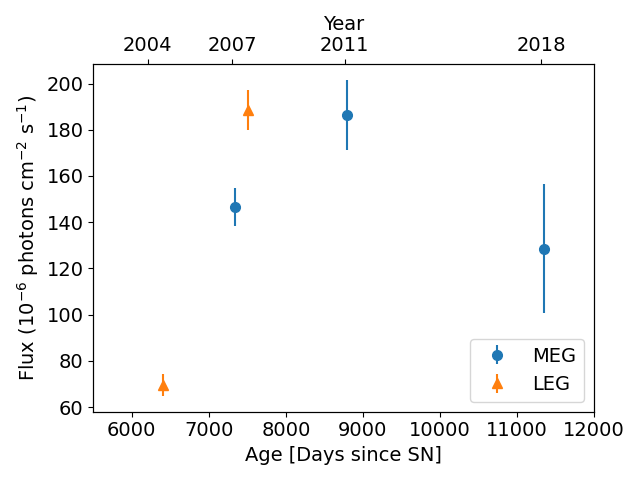} \\
   (e) Ne (Ly$\alpha$) & (f) Ne (He$\alpha$) \\  
\end{tabular}
\caption{Temporal flux variations of some strong emission lines of Si (a,b), Mg (d,e), Ne (g,h) ions based on  X-ray grating spectra (MEG/HEG $\pm$1, LEG $\pm$1) of SNR 1987A.\label{fig:4}}

\end{figure*}
In Table \ref{tab:3}, based on these line profile fits, we present our measured line emission fluxes. Our estimated fluxes for LETG 2004, HETG 2007, and LETG 2007 observations are consistent with the published values \citep{2005ApJ...628L.127Z, 2008ApJ...676L.131D, 2009ApJ...692.1190Z} within statistical uncertainties. In Figure \ref{fig:4}, we show these temporal variations of He$\alpha$ and Ly$\alpha$ line fluxes for Si, Mg and Ne in 2004 - 2018. Between 2011 and 2018, Si Ly$\alpha$ line flux has increased $\sim$51$\%$ while Si He$\alpha$ line flux has stayed roughly constant. On the other hand, Mg He$\alpha$ line flux has decreased by $\sim$37$\%$ while Mg Ly$\alpha$ line flux has stayed roughly constant. 
\begin{figure*}[h!]
\centering
\begin{tabular}{cc}
   \includegraphics[width=0.5\textwidth]{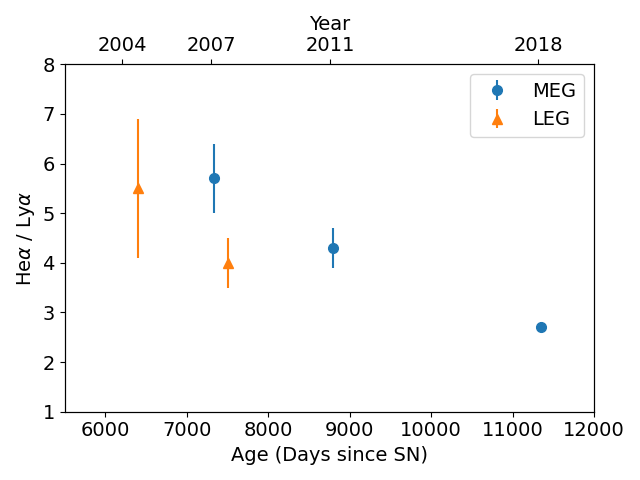} &
   \includegraphics[width=0.5\textwidth]{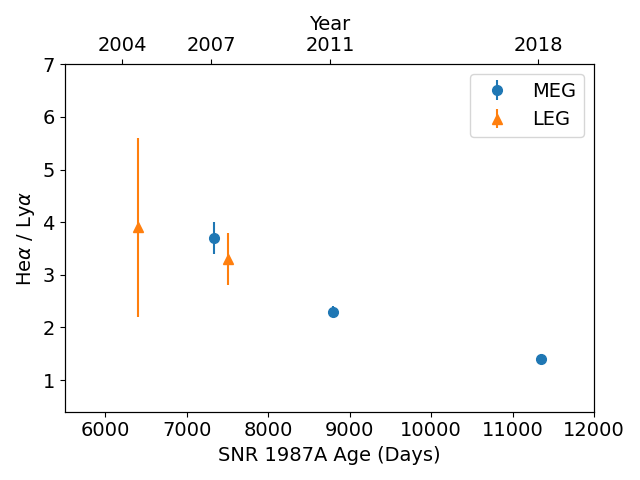}\\
   (a) Si (He$\alpha$/Ly$\alpha$) & (b) Mg (He$\alpha$/Ly$\alpha$) \\

\end{tabular}
\caption{Temporal changes of He$\alpha$/Ly$\alpha$ line flux ratios for (a) Si (Si XIII / Si XIV) and (b) Mg (Mg XI / Mg XII) based on high-resolution \textit{Chandra} gratings spectra of SNR 1987A. \label{fig:5}}
\end{figure*}
These changing trends of the He$\alpha$ and Ly$\alpha$ line fluxes of the strongest line profiles (Si and Mg) can also be quantified with the He$\alpha$/Ly$\alpha$ line flux ratios.  For both Si and Mg, He$\alpha$/Ly$\alpha$ line flux ratios have significantly decreased (by $\sim$40$\%$) between 2011 and 2018 (Figure \ref{fig:5}). Thus, between 2004 and 2018, line fluxes and line flux ratios (Si, Mg: He$\alpha$/Ly$\alpha$) have undergone statistically significant changes.\\

\section{Discussion \label{4}} 

\subsection{Temporal Evolution of X-ray Emission Spectrum \label{4.1}}

\begin{figure*}[h!]
\centering
\begin{tabular}{cc}
\includegraphics[width=0.5\textwidth]{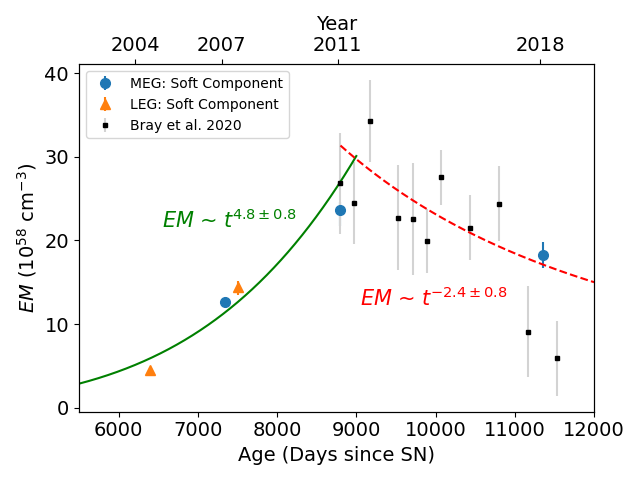} &
\includegraphics[width=0.5\textwidth]{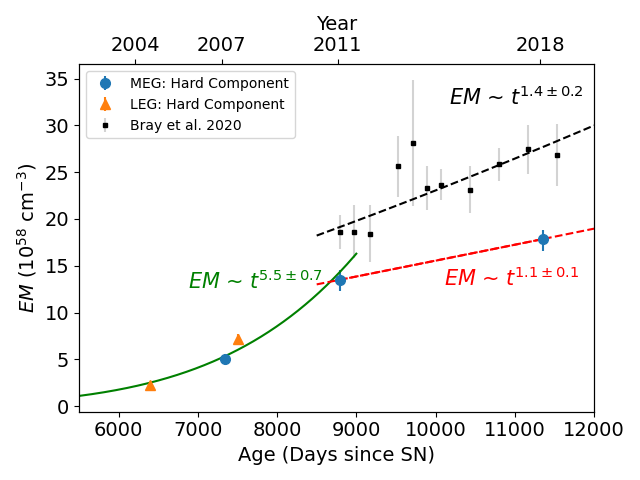}\\
 (a) Volume Emission Measure: Soft Shock Component & (b) Volume Emission Measure: Hard Shock Component \\
\end{tabular}

\caption{Temporal variation of \textit{EM} for soft (a) and hard (b) components in 2004 - 2018. The best-fit PL models are overlaid in panels (a) and (b). Statistical uncertainties (90$\%$ C.L.) are shown in both panels. \label{fig:6}}

\end{figure*}

Combining information from all five deep \textit{Chandra} gratings spectroscopic data of SNR 1987A, including our new HETG 2018 data, we explore the X-ray spectral evolution of SNR 1987A and changing thermal conditions of the X-ray emitting plasma over the 14 year period from 2004 to 2018. Volume Emission Measure (\textit{EM} $\sim n^{2}_{e} V$, where $n_{e}$ is the electron density and $V$ is the X-ray emitting volume) is considered to be a tracer of density of the shocked gas. Based on our broadband spectral model fits (Table \ref{tab:2}), we show the temporal changes in \textit{EM} for the best-fit soft and hard component spectral models (Figure \ref{fig:6}). While the \textit{EM} increase for both the soft and hard components between 2004 and 2007 are consistent (within statistical uncertainties) with results from previous works \citep{2006ApJ...645..293Z,2009ApJ...692.1190Z}, we note significant deviations in this trend since 2011. The soft component \textit{EM} has declined (by $\sim$23$\%$) between 2011 and 2018 (Figure \ref{fig:6}a), while the hard component \textit{EM} continues to increase (by $\sim$32$\%$), but at a slower rate than before 2011 (Figure \ref{fig:6}b). 

While the changes in \textit{EM} after 2011 are intriguing, we note that it is based on a minimal sample of our measurements, covering only two epochs (HETG 2011, HETG 2018) with a relatively long time separation of 7 years between them. To complement this small number of measurements, we add the published \textit{EM} measurements by \cite{2020ApJ...899...21B} in our sample. Based on similar two-component spectral model fits to our annual \textit{Chandra} HETG monitoring observation data of SNR 1987A, \cite{2020ApJ...899...21B} measured \textit{EM}s for the soft and hard component emission models. Their data were taken with shorter exposure times ($\sim$50 - 70 ks), but covered a similar time period (from 2011 to 2019) to that of our interest with a significantly higher cadence of observations at eleven epochs. In Figure \ref{fig:6}, we overlay \textit{EM} values measured by \cite{2020ApJ...899...21B}. We find that the soft component \textit{EM}s measured by \cite{2020ApJ...899...21B} between 2011 and 2019 are well aligned with our measured decline, filling the gap between 2011 and 2018. While the statistical uncertainties are large due to poor count statistics in \cite{2020ApJ...899...21B} measurements, they provide us confidence for the declining soft component \textit{EM} since 2011. The hard component \textit{EM} measurements by \cite{2020ApJ...899...21B} also show a similar trend to our measurements (Figure \ref{fig:6}b). Their hard component \textit{EM} values are systematically higher (by $\sim$30$\%$) than our corresponding measurements of 2011 and 2018. We find that this discrepancy is the model-dependent systematics between \cite{2020ApJ...899...21B} and this work (e.g., differences among plasma models adopted for the soft component, fitting the foreground absorption column, adopted versions of CALDBs, etc.). Applying exactly the same spectral model fits (as those we use in this work) for the \cite{2020ApJ...899...21B} data, we confirmed that such a discrepancy in the hard component \textit{EM}s \citep[between this work and][]{2020ApJ...899...21B} are removed.

Based on these \textit{EM} variations, we estimate the density profiles of the X-ray emitting gas. We construct a toy model in the form of a simple powerlaw (PL), $EM \sim t^{\beta}$ (where $\beta$ is the PL index). We fit the temporal changes of the soft and hard component \textit{EM}s with this simple PL model (Figure \ref{fig:6}). The best-fit PL indices for the soft component \textit{EM} variation are $\beta =$ 4.8 $\pm$ 0.8 until 2011 and $\beta =$ -2.4 $\pm$ 0.8 after 2011, indicating the clear turn-over since 2011. For the hard component \textit{EM} variation, $\beta =$ 5.5 $\pm$ 0.7 until 2011 and $\beta =$ 1.1 $\pm$ 0.1 afterwards. While the change in the PL slope for the hard component is less pronounced than for the soft component, the decrease in the PL index is evident for the hard component \textit{EM} after 2011. Based on our current physical picture with previous \textit{Chandra} spectral analyses \citep[e.g.,][]{2002ApJ...567..314P, 2004ApJ...610..275P, 2006ApJ...646.1001P, 2009ApJ...692.1190Z, 2006ApJ...645..293Z, 2008ApJ...676L.131D, 2013ApJ...764...11H, 2016ApJ...829...40F, 2020ApJ...899...21B}, the soft component is dominated by emission from the shocks interacting with high density clumps in the ER, and the hard component may represent a combination of shock interactions with lower density inter-clump regions in the ER and less dense CSM (that may include regions outside of the ER). Thus, the X-ray emitting volumes represented by the soft and hard components could be different, and we estimate their associated density profiles independently.

As the X-ray emission characterized by the soft component is dominated by emission from the dense ER, we assume the soft component X-ray emitting volume approximately to be the volume of the ER. Early \textit{HST} images of SN 1987A show that the ER has a thin disk-like ring geometry \citep{1995ApJ...439..730P}. The volume for such a disk-like ring can be expressed as $V \sim h \times (r_{out}^{2} - r_{in}^{2})$, where $h$ represents the vertical thickness (height) of the disk, and $r_{in}$ and $r_{out}$ are the inner and outer radii of the disk, respectively. The height $h$ of the ER was estimated to be confined within $\pm$30$^{\circ}$ of the equatorial plane based on modeling of the Ly$\alpha$ line in the \textit{HST} spectra \citep{1998ApJ...509L.117M, 2003ApJ...593..809M}. As no significant evolution of this thickness is observed, we can assume that $h$ has been relatively constant. The soft X-ray spectrum of SNR 1987A is dominated by emission from a disk-like volume of the ER \citep[e.g.,][]{2005ApJ...628L.127Z, 2006ApJ...645..293Z}. Thus, we consider that the X-ray emission originates from a disk-like ring for which the emission volume would be proportional to $r_{out}^{2} - r_{in}^{2}$ $\sim$ $r^{2}$. Assuming $n_{e} \sim r^{\nu}$ (where \textit{r} is the distance from the SN site, and $\nu$ is the density profile index) and $V_{soft} \sim r^{2}$, \textit{EM} for the soft component can be expressed as $EM_{soft} \sim r^{2\nu + 2}$. From morphological studies of ACIS images of SNR 1987A, the expansion rate of the X-ray ring is linear in time ($r \sim t$) since $\sim$2004 \citep{2009ApJ...703.1752R, 2016ApJ...829...40F}. Since all our epochs in this work (2004 - 2018) fall within this linear regime, the volume emission measure is $EM \sim t^{\beta} \sim t^{2\nu+2}$, and thus, $\beta = 2\nu + 2$. Our best-fit values of $\beta$ for the soft component thus indicate the associated density profiles of $n_{e} \sim r^{1.4 \pm 0.4}$ in 2004 - 2011 and $n_{e} \sim r^{-2.2 \pm 0.4}$ in 2011 - 2018. While our PL analysis with a disk-like emission volume might be an over-simplification of a physically more complicated geometry, a significant change in the density profile of the shocked gas responsible for the soft component X-ray emission since 2011 is evident.

\begin{figure}[htbp]
\centering
\begin{tabular}{c}
\includegraphics[width=0.5\textwidth]{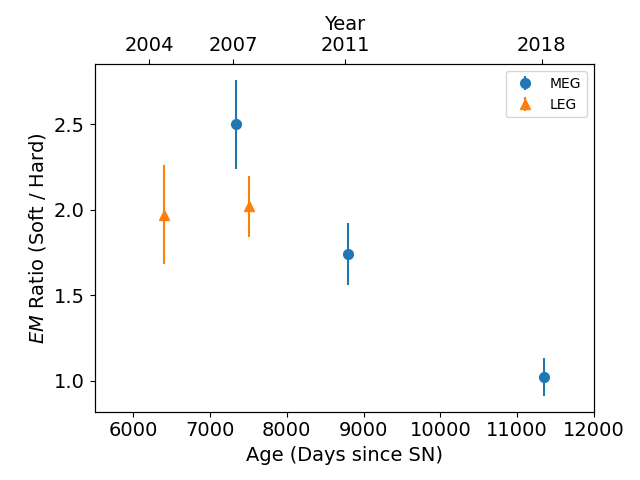}\\
  Volume Emission Measure Ratios: Soft / Hard \\
\end{tabular}
\caption{Temporal evolution of the \textit{EM} ratio between the soft and hard components in 2004 - 2018. \label{fig:7}}
\end{figure}

In Figure \ref{fig:7} we present the time evolution of the the soft-to-hard component \textit{EM} ratio. It shows a significant decrease (by a factor of $\sim$2) in 2018, compared to those in 2004 - 2011. Such a drop indicates that the soft X-ray emission from the shocked dense clumpy gas in the ER is less significant in 2018 compared to 2004 - 2011. We note that our derived density profiles associated with the soft component post 2011 ($n_{e} \sim r^{-2.2 \pm 0.4}$) are in plausible agreement with the standard density profile of red supergiant (RSG) winds ($\rho \sim r^{-2}$). Thus, the X-ray emission associated with the soft shock component appears to have increasing contributions from the low-density CSM (beyond the dense ER) since 2011, probably created by the massive progenitor's stellar winds in its RSG stage before the SN explosion.

The observed X-ray spectrum fitted by the hard component shock model has been likely dominated by the emission from the low-density inter-clump regions in the ER. In our recent \textit{Chandra} observations, this hard component X-ray emission may also include some contribution from the shocked gas beyond the ER \citep{2016ApJ...829...40F, Sun_2021, 2021ApJ...916...76A}. As a zeroth-order approximation, we consider a spherical shell-like volume with a radius of the blast wave ($r_{b}$) for the hard component X-ray emitting gas. In such a case, the emission volume for the hard component is approximated to be $V_{hard} \sim r_{b}^{3}$. While changes in the temporal variation of the hard component \textit{EM} (after 2011) are less prominent than those in the soft component \textit{EM}, a single PL model fit significantly underestimates the measured hard component \textit{EM} (by $\sim$60$\%$) in 2018. We find that a two-component PL fit is required to adequately fit the entire \textit{EM} variation including the 2018 data (e.g., $\chi_{\nu}^{2} \sim$17 for the single PL model fit, and $\chi_{\nu}^{2} \sim$3 for the two-component PL fit, Figure \ref{fig:6}b). Thus, we consider the two-component PL model fit to adequately describe the temporal variation of the hard component \textit{EM} between 2004 and 2018, with the break around 2011, similar to its softer counterpart. As discussed in Section \ref{4.1}, we note that there are systematic differences between hard component \textit{EM} values between deep \textit{Chandra} grating (our work) and monitoring grating spectra \citep{2020ApJ...899...21B}. Based on the hard component \textit{EM} measurements by \cite{2020ApJ...899...21B}, we obtain the best-fit PL index $\beta =$ 1.4 $\pm$ 0.2, which is in general agreement (within statistical uncertainties) with that fitted to our deep \textit{Chandra} HETG spectra taken in 2011 and 2018. Thus, best-fit values of $\beta$ for the hard component \textit{EM} evolution based on measurements with the deep HETG/LETG observations ($\beta =$ 5.5 $\pm$ 0.7 in 2004 - 2011 and $\beta =$ 1.1 $\pm$ 0.1 in 2011 - 2018) and the annual monitoring grating observations ($\beta =$ 1.4 $\pm$ 0.2 in 2011 - 2019) both indicate that the shocks responsible for the hard component have been interacting with less dense CSM since 2011.

For our assumed simple spherical geometry ($V_{hard} \sim r_{b}^{3}$) for the X-ray emitting gas outside the ER,  we consider self-similar solutions for the SNR dynamics \citep{1982ApJ...258..790C}. The radius of the blast wave in such a solution can be expressed as $r_{b} \sim t^{(n-3)/(n-s)}$, where $n$ is the PL index for the SN ejecta density profile (i.e., $\rho_{ejecta} \sim r_{b}^{-n}$) and $s$ is the PL index for the density profile of the CSM (i.e., $\rho_{CSM} \sim r_{b}^{-s}$). For SN 1987A, we assume $n$ = 9 \citep{1989ApJ...347..771E, 1993A&A...274..883S, 1997ApJ...476L..31B, 1997ApJ...477..281B}. For the two representative cases $s$ = 0 (constant density) and $s$ = 2 (RSG wind density), the radius of the blast wave would be $r_{b} \sim t^{2/3}$ (constant density) and $r_{b} \sim t^{6/7}$ (RSG wind density). With our adopted spherical X-ray emitting  geometry, the hard component \textit{EM} is \textit{EM$_{hard}$} $\sim \rho_{CSM}^{2} V_{hard}$ $\sim \rho_{CSM}^{2} r_{b}^{3}$. Thus, $EM_{hard} \sim t^{2}$ and $EM_{hard} \sim t^{-6/7}$ are inferred for the two representative cases of constant and RSG wind densities for the interacting CSM, respectively. Our derived PL index post 2011 ($\beta =$ 1.1 $\pm$ 0.1) for the hard component is in between what we expect for the constant density and the RSG wind-density cases, suggesting a transitionary phase for the blast wave interacting with the ER densities and lower ambient CSM densities outside it. A new phase of SNR 1987A's evolution, in which the shock starts interacting with the low-density CSM beyond the dense ER, has been proposed to have started at around $\sim$2016 ($\sim$10400 days) based on our previous \textit{Chandra} monitoring data \cite{2016ApJ...829...40F}. \textit{HST} observations show that the optical emission from the ER peaked around around 2009 ($\sim$8000 days) and has been fading since then, signalling that the blast wave has started to leave the ER \citep{2015ApJ...806L..19F}. While the physical processes leading to X-ray and optical emission might be different, our estimated ``switching" period at around 2011 ($\sim$8500 days) for both soft and hard components (Figure \ref{fig:6}) is in plausible agreement with those previously suggested phase transitions.

\begin{figure*}[h!]
\centering
\begin{tabular}{cc}
  \includegraphics[width=0.5\textwidth]{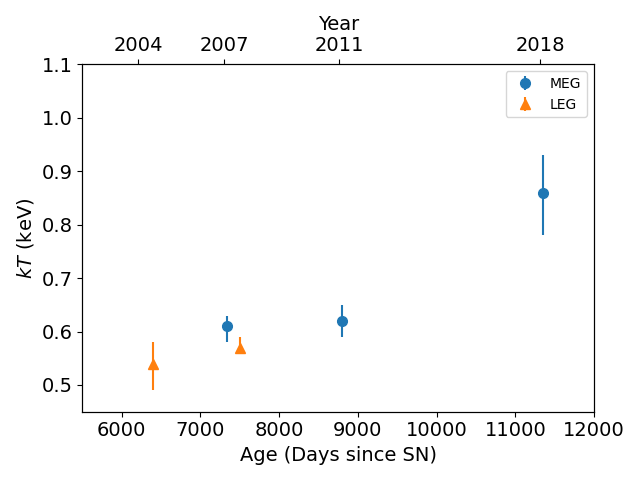} & \includegraphics[width=0.5\textwidth]{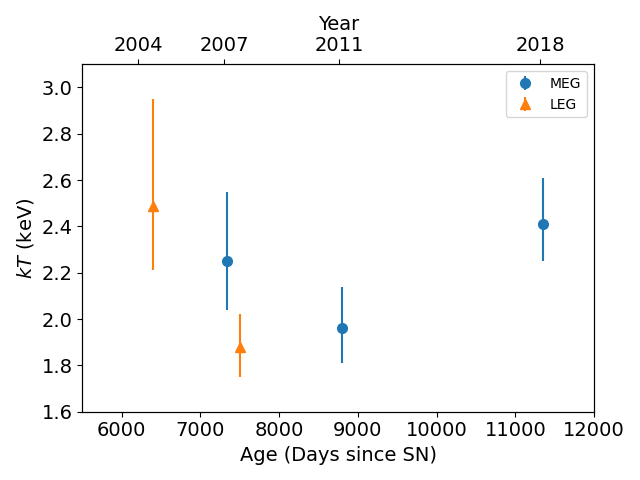}   \\
(a) Electron Temperature: Soft Shock Component & (b) Electron Temperature: Hard Shock Component \\[6pt]
\end{tabular}
\caption{Evolution of the post-shock electron temperature (\textit{kT}) of SNR 1987A for (a) the soft and (b) the hard components. Statistical uncertainties (90$\%$ C.L.) are shown in both panels.\label{fig:8}}  

\end{figure*}

Based on the broadband spectral model fits (Table \ref{tab:2}), we show the evolution of electron temperatures (\textit{kT}) between 2004 - 2018 (Figure \ref{fig:8}). The soft component electron temperature has clearly increased between 2011 and 2018 (by $\sim$38$\%$). While it is statistically marginal due to large uncertainties, the hard component electron temperature shows a similar increase ($\sim$23$\%$) during this period. In contrast, such a large increase in \textit{kT} was not evident in 2004 - 2011 for either the soft or hard components. Similar temporal changes of \textit{kT} between 2011 and 2018 have been reported in the literature based on the annual monitoring data of SNR 1987A with \textit{Chandra} HETG \citep{2020ApJ...899...21B}. Our results in 2004 - 2007 for both the soft and hard components are consistent (within statistical uncertainties) with those previously reported by \cite{2009ApJ...692.1190Z} for the same epochs. Such increases in \textit{kT} for both the soft and hard components in the new HETG 2018 data are in plausible agreement with the overall decreasing density of the shocked gas since 2011.

\begin{figure*}[htbp]
\centering
\begin{tabular}{cc}
  \includegraphics[width=0.5\textwidth]{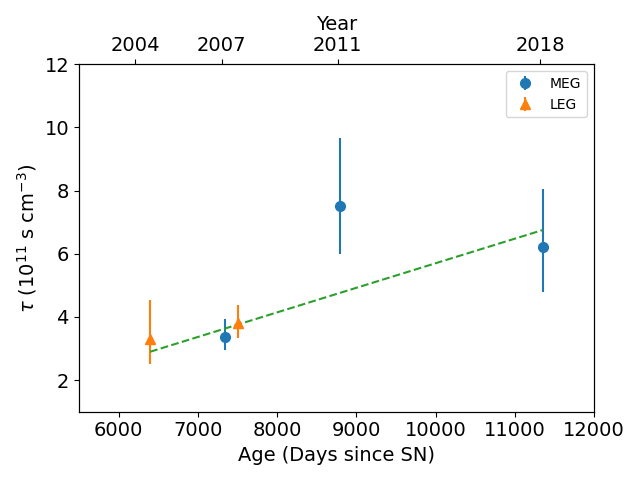} & \includegraphics[width=0.5\textwidth]{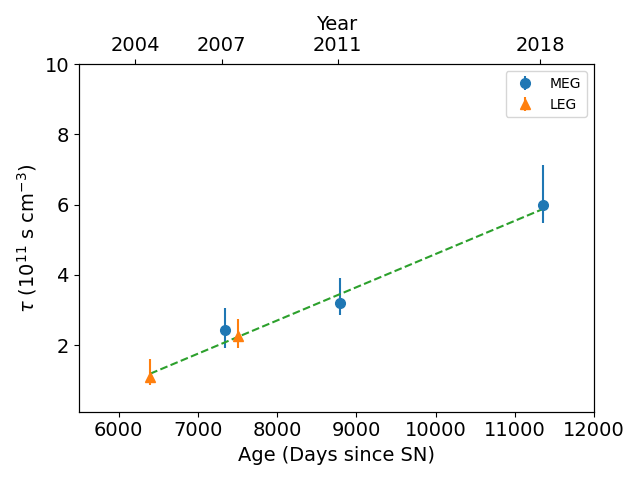}   \\
(a) Ionization Age: Soft Shock Component & (b) Ionization Age: Hard Shock Component  \\[6pt]
\end{tabular}
\caption{Evolution of the ionization age ($\tau = n_{e}t$) of SNR 1987A for (a) the soft and (b) the hard components. A best-fit linear model (dashed line) is overlaid in (a) and (b). Statistical uncertainties (90$\%$ C.L.) are shown in both panels. \label{fig:9}}
\end{figure*}

We present the evolution of the ionization age over the same period in Figure \ref{fig:9}. The soft component ionization age increases between 2004 and 2011 as the dense gas is shocked within the ER. While statistical uncertainties are large, the soft component ionization age appears to have leveled off or is even marginally decreasing since 2011 (Figure \ref{fig:9}a). These results generally support the decreasing density for the shocked gas responsible for the soft component. Such a leveling off or a marginal decrease of the soft component ionization age since 2011 may also indicate that some of the densest material in the ER (representing the highest $\tau$) becomes cold (possibly through radiative cooling), that its emission falls out of the X-ray band, making only the material close to the shock front dominate the observed X-ray spectrum.  The hard component ionization age increases linearly (by $\sim$87$\%$) from 2011 to 2018 (Figure \ref{fig:9}b). Thus for both \textit{kT} and ionization age, the soft and hard components are becoming less distinguishable since 2011. Such a ``merging" trend could be a combined effect of increasing contribution from the low density shocked gas around the ER (beyond and/or above and below the ER) and of the gradual thermalization of the shocked plasma of the ER due to Coulomb collisions. Future high-resolution X-ray spectroscopic observations would be required to adequately trace and constrain the changing physical conditions in the evolution of SNR 1987A.\\

\subsection{Emission Measure Distribution: Magnetohydrodynamic (MHD) Simulations \label{4.2}}

We compare the results from our \textit{Chandra} HETG spectral analysis of SNR 1987A with the 3-D MHD simulations by \cite{2019A&A...622A..73O}. It is worth noting that these MHD simulations are not the result of fitting to the X-ray data rather they have been constrained by X-ray data collected before 2016. In particular, the model parameters have been explored  within ranges of values inferred from optical and X-ray observations, performing an iterative process of trial and error to converge on model parameters that best reproduce the X-ray light curves (soft and hard bands) and moderate-resolution spectra of SN 1987A \citep[see][]{2015ApJ...810..168O, 2019A&A...622A..73O, 2020A&A...636A..22O}. Then the X-ray emission has been synthesized from the best model by adopting the approach described in a number of papers \citep[e.g.,][]{2015ApJ...810..168O, 2019NatAs...3..236M} and based on the NEI emission model \verb|vpshock|, available in the XSPEC package. Thus the model is not fully independent of the X-ray data (having been constrained with them) but it provides a single 3-D physical (not phenomenological) model that reproduces self-consistently the X-ray light curves, spectra (even high-resolution dispersed spectra not used to constrain the model) and morphology of the remnant at all epochs (even after 2016, the date of the last X-ray observations used to constrain the model), since the SN event. 

\begin{figure*}[!htbp]
\centering
\begin{tabular}{c}
\includegraphics[width=0.97\textwidth,height=0.97\textheight,keepaspectratio]{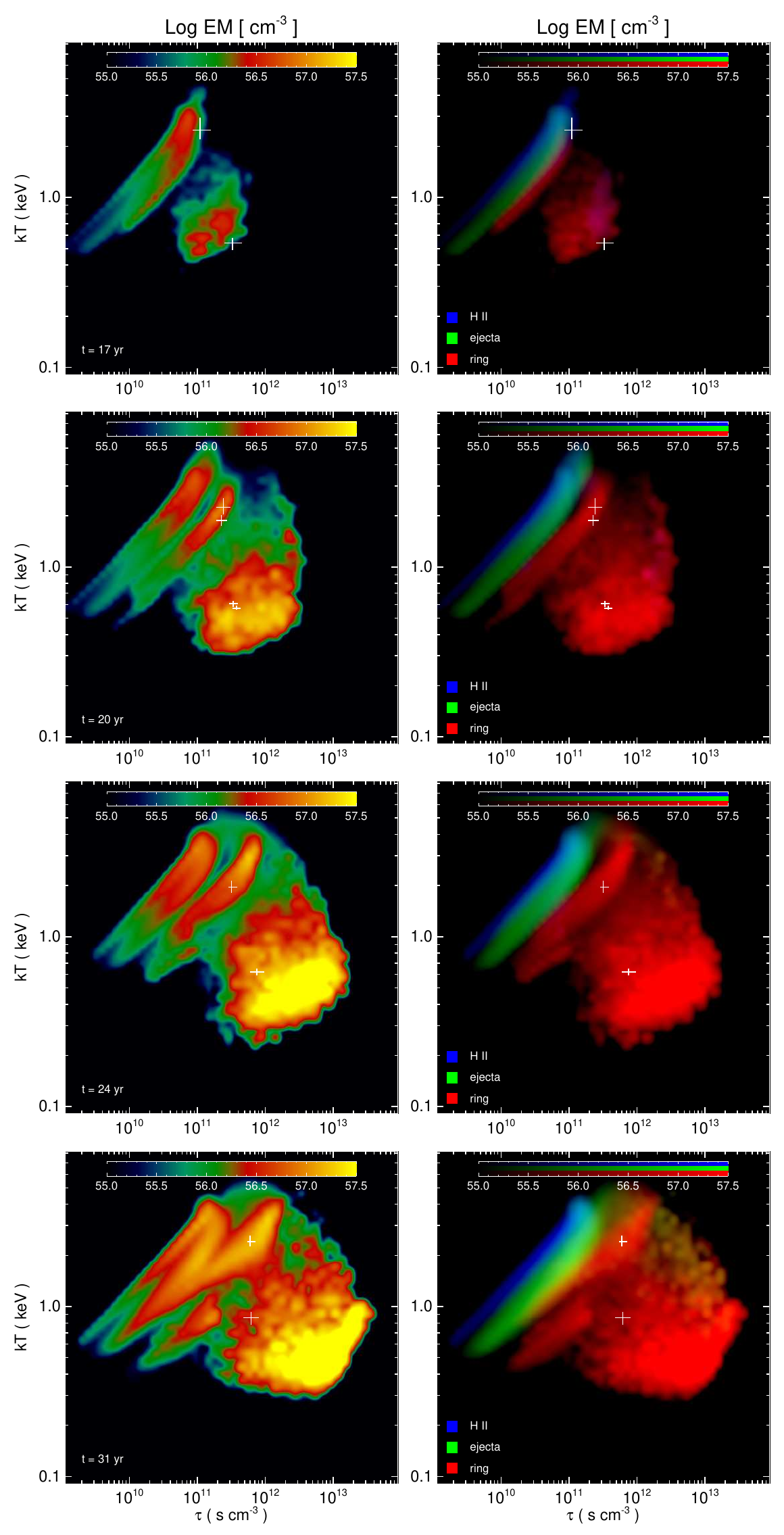}
\end{tabular}
\caption{Left panels show the \textit{EM} distributions from a 3-D MHD simulation \citep{2019A&A...622A..73O} for epochs from 2004 (\textit{t} = 17 yr) to 2018 (\textit{t} = 31 yr). In right panels, the three colors show the contribution to \textit{EM} from the different shocked plasma components, namely the H II region (blue), ejecta (green), and the dense ER or ring (red). Our measured \textit{kT} and $\tau$ (and their uncertainties) for the soft and hard components are marked with two white crosses in each panel. For 2007 (\textit{t} = 20 yr), two sets of measurements (HETG 2007, LETG 2007) with four white crosses have been overlaid.\label{fig:10}}

\end{figure*}

In the left panels of Figure \ref{fig:10}, we show the simulated \textit{EM} distribution relative to the electron temperature and ionization age of the shocked gas in SNR 1987A from 2004 (\textit{t} = 17 yr) to 2018 (\textit{t} = 31 yr). The contributions from each of the three assumed components (H II region, ER, and SN ejecta) are shown in the right panels of Figure \ref{fig:10}. The model predicts a peak of \textit{EM} for the shocked clumps of the ER at higher ionization age and lower temperature than the isothermal components fitting the observed spectra. This discrepancy might have been caused in part by our simple approach in the broadband spectral model fits, where we fit the observed spectra with only two characteristic components rather than the physically more realistic ``continuous" distributions of electron temperatures. A continuous distribution of electron temperatures have been considered to describe the X-ray spectrum of SNR 1987A in previous works \citep[\textit{Chandra}:][\textit{XMM-Newton}: \citealp{2021ApJ...916...76A}]{2006ApJ...645..293Z, 2009ApJ...692.1190Z}. These models, even if they allow continuous temperature distributions, show peaks around the characteristic temperatures of the discrete models. The peak of the modeled \textit{EM} gradually shifts to higher \textit{kT} and $\tau$ as the shock evolves. This is in general agreement with the observed increases in \textit{kT} and $\tau$ for both of the soft and hard component X-ray emission spectra, indicating the progressive equilibration of the shocked gas. Increasing \textit{kT} from the MHD models and observed decreasing \textit{EM} ratios between the soft and hard components since 2011 (Figure \ref{fig:9}) suggest that the CSM responsible for the soft component is less dense in 2018 than before. This also supports the scenario of the CSM associated with the soft component emission being less clumpy than before. 

In 2004 (17 yr after the SN) the hard component \textit{kT} and $\tau$ as observed with \textit{Chandra} data (the upper white cross in the upper right panel of Figure \ref{fig:10}) are in between two local peaks of the MHD modeled \textit{EM} distribution, indicating contributions from both the shocked H II region (blue in the upper right panel of Figure \ref{fig:10}) and the low-density component of the shocked ER region (red in the upper right panel of Figure \ref{fig:10}). There is also some contribution expected from the shocked ejecta (in green: upper right panel) component, though at this stage of SNR 1987A evolution (\textit{t} = 17 yr), this component may include only the ejected mantle, which is not metal-rich. As the blast wave evolves, the second local peak (low-density component of the shocked ER) gradually becomes dominant over the first peak (shocked H II region), due to the increase in the amount of the shocked ER gas. In fact, from 2004 (\textit{t} = 17 yr) to 2011 (\textit{t} = 24 yr), the hard component \textit{kT} in the MHD models slightly decreases because it is most affected by the second peak (the \textit{kT} component migrates to fit the second peak). At later times, once the ER is fully shocked, no new freshly shocked ER material contributes to the \textit{EM}. Hence the peak of the \textit{EM} shifts to higher values of \textit{kT} due to the plasma thermalization through Coulomb collisions. Thus, \textit{EM} maps from MHD simulations in 2018 (\textit{t} = 31 yr: lower left and right panels) suggest that the blast wave has started propagating beyond the ER. These maps are consistent with previous findings of \cite{2016ApJ...829...40F},  in the X-ray band, of \cite{2018ApJ...867...65C} in the radio band, and of \cite{2015ApJ...806L..19F} and \cite{2019ApJ...886..147L} in the optical band, showing that the blast wave has propagated beyond the ER. These effects in the X-ray band have also been corroborated by recent results from \cite{Sun_2021}, and \cite{2021ApJ...916...76A}. The ionization age gradually increases during this time period, due to the progressive equilibration of the shocked plasma.

These \textit{EM} distribution maps based on our MHD calculations suggest that our characteristic soft component \textit{kT} represents emission from the shocked high density clumps of the ER, whereas emission from the low density component of the shocked ER gas is mainly responsible for the hard component (although a contribution from the shocked H II region is also evident in 2004; \textit{t} = 17 yr). Shifting of the \textit{EM} peak towards higher \textit{kT} is consistent with the scenario of the blast wave propagating into a low-density region beyond the dense ER. Thus, these results from the 3-D MHD calculations support our conclusions on the changing density profiles of the X-ray emitting gas as we discussed in Section \ref{4.1}. Unveiling matter beyond the ER is important to characterize the structure of the CSM sculpted by the winds of the progenitor star and thus to trace the final phases of its evolution.\\

\subsection{Line Flux Ratios \label{4.3}}

The ionic He$\alpha$/Ly$\alpha$ line flux ratios are functions of both the electron temperatures (\textit{kT}) and ionization ages ($\tau$).
\begin{figure*}[h!]
\centering
\begin{tabular}{cc}
   \includegraphics[width=0.5\textwidth]{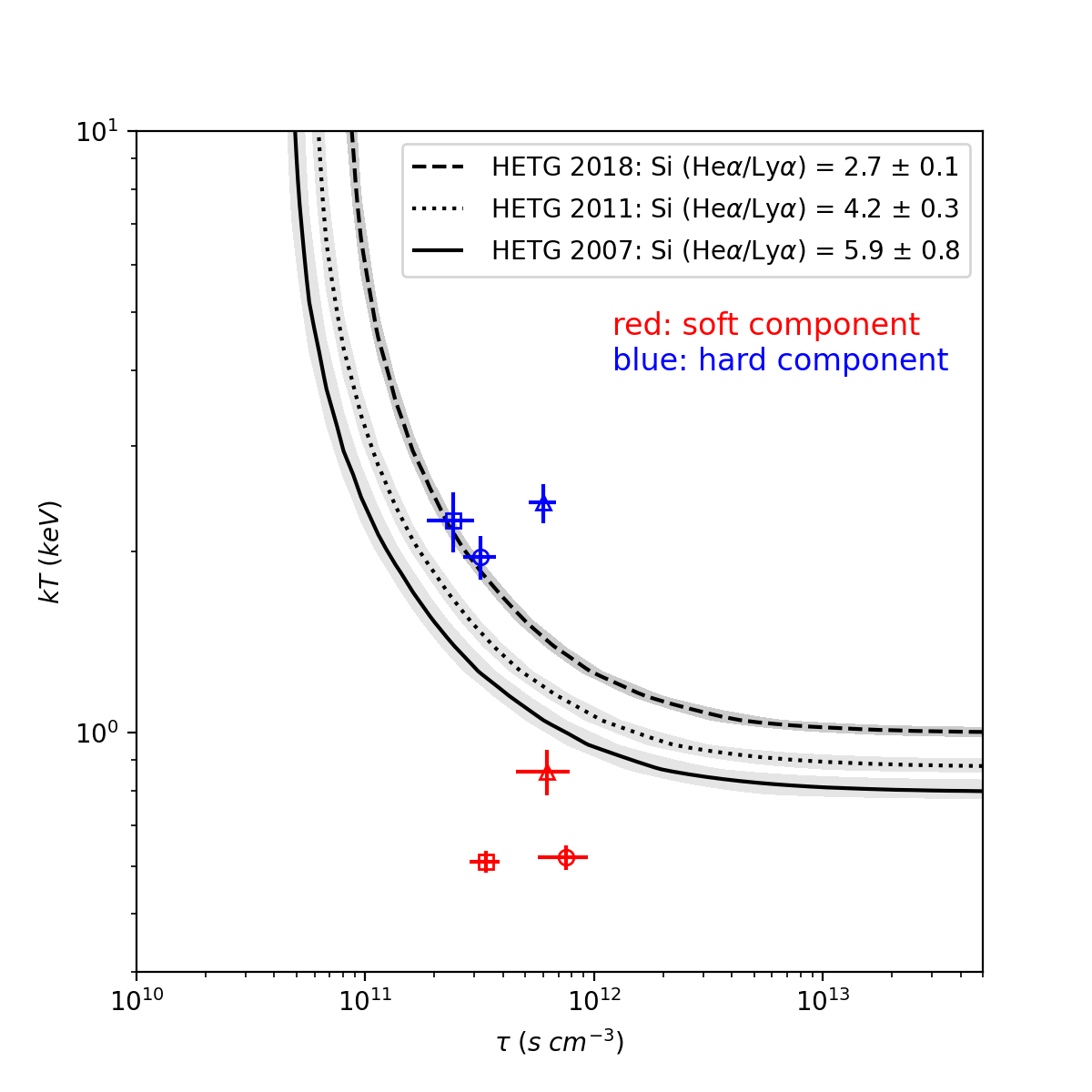}&
   \includegraphics[width=0.5\textwidth]{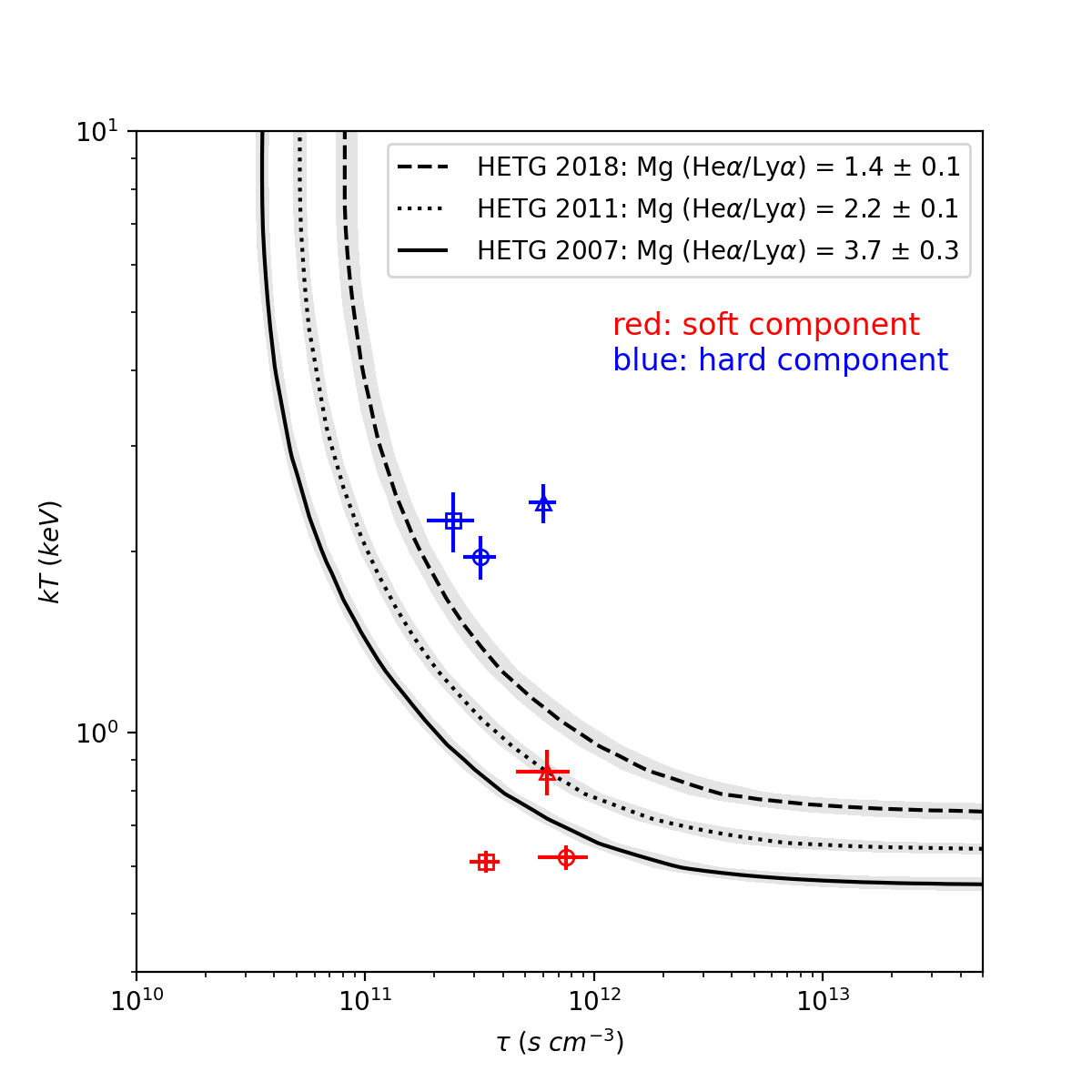} \\
   (a) Si: Isolines  & (b) Mg: Isolines \\[3pt]

\end{tabular}
\caption{Isolines of He$\alpha$/Ly$\alpha$ line flux ratios for (a) Si and (b) Mg from single plane-parallel shock model with a foreground absorption column ($N_\mathrm{H} =$ 2.17 $\times$ 10$^{21}$ cm$^{-2}$, Table \ref{fig:2}), which correspond to the observed ratios in HETG 2018 (dashed), HETG 2011 (dotted) and HETG 2007 (solid) data sets. The shaded region (grey) around each isoline represents the 1$\sigma$ uncertainty. Based on broadband spectral model fit results (Table \ref{tab:2}), combinations of best-fit \textit{kT} and $\tau$ for both soft (red) and hard (blue) components of HETG 2018 (open triangles),  HETG 2011 (open circles), and HETG 2007 (open squares) are overlaid over both panels. \label{fig:11}}

\end{figure*}
 It has been shown that a single shock component cannot explain the observed line ratios from SNR 1987A spectra \citep{2005ApJ...628L.127Z} as these are products of a distribution of shocks with a range of ionization ages and post-shock electron temperatures \citep{2006ApJ...645..293Z, 2008ApJ...676L.131D, 2009ApJ...692.1190Z}. 
 Utilizing a simple NEI plane-parallel shock model with a foreground absorbing column ($N_\mathrm{H} =$ 2.17 $\times$ 10$^{21}$ cm$^{-2}$, Table \ref{tab:2}), we show the ranges of electron temperature and ionization age which are consistent with observed He$\alpha$/Ly$\alpha$ line flux ratios for Si and Mg (Figure \ref{fig:11}). For these comparisons with observed line flux ratios, we have considered HETG 2007, HETG 2011, and HETG 2018.
 
 Based on broadband spectral model fits (Table \ref{tab:2}), the observed range of $\tau$ is $\sim$1$\times$10$^{11}$ - 8$\times$10$^{11}$ s cm$^{-3}$. In this range, we note that the \textit{kT}s corresponding to the observed line flux ratios for any given $\tau$ have steadily increased from 2007 to 2018 for both Si and Mg. In Figure \ref{fig:11}, except for a very narrow range of $\tau$ (around $\sim$1$\times$10$^{11}$ s cm$^{-3}$), the allowed \textit{kT}s for X-ray emitting plasma are likely to be in the range $\sim$0.5 - 2.0 keV (HETG 2007), $\sim$0.6 - 2.0 keV (HETG 2011) and $\sim$0.8 - 2.5 keV (HETG 2011). While there are significant overlaps in the allowed \textit{kT}s for HETG 2007, HETG 2011 and HETG 2018, these ranges have consistently shifted to higher values from 2007 to 2018. This is consistent with the broadband spectral model fits showing a continuous increase of \textit{kT} from 2007 to 2018 (Figure \ref{fig:8}). For a direct comparison, we overlay the best-fit soft and hard component \textit{kT} and $\tau$ from Table \ref{tab:2} over Si and Mg isolines in HETG 2007, HETG 2011, and HETG 2018 (Figure \ref{fig:11}a,b). We note that for all three epochs, the locus of temperatures and ionization ages suggested by the single NEI component is in between the values of \textit{kT} and $\tau$ measured with the more detailed two-component shock model fits. Allowed electron temperatures derived from line flux ratios for HETG 2007, HETG 2011, and HETG 2018 are consistent with what we find from broadband spectral model fits (Table \ref{tab:2}). Thus, our observed He$\alpha$/Ly$\alpha$ line flux ratios are in plausible agreement with the overall increase in the post-shock electron temperatures.\\
 
\subsection{Shock Kinematics \label{4.4}}

Adopting a similar approach to that used for the previous \textit{Chandra} gratings data of SNR 1987A \citep{2005ApJ...628L.127Z}, we simultaneously fit the widths of individual lines detected in both grating dispersion arms (positive and negative). We assume the same physical characteristics but different line broadening parameters (due to spatial-spectral effects of the extended source along the dispersion direction) between the positive and negative arm spectra. As a first approximation, we assume that the line profiles are Gaussian and have two sources of broadening: (1) angular extent of SNR 1987A, and (2) the bulk motion of the shocked gas. In general, lines for a given element in higher ionization states form at shorter wavelengths. Also, the higher ionization states are more populated at higher plasma temperatures. The higher derived plasma temperatures may be considered to be associated with faster shock velocities. Thus, we adopt a stratification in the line-formation zone. We include this broadening from bulk motion as $2z_{0}(\lambda / \lambda_0)^{\alpha} \lambda$, where $z_{0}$, a redshift-like parameter fits the line broadening at some fiducial wavelength, $\lambda_0$, and $\alpha$ measures the degree of stratification. The total line broadening can be expressed as 
\begin{equation} \label{eq:1}
    \Delta \lambda_{total} = 2\Delta \lambda_{0} \pm 2z_{0}(\lambda / \lambda_0)^{\alpha} \lambda 
\end{equation}
\begin{figure*}[hbp]
\centering
\begin{tabular}{cc}
\includegraphics[width=0.5\textwidth]{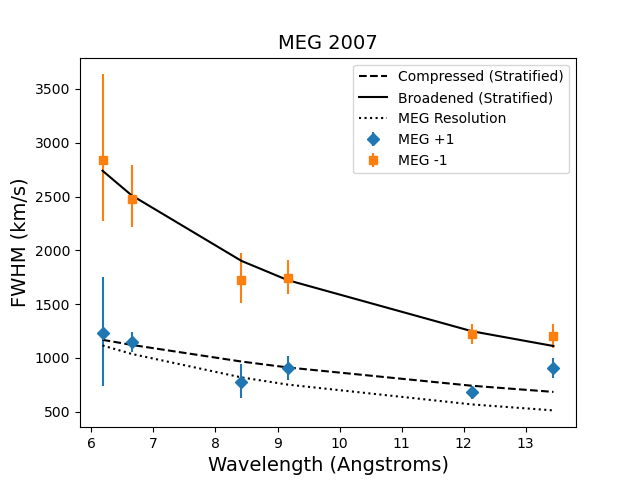} &
\includegraphics[width=0.5\textwidth]{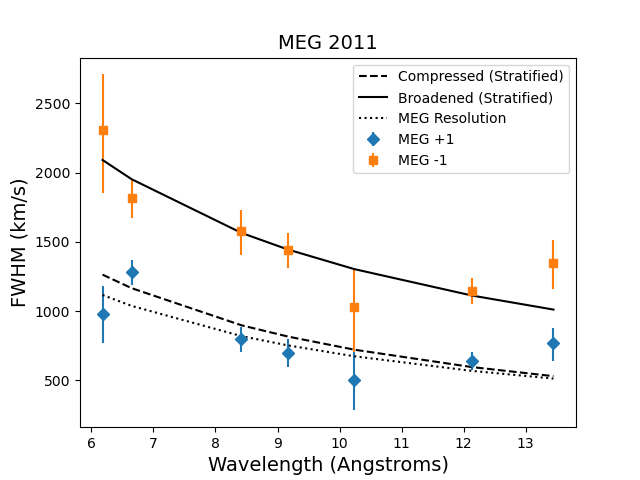} \\
 (a) & (b) \\
\end{tabular}
\begin{tabular}{c}
\includegraphics[width=0.5\textwidth]{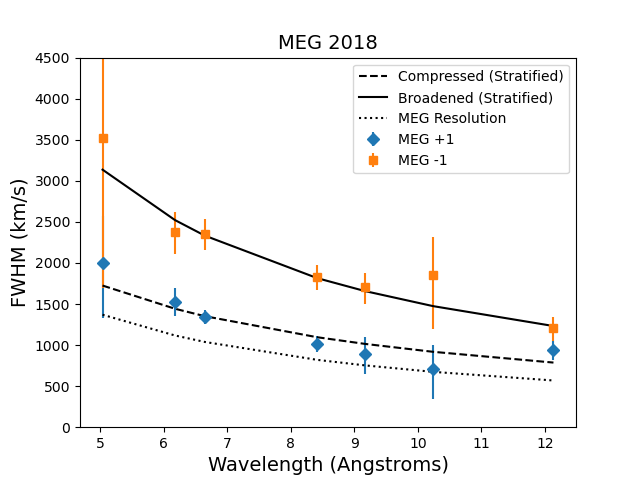}\\
 (c) \\
\end{tabular}
\caption{Measured line widths (FWHM) for the positive (diamonds) and negative (squares) arms for (a) MEG 2007, (b) MEG 2011, and (c) MEG 2018. The dotted curve represents the resolving power of the MEG. The dashed and solid curves represent compressed and broadened best-fit models, respectively for the shock stratified case. \label{fig:12}}
\end{figure*}
Here, the first term (2$\Delta \lambda_{0}$) accounts for the broadening caused by the angular extent of SNR 1987A, which is independent of wavelength, while the second term ($\pm 2z_{0}(\lambda / \lambda_0)^{\alpha} \lambda$) accounts for broadening (or compression) by bulk motion of the gas (broadening in the MEG -1 arm and compression in the MEG +1 arm due to the 45 $^{\circ}$-tilted geometry of the ER along the line of sight). We fit this line-broadening model (Equation \ref{eq:1}) to the measured FWHM (Section \ref{3.2}) as a function of wavelength for both the positive and negative gratings spectra. Free parameters of this fit are $\Delta \lambda_{0}$, $z_{0}$, and $\alpha$. An additional source of broadening due to the shock-heating of heavy ions was reported in \cite{2019NatAs...3..236M}. As the effects of such a thermal broadening are significantly smaller than those of the bulk motion of the gas \citep[see][]{2019NatAs...3..236M}, as a first approximation, we do not include it in this model. We discuss the effects of thermal broadening in Section \ref{4.5}.

Our best-fit parameters (with 1$\sigma$ errors) are: $\Delta \lambda_{0}$ =  0.0172 $\pm$ 0.0006 \AA{}, $z_{0}$ = 0.00049 $\pm$ 0.00007, $\alpha$ = -0.7 $\pm$ 0.5 (MEG 2011); $\Delta \lambda_{0}$ = 0.0204 $\pm$ 0.0004 \AA{}, $z_{0}$ = 0.00047 $\pm$ 0.00006, $\alpha$ = -1.3 $\pm$ 0.3 (MEG 2018). Using the grating dispersion of 0.0226 \AA{} arcsec$^{-1}$ for MEG, we derive the source ``half-size" $\theta_{R}$ to be 0$^{\prime\prime}$.76 $\pm$ 0$^{\prime\prime}$.01 (MEG 2011) and 0$^{\prime\prime}$.90 $\pm$ 0$^{\prime\prime}$.01 (MEG 2018). These angular extents are in good agreement (within statistical uncertainties) with those measured in the X-ray imaging analysis of the ACIS data taken in 2011 ($\sim$0$^{\prime \prime}$.79; \citealp{2016ApJ...829...40F}) and in 2018 ($\sim$0$^{\prime \prime}$.84; Ravi et al. 2021 in prep). We also apply this method for the MEG 2007 data, and obtain consistent (within statistical uncertainties) values of $\Delta \lambda_{0}$, $z_{0}$, and $\alpha$ when compared with the previously published values \citep{2009ApJ...692.1190Z}. For comparing shock velocities measured with emission line widths, we consider the three available sets of MEG ($\pm$1) spectra as representative data sets from 2007 till 2018, respectively. Best-fit model plots to line widths of MEG 2007, MEG 2011, and MEG 2018 are shown in Figure \ref{fig:12}.

\begin{figure}[h!]
\centering
\includegraphics[width=0.5\textwidth]{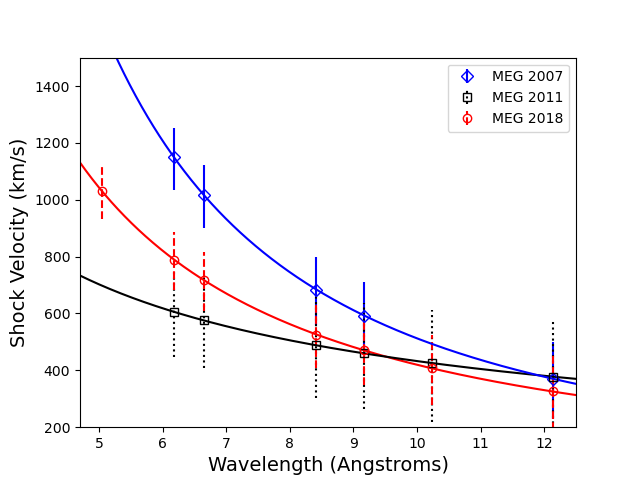} \\
\caption{Estimated shock velocities encountered by the X-ray emitting gas as a function of wavelength. Shock velocities measured from line widths of strong lines in the grating X-ray spectra of MEG 2007 (open blue diamonds), MEG 2011 (open black squares), and MEG 2018 (open red circles) are shown. Associated 1$\sigma$ uncertainties are represented as solid lines (MEG 2007), dotted lines (MEG 2011), and dashed lines (MEG 2018).  \label{fig:13}}

\end{figure}

Based on these best-fit values for $z_{0}$ and $\alpha$, we estimate the radial shock velocities encountered by the X-ray emitting gas in 2007 - 2018. For the simple case of a strong shock with an adiabatic index $\gamma$ = 5/3, the shock velocity ($v_{s}$) is related to the bulk velocity ($v_{b}$) of the shock-heated gas as $v_{s}$= (4/3)$v_{b}$. Assuming a radially expanding circular ring with velocity $v_{b}$, we adopt an average value for the azimuthal $\phi$ ($\sin \phi_{mean}$ = $2/\pi$, where 0$\leq \phi \leq\pi$/2) and inclination angle, \textit{i} = 45$^{\circ}$, to account for the geometric effects from viewing angles. Then, the shock velocity as a function of wavelength is
\begin{equation} \label{eq:2}
    v_{s} = \frac{4}{3}\times 2z_{0} \times 1.10c \times (\lambda/\lambda_0)^{\alpha}
\end{equation}  
where \textit{c} is the velocity of light. $\lambda_{0}$ is a fiducial wavelength in our broadband range, for which we choose $\lambda_{0} =$ 10 \AA{} (roughly a mean between 4.5 and 15 \AA{}). The derived average shock velocities are: $\sim$431$(\lambda/10)^{-0.7}$ km s$^{-1}$ (MEG 2011) and $\sim$413$(\lambda/10)^{-1.3}$ km s$^{-1}$ (MEG 2018). In Figure \ref{fig:13}, we show the estimated shock velocity profile as a function of wavelength, plotted for MEG 2007, MEG 2011, and MEG 2018. Despite relatively large errors, the average shock velocities decreased between 2007 and 2011, which is a continuation of the trend detected in 2004 - 2007 \citep{2009ApJ...692.1190Z}. The shock velocities increase between 2011 and 2018, particularly at short wavelengths ($\lambda$ $<$ 9 \AA{}) for Mg, Si, and S ions (Figure \ref{fig:13}). As an example, shock velocities encountering Si ions (6 - 7 \AA{}) are in the range: $\sim$553 - 616 km s$^{-1}$ (MEG 2011) and $\sim$656 - 802 km s$^{-1}$ (MEG 2018). These faster shocks since 2011 are consistent with the increase in the post-shock electron temperature (see Section \ref{4.1}). A caveat in this method is that, while we are assuming a single bulk velocity from each line, these lines more likely form from multi-velocity gas flows.
 
For a fast moving shock (with the velocity $v_{s}$) entering a stationary packet of gas, the immediate post-shock ion temperature ($kT_{i}$) can be expressed as $kT_{i} \approx 1.4(v_{s}/1000$ km s$^{-1})^2$ keV, where the assumed mean molecular mass of the gas ($\mu$) is 0.72 for the abundances of SNR 1987A. Based on the shock velocity ranges encountered by Si ions, their post shock ion temperatures are: 0.42 - 0.53 keV (MEG 2011) and 0.60 - 0.90 keV (MEG 2018). In the non-equilibrium plasma condition, we find that these ion temperatures from our simple line profile modeling are still significantly lower than our estimated post-shock electron temperature ranges (from broadband spectral model fits): $\sim$0.6 - 1.9 keV (MEG 2011) and $\sim$0.8 - 2.4 keV (MEG 2018). This discrepancy suggests that emission from reflected shocks in the ER (heating the inter-clump regions multiple times) may significantly contribute in the observed X-ray spectrum as proposed by \cite{2009ApJ...692.1190Z}. The increase in the ion temperatures from 2011 to 2018 may indicate that emission from the shocked low-density gas outside of the ER contribute in the observed X-ray spectrum more in 2018 than in 2011.\\

\subsection{Thermal Broadening of X-ray Emission Lines \label{4.5}}
Based on the high-resolution HETG (MEG) spectrum (taken in 2011) and hydrodynamic (HD) models of SNR 1987A \citep{2015ApJ...810..168O}, evidence for thermal broadening of X-ray emission lines was reported  \citep{2019NatAs...3..236M}. Thermal broadening can be modeled by a Gaussian, whose width is directly proportional to the ion temperature and inversely proportional to the ion mass. We assume that the ion temperature is equal to the product of atomic mass of ions and proton temperature \citep{2019NatAs...3..236M}. This is because at the shock front, the heavy ions have been heated up to a temperature proportional to their mass. In such a case, the thermal motion of all ions should be relatively the same, regardless of their mass. Thus for simplicity, we assume a similar functional form to that of our Doppler broadening term, for which the thermal broadening parameter ($z_{1}$) is included as an additional broadening term. We modify our model (Equation \ref{eq:1}) to include thermal broadening, and thus, the overall broadening is expressed as:

\begin{equation} \label{eq:3}
    \Delta \lambda_{total} = \sqrt{(2\Delta \lambda_{0} \pm 2z_{0}(\lambda / \lambda_0)^{\alpha} \lambda )^{2} + (2z_{1}(\lambda / \lambda_0)^{\alpha}) \lambda)^{2}}
\end{equation} 
where the plus (minus) sign refers to the negative (positive) order of the MEG spectrum. Free parameters in the fit are $\Delta \lambda_{0}$, $z_{0}$, $z_{1}$, and $\alpha$. We define the broadening due to the third term in Equation \ref{eq:3} per \AA{} (i.e., $2z_{1}(\lambda / \lambda_0)^{\alpha}$) as the thermal broadening contribution. Observed line widths are significantly broader in the negative dispersion arm than in their positive counterpart (see Section \ref{3.2}). Therefore, the relative contribution of thermal broadening (which is the same in the two orders) is higher in the positive arm dispersion spectrum. \cite{2019NatAs...3..236M} showed that, because of this effect, thermal broadening can be detected with high statistical significance in the positive order (MEG +1), while it is barely needed in the negative order (MEG -1).

\begin{figure*}[h!]
\begin{center}
\begin{tabular}{cc}
\hbox{\hspace{-0.5cm}\includegraphics[width=0.48\textwidth]{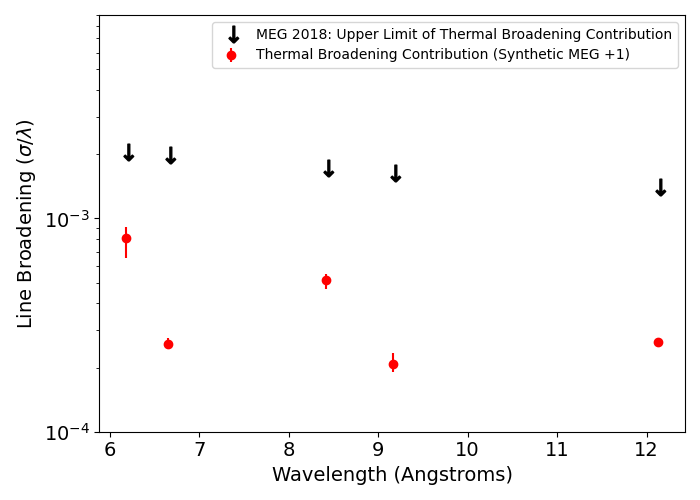}} &
\includegraphics[width=0.492\textwidth]{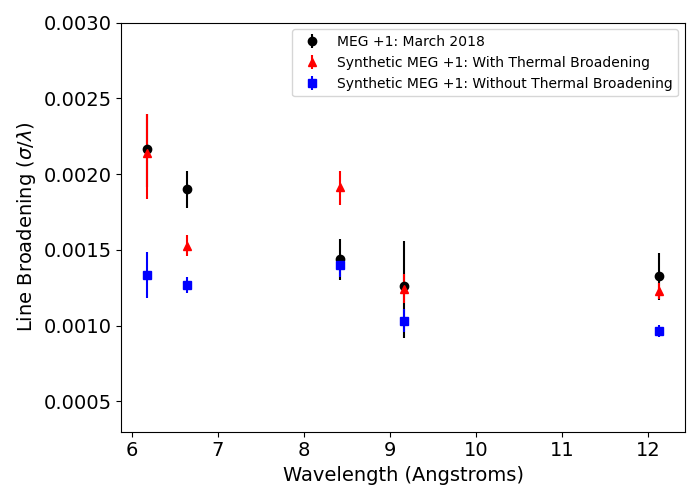} \\
(a) Thermal Broadening Contribution: Data v/s Synthetic & (b) Width Comparison: Data v/s Synthetic \\[3 pt]
\end{tabular}
\caption{(a) Upper limits (90$\%$ C.L.) of thermal broadening contribution at individual spectral line wavelengths are shown as black arrows. Predicted thermal broadening contribution (difference between widths of spectral lines with and without thermal broadening) for individual synthetic spectral lines in 2018 (red circles). (b) Individual spectral line widths measured in our synthetic spectra simulated for 2018 (MEG +1). Red triangles and blue squares are line widths with and without thermal broadening, respectively. Our measured line widths based on the MEG +1 spectrum observed in 2018 are shown with black circles.  \label{fig:14}}
\end{center}
\end{figure*}

In agreement with these findings, we find that thermal broadening is not statistically needed when including both the positive and negative first-order dispersed MEG 2018 spectra in our fits with the model described by Equation \ref{3}. Thus, we place an upper limit of $z_{1} <$ 0.0018 (90 $\%$ C.L.) on the thermal broadening parameter with $\alpha$ $\sim$ -0.57. The wavelength-dependent upper limits on the thermal broadening contribution in 2018 are expressed as 0.0036 $(\lambda / 10)^{-0.57}$. In Figure \ref{fig:14}a, we show the upper limits on thermal broadening in the observed individual lines based on our MEG 2018 data. The assumed thermal broadening in our synthetic spectrum (as described below) is lower than the observed upper limits in 2018, which is consistent with the non-detection of thermal broadening in 2018. 

Based on a similar approach to that adopted in \cite{2019NatAs...3..236M}, we synthesized MEG +1 X-ray spectra of SNR 1987A as of 2018 from the HD simulation presented in \cite{2015ApJ...810..168O}. In this HD simulation, the thermal condition of the shocked gas is calculated self-consistently in each cell of the spatial domain during the whole evolution from the SN event to the remnant interacting with the inhomogeneous CSM. For comparisons, we considered two different cases of including and excluding the thermal broadening effects in the synthesis of the X-ray spectra. We note that these HD models predicted a significant  contribution from the reverse-shocked high velocity ejecta in the synthetic X-ray emission spectrum of SNR 1987A by 2018. This ejecta contribution resulted in previously-unseen broad wings in our synthetic spectral lines calculated for 2018, even without the thermal broadening effects. However, because our deep HETG spectrum  taken in 2018 shows no evidence for enhanced elemental abundances (see Section \ref{3.1}), SNR 1987A has unlikely arrived at such an ejecta-dominated phase yet. Thus, we fit these extra broad wing features in the synthetic spectra of 2018 with a separate Gaussian component. We exclude this extra Gaussian component in our discussion. 

Figure \ref{fig:14}b shows the comparisons between the observed line widths and the synthetic spectral line widths (with and without thermal broadening), for MEG +1 in 2018. We note that including thermal broadening in the HD simulated spectrum better describes these observed line widths (in agreement with \cite{2019NatAs...3..236M}, see Figure \ref{fig:14}b, with an exception at $\sim$8.5 \AA{}). Detailed studies of thermal broadening contributions will be important for the forthcoming generation of X-ray spectrometers based on microcalorimeters.\\

\subsection{Metal Abundances \label{4.6}}

HD/MHD models of SN 1987A suggest that the contribution of the reverse-shocked outer layers of ejecta would soon ($\sim$35 years after the SN) become dominant in the soft (0.5 - 2 keV) X-ray band \citep{2015ApJ...810..168O, 2019A&A...622A..73O, 2020A&A...636A..22O}. These models predict that such an ejecta-dominated phase will be marked by a significant change in the observed X-ray spectrum, evidently showing strong enhancements in the individual line fluxes and thus in the metal abundances (compared to those of the ER emission). Our estimated metal abundances based on the broadband spectral model fits are consistent (within statistical uncertainties) with those for the ER as measured in previous works \citep{2006ApJ...645..293Z, 2008ApJ...676L.131D, 2009ApJ...692.1190Z}. Thus, we conclude that the onset of such an ejecta-dominated era in the evolution of SNR 1987A has not been ensued as of 2018. We note that the detection of Fe K emission lines possibly from the reverse-shocked overabundant ejecta has been suggested based on earlier \textit{XMM-Newton} observations \citep{Sun_2021}. We cannot confirm the detection of such Fe K lines based on our deep \textit{Chandra} HETG observations due to the low responses of the HETG spectrometer in the $\sim$6 keV energy band.\\

\subsection{Hard X-rays from SNR 1987A \label{4.7}}
Recently, \textit{NuSTAR} observations of SNR 1987A showed evidence for hard X-ray emission up to \textit{E} $\sim$ 20 keV \citep{2021ApJ...916...76A,Greco_2021}. The \textit{NuSTAR}-detected hard X-ray spectrum may be attributed to either nonthermal \citep[PL with photon index $\Gamma \sim$ 2.5,][]{Greco_2021} or thermal \citep[\textit{kT} $\sim$ 4 keV,][]{2021ApJ...916...76A} origins. Our \textit{Chandra} gratings data do not have significant count statistics at \textit{E} \small$\geq$ 5 keV and a third hard component X-ray emission is not statistically required to model our \textit{Chandra} HETG/LETG spectra of SNR 1987A. In fact, the contribution from the third component (either the nonthermal PL component with $\Gamma \sim$ 2.5 in \citealp{Greco_2021} or the thermal component with \textit{kT} $\sim$ 4 keV in \citealp{2021ApJ...916...76A}) in our observed \textit{Chandra} HETG/LETG bandpass is negligible ($\leq$ 4$\%$ of the observed flux in the 0.5 - 5 keV band), and thus it does not affect our conclusions. Discussing the true nature of the \textit{NuSTAR}-detected hard X-ray emission at \textit{E} \small$\geq$ 10 keV from SNR 1987A is beyond the scope of this work. Nonetheless, we note that electron temperatures estimated for the soft X-ray spectrum (\textit{E} $\sim$ 0.5 - 8 keV) in \citealp{Greco_2021} (\textit{kT} $\sim$ 0.6 keV and 2.6 keV) are in better agreement with our results than those in \citealp{2021ApJ...916...76A} (\textit{kT} $\sim$ 0.5 keV and 0.9 keV), although these comparisons need to be taken with cautions due to different spectral modeling among them (i.e., two-component model fits in this work vs three-component models in \citealp{Greco_2021} and \citealp{2021ApJ...916...76A}).\\

\section{Conclusions \label{5}} 

Based on our deep ($\sim$312 ks) \textit{Chandra} HETG observation taken in March 2018, we perform a detailed spectral analysis of the high-resolution dispersed X-ray spectrum of SNR 1987A. We also re-analyze the archival \textit{Chandra} HETG and LETG data taken in 2004, 2007, and 2011 with similarly deep exposures ($\sim$170 - 350 ks). Combining these results, we present the spectral evolution of the X-ray emitting hot gas in SNR 1987A from 2004 to 2018 as follows:
\begin{itemize}

    \item Changing volume emission measures derived from our broadband spectral model fits for both of the soft component and the hard component emission suggest a decreasing density profile for the X-ray emitting gas since 2011. Our estimated radial density profiles are in plausible agreement with a transitionary phase of the SNR evolution, as the shock leaves the ER and propagates into a red supergiant wind that was created by the massive progenitor star shortly before the SN explosion.
    \item Our broadband fits of the two-component shock models to the X-ray grating spectra of SNR 1987A show that the average post-shock electron temperatures for the soft component, which had stayed nearly constant ($\sim$0.6 keV) until 2011, has increased (by $\sim$38$\%$) between 2011 and 2018. A cooling trend observed for the average post-shock electron temperatures for the hard component, which continued until 2007 has been reversed, and the hard component electron temperature has increased (by $\sim$28$\%$) between 2007 and 2018. These results indicate that shocks responsible for both of the soft and hard component X-ray emission is interacting with less dense CSM in 2018 than before. 
    \item Decreasing He$\alpha$ to Ly$\alpha$ line flux ratios and increasing shock velocities derived from the individual line profiles of Si and Mg ions are consistent with the increasing electron temperatures as inferred in our broadband spectral model fits. 
    \item We observe no significant abundance evolution as of 2018. This indicates that the observed X-ray spectrum of SNR 1987A is still dominated by emission from the shocked ER with some contributions from the shocked CSM beyond it as of 2018. The contribution from the reverse-shocked metal-rich SN ejecta in the observed X-ray spectrum is not significant yet. 

\end{itemize}

\section*{}
 We thank the anonymous referee for helpful comments to improve this manuscript. This work was supported in part by NASA through \textit{Chandra} grant G08-19057X. SO and MM acknowledge financial contribution from the PRIN INAF 2019 grant ``From massive stars to supernovae and supernova remnants: driving mass, energy and cosmic rays in our Galaxy" and the INAF mainstream program ``Understanding particle acceleration in galactic sources in the CTA era."


\bibliography{draft}

\begin{thebibliography}{}
\expandafter\ifx\csname natexlab\endcsname\relax\def\natexlab#1{#1}\fi
\providecommand{\url}[1]{\href{#1}{#1}}
\providecommand{\dodoi}[1]{doi:~\href{http://doi.org/#1}{\nolinkurl{#1}}}
\providecommand{\doeprint}[1]{\href{http://ascl.net/#1}{\nolinkurl{http://ascl.net/#1}}}
\providecommand{\doarXiv}[1]{\href{https://arxiv.org/abs/#1}{\nolinkurl{https://arxiv.org/abs/#1}}}

\bibitem[{{Alp} {et~al.}(2021){Alp}, {Larsson}, \&
  {Fransson}}]{2021ApJ...916...76A}
{Alp}, D., {Larsson}, J., \& {Fransson}, C. 2021, \apj, 916, 76,
  \dodoi{10.3847/1538-4357/ac052d}

\bibitem[{{Anders} \& {Grevesse}(1989)}]{1989GeCoA..53..197A}
{Anders}, E., \& {Grevesse}, N. 1989, \gca, 53, 197,
  \dodoi{10.1016/0016-7037(89)90286-X}

\bibitem[{{Arendt} {et~al.}(2016){Arendt}, {Dwek}, {Bouchet}, {Danziger},
  {Frank}, {Gehrz}, {Park}, \& {Woodward}}]{2016AJ....151...62A}
{Arendt}, R.~G., {Dwek}, E., {Bouchet}, P., {et~al.} 2016, \aj, 151, 62,
  \dodoi{10.3847/0004-6256/151/3/62}

\bibitem[{{Arendt} {et~al.}(2020){Arendt}, {Dwek}, {Bouchet}, {John Danziger},
  {Gehrz}, {Park}, \& {Woodward}}]{2020ApJ...890....2A}
---. 2020, \apj, 890, 2, \dodoi{10.3847/1538-4357/ab660f}

\bibitem[{{Arnaud}(1996)}]{1996ASPC..101...17A}
{Arnaud}, K.~A. 1996, in Astronomical Society of the Pacific Conference Series,
  Vol. 101, Astronomical Data Analysis Software and Systems V, ed. G.~H.
  {Jacoby} \& J.~{Barnes}, 17

\bibitem[{Arnett {et~al.}(1989)Arnett, Bahcall, Kirshner, \&
  Woosley}]{doi:10.1146/annurev.aa.27.090189.003213}
Arnett, W.~D., Bahcall, J.~N., Kirshner, R.~P., \& Woosley, S.~E. 1989, Annual
  Review of Astronomy and Astrophysics, 27, 629,
  \dodoi{10.1146/annurev.aa.27.090189.003213}

\bibitem[{{Asplund} {et~al.}(2009){Asplund}, {Grevesse}, {Sauval}, \&
  {Scott}}]{2009ARA&A..47..481A}
{Asplund}, M., {Grevesse}, N., {Sauval}, A.~J., \& {Scott}, P. 2009, \araa, 47,
  481, \dodoi{10.1146/annurev.astro.46.060407.145222}

\bibitem[{{Borkowski} {et~al.}(1997{\natexlab{a}}){Borkowski}, {Blondin}, \&
  {McCray}}]{1997ApJ...476L..31B}
{Borkowski}, K.~J., {Blondin}, J.~M., \& {McCray}, R. 1997{\natexlab{a}},
  \apjl, 476, L31, \dodoi{10.1086/310487}

\bibitem[{{Borkowski} {et~al.}(1997{\natexlab{b}}){Borkowski}, {Blondin}, \&
  {McCray}}]{1997ApJ...477..281B}
---. 1997{\natexlab{b}}, \apj, 477, 281, \dodoi{10.1086/303691}

\bibitem[{{Borkowski} {et~al.}(2001){Borkowski}, {Lyerly}, \&
  {Reynolds}}]{2001ApJ...548..820B}
{Borkowski}, K.~J., {Lyerly}, W.~J., \& {Reynolds}, S.~P. 2001, \apj, 548, 820,
  \dodoi{10.1086/319011}

\bibitem[{{Bray} {et~al.}(2020){Bray}, {Burrows}, {Park}, \&
  {Ravi}}]{2020ApJ...899...21B}
{Bray}, E., {Burrows}, D.~N., {Park}, S., \& {Ravi}, A.~P. 2020, \apj, 899, 21,
  \dodoi{10.3847/1538-4357/ab9c9e}

\bibitem[{{Brinkman} {et~al.}(2000){Brinkman}, {Gunsing}, {Kaastra}, {van der
  Meer}, {Mewe}, {Paerels}, {Raassen}, {van Rooijen}, {Br{\"a}uninger},
  {Burkert}, {Burwitz}, {Hartner}, {Predehl}, {Ness}, {Schmitt}, {Drake},
  {Johnson}, {Juda}, {Kashyap}, {Murray}, {Pease}, {Ratzlaff}, \&
  {Wargelin}}]{2000ApJ...530L.111B}
{Brinkman}, A.~C., {Gunsing}, C.~J.~T., {Kaastra}, J.~S., {et~al.} 2000, \apjl,
  530, L111, \dodoi{10.1086/312504}

\bibitem[{{Burrows} {et~al.}(1995){Burrows}, {Krist}, {Hester}, {Sahai},
  {Trauger}, {Stapelfeldt}, {Gallagher}, {Ballester}, {Casertano}, {Clarke},
  {Crisp}, {Evans}, {Griffiths}, {Hoessel}, {Holtzman}, {Mould}, {Scowen},
  {Watson}, \& {Westphal}}]{1995ApJ...452..680B}
{Burrows}, C.~J., {Krist}, J., {Hester}, J.~J., {et~al.} 1995, \apj, 452, 680,
  \dodoi{10.1086/176339}

\bibitem[{{Burrows} {et~al.}(2000){Burrows}, {Michael}, {Hwang}, {McCray},
  {Chevalier}, {Petre}, {Garmire}, {Holt}, \& {Nousek}}]{2000ApJ...543L.149B}
{Burrows}, D.~N., {Michael}, E., {Hwang}, U., {et~al.} 2000, \apjl, 543, L149,
  \dodoi{10.1086/317271}

\bibitem[{{Canizares} {et~al.}(2000){Canizares}, {Huenemoerder}, {Davis},
  {Dewey}, {Flanagan}, {Houck}, {Markert}, {Marshall}, {Schattenburg},
  {Schulz}, {Wise}, {Drake}, \& {Brickhouse}}]{2000ApJ...539L..41C}
{Canizares}, C.~R., {Huenemoerder}, D.~P., {Davis}, D.~S., {et~al.} 2000,
  \apjl, 539, L41, \dodoi{10.1086/312823}

\bibitem[{{Cendes} {et~al.}(2018){Cendes}, {Gaensler}, {Ng}, {Zanardo},
  {Staveley-Smith}, \& {Tzioumis}}]{2018ApJ...867...65C}
{Cendes}, Y., {Gaensler}, B.~M., {Ng}, C.~Y., {et~al.} 2018, \apj, 867, 65,
  \dodoi{10.3847/1538-4357/aae261}

\bibitem[{{Chevalier}(1982)}]{1982ApJ...258..790C}
{Chevalier}, R.~A. 1982, \apj, 258, 790, \dodoi{10.1086/160126}

\bibitem[{{Dewey} {et~al.}(2012){Dewey}, {Dwarkadas}, {Haberl}, {Sturm}, \&
  {Canizares}}]{2012ApJ...752..103D}
{Dewey}, D., {Dwarkadas}, V.~V., {Haberl}, F., {Sturm}, R., \& {Canizares},
  C.~R. 2012, \apj, 752, 103, \dodoi{10.1088/0004-637X/752/2/103}

\bibitem[{{Dewey} {et~al.}(2008){Dewey}, {Zhekov}, {McCray}, \&
  {Canizares}}]{2008ApJ...676L.131D}
{Dewey}, D., {Zhekov}, S.~A., {McCray}, R., \& {Canizares}, C.~R. 2008, \apjl,
  676, L131, \dodoi{10.1086/587549}

\bibitem[{Dickey \& Lockman(1990)}]{doi:10.1146/annurev.aa.28.090190.001243}
Dickey, J.~M., \& Lockman, F.~J. 1990, Annual Review of Astronomy and
  Astrophysics, 28, 215, \dodoi{10.1146/annurev.aa.28.090190.001243}

\bibitem[{{Eastman} \& {Kirshner}(1989)}]{1989ApJ...347..771E}
{Eastman}, R.~G., \& {Kirshner}, R.~P. 1989, \apj, 347, 771,
  \dodoi{10.1086/168168}

\bibitem[{{Fitzpatrick} \& {Walborn}(1990)}]{1990AJ.....99.1483F}
{Fitzpatrick}, E.~L., \& {Walborn}, N.~R. 1990, \aj, 99, 1483,
  \dodoi{10.1086/115432}

\bibitem[{{France} {et~al.}(2011){France}, {McCray}, {Penton}, {Kirshner},
  {Challis}, {Laming}, {Bouchet}, {Chevalier}, {Garnavich}, {Fransson}, {Heng},
  {Larsson}, {Lawrence}, {Lundqvist}, {Panagia}, {Pun}, {Smith}, {Sollerman},
  {Sonneborn}, {Sugerman}, \& {Wheeler}}]{2011ApJ...743..186F}
{France}, K., {McCray}, R., {Penton}, S.~V., {et~al.} 2011, \apj, 743, 186,
  \dodoi{10.1088/0004-637X/743/2/186}

\bibitem[{{Frank} {et~al.}(2016){Frank}, {Zhekov}, {Park}, {McCray}, {Dwek}, \&
  {Burrows}}]{2016ApJ...829...40F}
{Frank}, K.~A., {Zhekov}, S.~A., {Park}, S., {et~al.} 2016, \apj, 829, 40,
  \dodoi{10.3847/0004-637X/829/1/40}

\bibitem[{{Fransson} {et~al.}(1996){Fransson}, {Lundqvist}, \&
  {Chevalier}}]{1996ApJ...461..993F}
{Fransson}, C., {Lundqvist}, P., \& {Chevalier}, R.~A. 1996, \apj, 461, 993,
  \dodoi{10.1086/177119}

\bibitem[{{Fransson} {et~al.}(2015){Fransson}, {Larsson}, {Migotto}, {Pesce},
  {Challis}, {Chevalier}, {France}, {Kirshner}, {Leibundgut}, {Lundqvist},
  {McCray}, {Spyromilio}, {Taddia}, {Jerkstrand}, {Mattila}, {Smith},
  {Sollerman}, {Wheeler}, {Crotts}, {Garnavich}, {Heng}, {Lawrence}, {Panagia},
  {Pun}, {Sonneborn}, \& {Sugerman}}]{2015ApJ...806L..19F}
{Fransson}, C., {Larsson}, J., {Migotto}, K., {et~al.} 2015, \apjl, 806, L19,
  \dodoi{10.1088/2041-8205/806/1/L19}

\bibitem[{{Fruscione} {et~al.}(2006){Fruscione}, {McDowell}, {Allen},
  {Brickhouse}, {Burke}, {Davis}, {Durham}, {Elvis}, {Galle}, {Harris},
  {Huenemoerder}, {Houck}, {Ishibashi}, {Karovska}, {Nicastro}, {Noble},
  {Nowak}, {Primini}, {Siemiginowska}, {Smith}, \&
  {Wise}}]{2006SPIE.6270E..1VF}
{Fruscione}, A., {McDowell}, J.~C., {Allen}, G.~E., {et~al.} 2006, in Society
  of Photo-Optical Instrumentation Engineers (SPIE) Conference Series, Vol.
  6270, Society of Photo-Optical Instrumentation Engineers (SPIE) Conference
  Series, ed. D.~R. {Silva} \& R.~E. {Doxsey}, 62701V,
  \dodoi{10.1117/12.671760}

\bibitem[{{Gabriel} \& {Jordan}(1969)}]{1969MNRAS.145..241G}
{Gabriel}, A.~H., \& {Jordan}, C. 1969, \mnras, 145, 241,
  \dodoi{10.1093/mnras/145.2.241}

\bibitem[{{Garmire} {et~al.}(2003){Garmire}, {Bautz}, {Ford}, {Nousek}, \&
  {Ricker}}]{2003SPIE.4851...28G}
{Garmire}, G.~P., {Bautz}, M.~W., {Ford}, P.~G., {Nousek}, J.~A., \& {Ricker},
  George~R., J. 2003, in Society of Photo-Optical Instrumentation Engineers
  (SPIE) Conference Series, Vol. 4851, X-Ray and Gamma-Ray Telescopes and
  Instruments for Astronomy., ed. J.~E. {Truemper} \& H.~D. {Tananbaum},
  28--44, \dodoi{10.1117/12.461599}

\bibitem[{Greco {et~al.}(2021)Greco, Miceli, Orlando, Olmi, Bocchino, Nagataki,
  Ono, Dohi, \& Peres}]{Greco_2021}
Greco, E., Miceli, M., Orlando, S., {et~al.} 2021, The Astrophysical Journal,
  908, L45, \dodoi{10.3847/2041-8213/abdf5a}

\bibitem[{{Gr{\"o}ningsson} {et~al.}(2008){Gr{\"o}ningsson}, {Fransson},
  {Leibundgut}, {Lundqvist}, {Challis}, {Chevalier}, \&
  {Spyromilio}}]{2008A&A...492..481G}
{Gr{\"o}ningsson}, P., {Fransson}, C., {Leibundgut}, B., {et~al.} 2008, \aap,
  492, 481, \dodoi{10.1051/0004-6361:200810551}

\bibitem[{{Haberl} {et~al.}(2006){Haberl}, {Geppert}, {Aschenbach}, \&
  {Hasinger}}]{2006A&A...460..811H}
{Haberl}, F., {Geppert}, U., {Aschenbach}, B., \& {Hasinger}, G. 2006, \aap,
  460, 811, \dodoi{10.1051/0004-6361:20066198}

\bibitem[{{Helder} {et~al.}(2013){Helder}, {Broos}, {Dewey}, {Dwek}, {McCray},
  {Park}, {Racusin}, {Zhekov}, \& {Burrows}}]{2013ApJ...764...11H}
{Helder}, E.~A., {Broos}, P.~S., {Dewey}, D., {et~al.} 2013, \apj, 764, 11,
  \dodoi{10.1088/0004-637X/764/1/11}

\bibitem[{{Heng} {et~al.}(2008){Heng}, {Haberl}, {Aschenbach}, \&
  {Hasinger}}]{2008ApJ...676..361H}
{Heng}, K., {Haberl}, F., {Aschenbach}, B., \& {Hasinger}, G. 2008, \apj, 676,
  361, \dodoi{10.1086/526517}

\bibitem[{{HI4PI Collaboration} {et~al.}(2016){HI4PI Collaboration}, {Ben
  Bekhti}, {Fl{\"o}er}, {Keller}, {Kerp}, {Lenz}, {Winkel}, {Bailin},
  {Calabretta}, {Dedes}, {Ford}, {Gibson}, {Haud}, {Janowiecki}, {Kalberla},
  {Lockman}, {McClure-Griffiths}, {Murphy}, {Nakanishi}, {Pisano}, \&
  {Staveley-Smith}}]{2016A&A...594A.116H}
{HI4PI Collaboration}, {Ben Bekhti}, N., {Fl{\"o}er}, L., {et~al.} 2016, \aap,
  594, A116, \dodoi{10.1051/0004-6361/201629178}

\bibitem[{{Hughes} {et~al.}(1998){Hughes}, {Hayashi}, \&
  {Koyama}}]{1998ApJ...505..732H}
{Hughes}, J.~P., {Hayashi}, I., \& {Koyama}, K. 1998, \apj, 505, 732,
  \dodoi{10.1086/306202}

\bibitem[{{Kalberla} {et~al.}(2005){Kalberla}, {Burton}, {Hartmann}, {Arnal},
  {Bajaja}, {Morras}, \& {P{\"o}ppel}}]{2005A&A...440..775K}
{Kalberla}, P.~M.~W., {Burton}, W.~B., {Hartmann}, D., {et~al.} 2005, \aap,
  440, 775, \dodoi{10.1051/0004-6361:20041864}

\bibitem[{{Larsson} {et~al.}(2019){Larsson}, {Fransson}, {Alp}, {Challis},
  {Chevalier}, {France}, {Kirshner}, {Lawrence}, {Leibundgut}, {Lundqvist},
  {Mattila}, {Migotto}, {Sollerman}, {Sonneborn}, {Spyromilio}, {Suntzeff}, \&
  {Wheeler}}]{2019ApJ...886..147L}
{Larsson}, J., {Fransson}, C., {Alp}, D., {et~al.} 2019, \apj, 886, 147,
  \dodoi{10.3847/1538-4357/ab4ff2}

\bibitem[{{Lundqvist} \& {Fransson}(1991)}]{1991ApJ...380..575L}
{Lundqvist}, P., \& {Fransson}, C. 1991, \apj, 380, 575, \dodoi{10.1086/170615}

\bibitem[{{Luo} \& {McCray}(1991)}]{1991ApJ...379..659L}
{Luo}, D., \& {McCray}, R. 1991, \apj, 379, 659, \dodoi{10.1086/170539}

\bibitem[{{Maggi} {et~al.}(2012){Maggi}, {Haberl}, {Sturm}, \&
  {Dewey}}]{2012A&A...548L...3M}
{Maggi}, P., {Haberl}, F., {Sturm}, R., \& {Dewey}, D. 2012, \aap, 548, L3,
  \dodoi{10.1051/0004-6361/201220595}

\bibitem[{{Maggi} {et~al.}(2016){Maggi}, {Haberl}, {Kavanagh}, {Sasaki},
  {Bozzetto}, {Filipovi{\'c}}, {Vasilopoulos}, {Pietsch}, {Points}, {Chu},
  {Dickel}, {Ehle}, {Williams}, \& {Greiner}}]{2016A&A...585A.162M}
{Maggi}, P., {Haberl}, F., {Kavanagh}, P.~J., {et~al.} 2016, \aap, 585, A162,
  \dodoi{10.1051/0004-6361/201526932}

\bibitem[{{Mattila} {et~al.}(2010){Mattila}, {Lundqvist}, {Gr{\"o}ningsson},
  {Meikle}, {Stathakis}, {Fransson}, \& {Cannon}}]{2010ApJ...717.1140M}
{Mattila}, S., {Lundqvist}, P., {Gr{\"o}ningsson}, P., {et~al.} 2010, \apj,
  717, 1140, \dodoi{10.1088/0004-637X/717/2/1140}

\bibitem[{McCray(1993)}]{doi:10.1146/annurev.aa.31.090193.001135}
McCray, R. 1993, Annual Review of Astronomy and Astrophysics, 31, 175,
  \dodoi{10.1146/annurev.aa.31.090193.001135}

\bibitem[{McCray \& Fransson(2016)}]{doi:10.1146/annurev-astro-082615-105405}
McCray, R., \& Fransson, C. 2016, Annual Review of Astronomy and Astrophysics,
  54, 19, \dodoi{10.1146/annurev-astro-082615-105405}

\bibitem[{{Miceli} {et~al.}(2019){Miceli}, {Orlando}, {Burrows}, {Frank},
  {Argiroffi}, {Reale}, {Peres}, {Petruk}, \& {Bocchino}}]{2019NatAs...3..236M}
{Miceli}, M., {Orlando}, S., {Burrows}, D.~N., {et~al.} 2019, Nature Astronomy,
  3, 236, \dodoi{10.1038/s41550-018-0677-8}

\bibitem[{{Michael} {et~al.}(1998){Michael}, {McCray}, {Pun}, {Borkowski},
  {Garnavich}, {Challis}, {Kirshner}, {Chevalier}, {Filippenko}, {Fransson},
  {Panagia}, {Phillips}, {Schmidt}, {Suntzeff}, \&
  {Wheeler}}]{1998ApJ...509L.117M}
{Michael}, E., {McCray}, R., {Pun}, C.~S.~J., {et~al.} 1998, \apjl, 509, L117,
  \dodoi{10.1086/311780}

\bibitem[{{Michael} {et~al.}(2002){Michael}, {Zhekov}, {McCray}, {Hwang},
  {Burrows}, {Park}, {Garmire}, {Holt}, \& {Hasinger}}]{2002ApJ...574..166M}
{Michael}, E., {Zhekov}, S., {McCray}, R., {et~al.} 2002, \apj, 574, 166,
  \dodoi{10.1086/340591}

\bibitem[{{Michael} {et~al.}(2003){Michael}, {McCray}, {Chevalier},
  {Filippenko}, {Lundqvist}, {Challis}, {Sugerman}, {Lawrence}, {Pun},
  {Garnavich}, {Kirshner}, {Crotts}, {Fransson}, {Li}, {Panagia}, {Phillips},
  {Schmidt}, {Sonneborn}, {Suntzeff}, {Wang}, \&
  {Wheeler}}]{2003ApJ...593..809M}
{Michael}, E., {McCray}, R., {Chevalier}, R., {et~al.} 2003, \apj, 593, 809,
  \dodoi{10.1086/376725}

\bibitem[{{Nymark} {et~al.}(2006){Nymark}, {Fransson}, \&
  {Kozma}}]{2006A&A...449..171N}
{Nymark}, T.~K., {Fransson}, C., \& {Kozma}, C. 2006, \aap, 449, 171,
  \dodoi{10.1051/0004-6361:20054169}

\bibitem[{{Orlando} {et~al.}(2015){Orlando}, {Miceli}, {Pumo}, \&
  {Bocchino}}]{2015ApJ...810..168O}
{Orlando}, S., {Miceli}, M., {Pumo}, M.~L., \& {Bocchino}, F. 2015, \apj, 810,
  168, \dodoi{10.1088/0004-637X/810/2/168}

\bibitem[{{Orlando} {et~al.}(2019){Orlando}, {Miceli}, {Petruk}, {Ono},
  {Nagataki}, {Aloy}, {Mimica}, {Lee}, {Bocchino}, {Peres}, \&
  {Guarrasi}}]{2019A&A...622A..73O}
{Orlando}, S., {Miceli}, M., {Petruk}, O., {et~al.} 2019, \aap, 622, A73,
  \dodoi{10.1051/0004-6361/201834487}

\bibitem[{{Orlando} {et~al.}(2020){Orlando}, {Ono}, {Nagataki}, {Miceli},
  {Umeda}, {Ferrand}, {Bocchino}, {Petruk}, {Peres}, {Takahashi}, \&
  {Yoshida}}]{2020A&A...636A..22O}
{Orlando}, S., {Ono}, M., {Nagataki}, S., {et~al.} 2020, \aap, 636, A22,
  \dodoi{10.1051/0004-6361/201936718}

\bibitem[{{Park} {et~al.}(2002){Park}, {Burrows}, {Garmire}, {Nousek},
  {McCray}, {Michael}, \& {Zhekov}}]{2002ApJ...567..314P}
{Park}, S., {Burrows}, D.~N., {Garmire}, G.~P., {et~al.} 2002, \apj, 567, 314,
  \dodoi{10.1086/338492}

\bibitem[{{Park} {et~al.}(2004){Park}, {Zhekov}, {Burrows}, {Garmire}, \&
  {McCray}}]{2004ApJ...610..275P}
{Park}, S., {Zhekov}, S.~A., {Burrows}, D.~N., {Garmire}, G.~P., \& {McCray},
  R. 2004, \apj, 610, 275, \dodoi{10.1086/421701}

\bibitem[{{Park} {et~al.}(2006){Park}, {Zhekov}, {Burrows}, {Garmire},
  {Racusin}, \& {McCray}}]{2006ApJ...646.1001P}
{Park}, S., {Zhekov}, S.~A., {Burrows}, D.~N., {et~al.} 2006, \apj, 646, 1001,
  \dodoi{10.1086/505023}

\bibitem[{{Park} {et~al.}(2005){Park}, {Zhekov}, {Burrows}, \&
  {McCray}}]{2005ApJ...634L..73P}
{Park}, S., {Zhekov}, S.~A., {Burrows}, D.~N., \& {McCray}, R. 2005, \apjl,
  634, L73, \dodoi{10.1086/498848}

\bibitem[{{Park} {et~al.}(2011){Park}, {Zhekov}, {Burrows}, {Racusin}, {Dewey},
  \& {McCray}}]{2011ApJ...733L..35P}
{Park}, S., {Zhekov}, S.~A., {Burrows}, D.~N., {et~al.} 2011, \apjl, 733, L35,
  \dodoi{10.1088/2041-8205/733/2/L35}

\bibitem[{{Plait} {et~al.}(1995){Plait}, {Lundqvist}, {Chevalier}, \&
  {Kirshner}}]{1995ApJ...439..730P}
{Plait}, P.~C., {Lundqvist}, P., {Chevalier}, R.~A., \& {Kirshner}, R.~P. 1995,
  \apj, 439, 730, \dodoi{10.1086/175213}

\bibitem[{{Pun} {et~al.}(2002){Pun}, {Michael}, {Zhekov}, {McCray},
  {Garnavich}, {Challis}, {Kirshner}, {Baron}, {Branch}, {Chevalier},
  {Filippenko}, {Fransson}, {Leibundgut}, {Lundqvist}, {Panagia}, {Phillips},
  {Schmidt}, {Sonneborn}, {Suntzeff}, {Wang}, \&
  {Wheeler}}]{2002ApJ...572..906P}
{Pun}, C. S.~J., {Michael}, E., {Zhekov}, S.~A., {et~al.} 2002, \apj, 572, 906,
  \dodoi{10.1086/340453}

\bibitem[{{Racusin} {et~al.}(2009){Racusin}, {Park}, {Zhekov}, {Burrows},
  {Garmire}, \& {McCray}}]{2009ApJ...703.1752R}
{Racusin}, J.~L., {Park}, S., {Zhekov}, S., {et~al.} 2009, \apj, 703, 1752,
  \dodoi{10.1088/0004-637X/703/2/1752}

\bibitem[{{Russell} \& {Dopita}(1992)}]{1992ApJ...384..508R}
{Russell}, S.~C., \& {Dopita}, M.~A. 1992, \apj, 384, 508,
  \dodoi{10.1086/170893}

\bibitem[{{Schenck} {et~al.}(2016){Schenck}, {Park}, \&
  {Post}}]{2016AJ....151..161S}
{Schenck}, A., {Park}, S., \& {Post}, S. 2016, \aj, 151, 161,
  \dodoi{10.3847/0004-6256/151/6/161}

\bibitem[{{Sturm} {et~al.}(2010){Sturm}, {Haberl}, {Aschenbach}, \&
  {Hasinger}}]{2010A&A...515A...5S}
{Sturm}, R., {Haberl}, F., {Aschenbach}, B., \& {Hasinger}, G. 2010, \aap, 515,
  A5, \dodoi{10.1051/0004-6361/200913317}

\bibitem[{Sun {et~al.}(2021)Sun, Vink, Chen, Zhou, Prokhorov, Pühlhofer, \&
  Malyshev}]{Sun_2021}
Sun, L., Vink, J., Chen, Y., {et~al.} 2021, The Astrophysical Journal, 916, 41,
  \dodoi{10.3847/1538-4357/ac033d}

\bibitem[{{Suzuki} {et~al.}(1993){Suzuki}, {Shigeyama}, \&
  {Nomoto}}]{1993A&A...274..883S}
{Suzuki}, T., {Shigeyama}, T., \& {Nomoto}, K. 1993, \aap, 274, 883

\bibitem[{{Willingale} {et~al.}(2013){Willingale}, {Starling}, {Beardmore},
  {Tanvir}, \& {O'Brien}}]{2013MNRAS.431..394W}
{Willingale}, R., {Starling}, R.~L.~C., {Beardmore}, A.~P., {Tanvir}, N.~R., \&
  {O'Brien}, P.~T. 2013, \mnras, 431, 394, \dodoi{10.1093/mnras/stt175}

\bibitem[{{Zhekov} {et~al.}(2005){Zhekov}, {McCray}, {Borkowski}, {Burrows}, \&
  {Park}}]{2005ApJ...628L.127Z}
{Zhekov}, S.~A., {McCray}, R., {Borkowski}, K.~J., {Burrows}, D.~N., \& {Park},
  S. 2005, \apjl, 628, L127, \dodoi{10.1086/432794}

\bibitem[{{Zhekov} {et~al.}(2006){Zhekov}, {McCray}, {Borkowski}, {Burrows}, \&
  {Park}}]{2006ApJ...645..293Z}
---. 2006, \apj, 645, 293, \dodoi{10.1086/504285}

\bibitem[{{Zhekov} {et~al.}(2009){Zhekov}, {McCray}, {Dewey}, {Canizares},
  {Borkowski}, {Burrows}, \& {Park}}]{2009ApJ...692.1190Z}
{Zhekov}, S.~A., {McCray}, R., {Dewey}, D., {et~al.} 2009, \apj, 692, 1190,
  \dodoi{10.1088/0004-637X/692/2/1190}

\bibitem[{{Zhekov} {et~al.}(2010){Zhekov}, {Park}, {McCray}, {Racusin}, \&
  {Burrows}}]{2010MNRAS.407.1157Z}
{Zhekov}, S.~A., {Park}, S., {McCray}, R., {Racusin}, J.~L., \& {Burrows},
  D.~N. 2010, \mnras, 407, 1157, \dodoi{10.1111/j.1365-2966.2010.16967.x}

\end{thebibliography}

\end{document}